\documentclass[letterpaper,english,reprint, notitlepage, aps,prb,floatfix,table,hyphenbreaks,hyphens]{revtex4-1}

\usepackage[T1]{fontenc}
\usepackage[latin9]{inputenc}
\usepackage{babel}
\usepackage{array}
\usepackage{verbatim}
\usepackage{longtable}
\usepackage{mathrsfs}
\usepackage{mathtools}
\usepackage{bm}
\usepackage{amsmath}
\usepackage{amssymb}
\usepackage{graphicx}
\usepackage[unicode=true,
 bookmarks=true,bookmarksnumbered=false,bookmarksopen=false,
 breaklinks=true,pdfborder={0 0 1},backref=false,colorlinks=false]
 {hyperref}
\hypersetup{
 pdfauthor={Zlatko K. Minev},
 breaklinks=true}
 \usepackage[hyphenbreaks]{breakurl}

\makeatletter

\pdfpageheight\paperheight
\pdfpagewidth\paperwidth

\providecommand{\tabularnewline}{\\}

\pdfoutput=1                    
\usepackage{etoolbox}			
\usepackage{titlesec}			
\usepackage[mathlines]{lineno}	
\usepackage{newtxmath}          

\usepackage{xcolor} 
\definecolor{HeaderColor}{rgb}{0.95, 0.95, 0.955}

\newcommand{\zkmDataTablePrep}{
	\setlength{\extrarowheight}{2pt}
	\setlength{\doublerulesep}{4pt}
	\addtolength{\tabcolsep}{4pt}
}

\DeclareMathOperator{\sign}{sign}
\DeclareMathOperator{\diag}{diag}
\DeclareMathOperator{\Diag}{Diag}

\usepackage{fancyhdr}
\usepackage{lipsum} 
\AtBeginDocument{\let\oldcontentsline\contentsline}
\newcommand{\notoccontentsline}[4]{}
\newcommand{\droptocpage}{\addtocontents{toc}{\let\protect\contentsline\protect\notoccontentsline}}
\newcommand{\incltocpage}{\addtocontents{toc}{\let\protect\contentsline\protect\oldcontentsline}}

\usepackage[titles]{tocloft}
\cftsetindents{part}{0.0in}{0.45in}  
\cftsetindents{section}{0.285in}{0.3in}
\cftsetindents{subsection}{0.4in}{0.35in}
\cftsetindents{subsubsection}{0.75in}{0.35in}

\makeatother

\begin{document}
\global\long\def\ket#1{\left|#1\right\rangle }%

\global\long\def\bra#1{\left\langle #1\right|}%

\global\long\def\braket#1#2{\left\langle #1\middle|#2\right\rangle }%

\global\long\def\ketbra#1#2{\left|#1\vphantom{#2}\right\rangle \left\langle \vphantom{#1}#2\right|}%

\global\long\def\kb#1#2{\left|#1\vphantom{#2}\right\rangle \left\langle \vphantom{#1}#2\right|}%

\global\long\def\braOket#1#2#3{\left\langle #1\middle|#2\middle|#3\right\rangle }%

\global\long\def\isdef{\coloneqq}%

\global\long\def\m#1{\mathrm{\mathbf{#1}}}%

\global\long\def\mh#1{\boldsymbol{\hat{\mathrm{#1}}}}%

\global\long\def\abs#1{\left|#1\right|}%

\global\long\def\normord#1{{:\mathrel{\mspace{1mu}#1\mspace{1mu}}:}}%

\global\long\def\diag{\operatorname{diag}}%

\global\long\def\Re{\operatorname{Re}}%

\global\long\def\ddt{\frac{\mathrm{d}}{\mathrm{d}t}}%

\global\long\def\phit{\m{\Phi}_{\mathrm{t}}}%

\global\long\def\phithat{\m{\hat{\Phi}}_{\mathrm{t}}}%

\global\long\def\phitd{\dot{\m{\Phi}}_{\mathrm{t}}}%

\global\long\def\phim{\m{\Phi}_{\mathrm{m}}}%

\global\long\def\phimhat{\m{\hat{\Phi}}_{\mathrm{m}}}%

\global\long\def\phimd{\dot{\m{\Phi}}_{\mathrm{m}}}%

\global\long\def\dashedph{s}%

\global\long\def\ejlin{\mathcal{E}_{j}^{\mathrm{lin}}}%

\global\long\def\ejnl{\mathcal{E}_{j}^{\mathrm{nl}}}%

\global\long\def\pml{p_{ml}}%

\title{Energy-participation quantization of Josephson circuits}
\author{Zlatko K.~Minev$^{1,*}$, Zaki~Leghtas$^{1,2}$, Shantanu O.~Mundhada$^{1,\dagger}$,
Lysander Christakis$^{1,\ddagger}$, Ioan M.~Pop$^{1,3}$, Michel
H.~Devoret$^{1}$}
\affiliation{$^{1}$Department of Applied Physics, Yale University, New Haven,
Connecticut 06511, USA}
\affiliation{$^{2}$Centre Automatique et Syst\`emes, Mines-ParisTech, PSL Research
University, 60 Bd Saint Michel, 75006 Paris, France}
\affiliation{$^{3}$IQMT, Karlsruhe~Institute~of~Technology, 76344 Eggenstein-Leopoldshafen,
Germany}
\affiliation{$^{*}$Current address: IBM T.J. Watson Research Center, Yorktown
Heights, New York 10598, USA; zlatko.minev@aya.yale.edu; www.zlatko-minev.com}
\affiliation{$^{\dagger}$Current address: Quantum Circuit Incorporated (QCI),
New Haven, CT 06511, USA}
\affiliation{$^{\ddagger}$Current address: Department of Physics, Princeton University,
Princeton, NJ 08540, USA}
\date{\today}
\begin{abstract}
Superconducting microwave circuits incorporating nonlinear devices,
such as Josephson junctions, are a leading platform for emerging quantum
technologies. Increasing circuit complexity further requires efficient
methods for the calculation and optimization of the spectrum, nonlinear
interactions, and dissipation in multi-mode distributed quantum circuits.
Here, we present a method based on the energy-participation ratio
(EPR) of a dissipative or nonlinear element in an electromagnetic
mode. The EPR, a number between zero and one, quantifies how much
of the mode energy is stored in each element. The EPRs obey universal
constraints and are calculated from one electromagnetic-eigenmode
simulation. They lead directly to the system quantum Hamiltonian and
dissipative parameters. The method provides an intuitive and simple-to-use
tool to quantize multi-junction circuits. We experimentally tested
this method on a variety of Josephson circuits, and demonstrated agreement
within several percents for nonlinear couplings and modal Hamiltonian
parameters, spanning five-orders of magnitude in energy, across a
dozen samples.
\end{abstract}
\maketitle

\makeatletter 

\renewcommand{\fnum@figure}{\textbf{Figure~\thefigure}}
\renewcommand{\fnum@table}{\textbf{Table~\thetable}}

\makeatother 
{
\titleformat*{\paragraph}{\bfseries\rmfamily}

\section{INTRODUCTION\label{sec:Introduction}}

\noindent Quantum information processing based on the control of microwave
electromagnetic fields in Josephson circuits is a promising platform
for both fundamental physics experiments and emerging quantum technologies
\citep{Devoret2013,GoogleSupremacy2019,Blais2020}. Key to the success
of this platform is the ability to quantitatively model the distributed
quantized electromagnetic modes of the system, their nonlinear interactions,
and their dissipation (see Fig.~1). This challenge is the subject
of intensifying interest \citep{Nigg2012,Bourassa2012,Solgun2014,Solgun2015,Smith2016,Gely2017,Malekakhlagh2017-Cutoff-Free,Pechal2017,Parra-Rodriguez2018,Parra-Rodriguez2018a,Ansari2018,Krupko2018,Malekakhlagh2018,Solgun2017,Petrescu2019,You2019-Koch,DiPaolo2019,Menke2021,Gely2020,Kyaw2020,Yan2020,Minev2021-lom,qiskit-metal-code},
as experimental architectures \citep{Barends2013,Minev2016,Brecht2016,Reagor2016-cavity,Gambetta2017,Rosenberg2017,Versluis2017,Naik2017,Krantz2019,Kjaergaard2019}
and nonlinear devices \citep{Josephson1962,Vijay2010,Kerman2010,Larsen2015,DeLange2015-nanowire,Janvier2015-atomic-point,Maleeva2018,Wang2018}
scale in both complexity and diversity.

In this paper, we introduce a circuit quantization method based on
the concept of the energy-participation ratio (EPR). We reduce the
quantization problem to answering the simple question: what fraction
of the energy of mode~$m$ is stored in element~$j$? This leads
to a constrained number between zero and one, the EPR, denoted~$p_{mj}$
\citep{Minev2019-Thesis}. This ratio is the key quantity that bridges
classical and quantum circuit analysis; we show it plays the primary
role in the construction of the system many-body Hamiltonian. Furthermore,
dissipation in the system is treated on equal footing by calculating
the EPR~$p_{ml}$ of lossy element~$l$ in mode~$m$.

The EPR method deviates from previous black-box quantization work
\citep{Nigg2012,Solgun2014,Solgun2015}, which uses the impedance-response
matrix, denoted~$Z_{jj^{\prime}}\left(\omega\right)$, where~$j$
and~$j^{\prime}$ index ports associated with nonlinear elements.
For all pairs of ports, the complex function~$Z_{jj^{\prime}}\left(\omega\right)$
is calculated from a finite-element (FE) driven simulation in the
vicinity of the eigenfrequency of every mode. Our method replaces
these steps with a more economical FE eigenmode simulation, from which
one extracts the energy participations~$p_{ml}$ and~$p_{mj}$,
needed to fully characterize both the dissipative and Hamiltonian
properties of the circuit.

To test the method, we compared EPR calculations of circuit parameters
to experimentally measured ones for 8 superconducting devices designed
with the EPR method, comprising a total of 15 qubits, 8 readout and
storage resonator modes, and one waveguide system. The results demonstrate
agreement for Hamiltonian parameters spanning over five orders of
magnitude in energy. Resonance frequencies were calculated to one
percent accuracy, large nonlinear interactions, such as anharmonicities
and cross-Kerr frequencies, to five percent, and small, nonlinear
interactions to ten percent. This level of accuracy is sufficient
for most current quantum information experiments.

\begin{figure*}
\centering{}\includegraphics[width=6.9in]{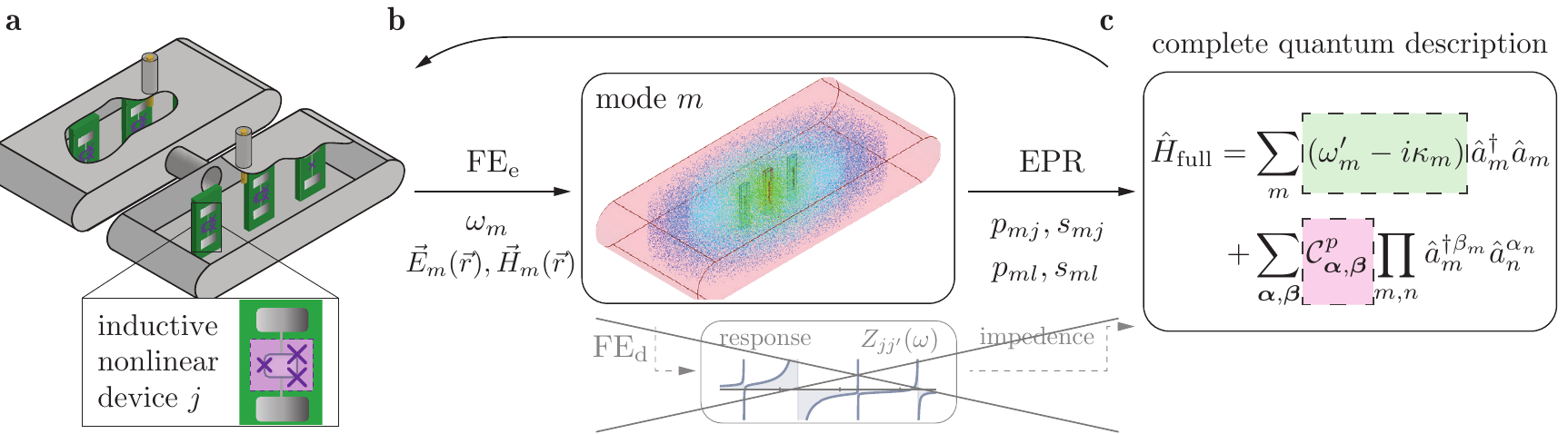}
\caption{\label{fig:ConceptFig}\textbf{Conceptual overview. a}~Illustration
of the physical model of an example quantum device, which comprises
two three-dimensional~(3D) cavities (grey enclosures), each housing
several qubit chips (green boxes). A close-up view of one of the chips
is depicted in the inset. The dotted box in the center of the chip
schematically outlines a non-linear inductive sub-circuit, referred
to as a \emph{Josephson dipole}.\textbf{ b}~Results of a finite-element
eigenmode analysis ($\mathrm{FE}_{\mathrm{e}}$) of the Josephson
circuit linearized about its equilibrium. The~$m$-th mode eigenfrequency
and electric and magnetic fields are~$\omega_{m}$, $\vec{E}_{m}\left(\vec{r}\right)$,
and~$\vec{H}_{m}\left(\vec{r}\right)$, respectively, where~$r$
denotes spatial position. Center inset:~$\left|\vec{E}_{m}\right|$
profile (red: high; blue: low) for the fundamental mode of one of
the 3D cavities. Additional FE driven simulations ($\mathrm{FE}_{\mathrm{d}}$)
are\emph{ }unnecessary; i.e., the impedance matrix~$Z_{jj'}\left(\omega\right)$
is not calculated. \textbf{c} The Hamiltonian~$\hat{H}_{\mathrm{full}}$,
which includes nonlinear interactions to arbitrary order (see Results
), is computed directly from the eigenanalysis via the EPRs~$p_{mj}$
and EPR signs~$s_{mj}=\pm1$ of the junctions,~$j$. Dissipative
contributions due to a lossy element~$l$ are similarly computed
from the loss EPRs~$p_{ml}$; for linear dissipation, the EPR signs~$s_{ml}$
are unnecessary. Direct extraction of Hamiltonian and dissipative
parameters from eigensolutions is unique to the EPR method. The geometry
of the classical model is modified in an iterative search for the
desired dissipative and Hamiltonian parameters (left-pointing arrow).}
\end{figure*}

\section{RESULTS AND DISCUSSION}

\begin{figure}
\centering{}\includegraphics[width=3.375in]{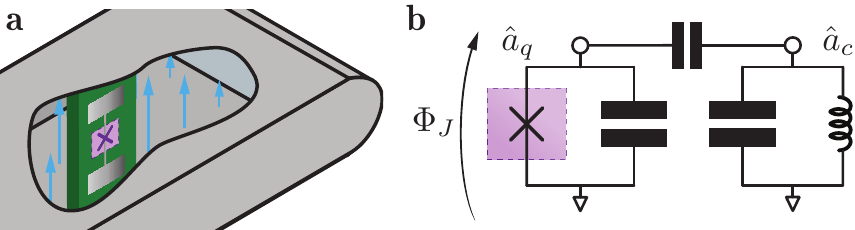}
\caption{\label{fig:simple-example-transmon-cavity} \textbf{Quantizing a simple
circuit.} \textbf{a}~Illustration of a~3D cavity enclosing a transmon
qubit chip. The cross symbol marks the location of a Josephson junction.
Vertical blue arrows depict the electric field~$\vec{E}_{m}\left(\vec{r}\right)$
of the fundamental cavity mode, TE$_{101}$. \textbf{b}~Equivalent
two-mode lumped-element- representation of the distributed circuit.
Operators~$\hat{a}_{q}$ and~$\hat{a}_{c}$ denote the qubit and
cavity mode operators, respectively.}
\end{figure}

\subsection{To quantize a simple circuit: qubit coupled to a cavity}

\textbf{}\label{sec:simple-circuit-example}\textbf{ }In this section,
we introduce the EPR method of quantum circuit design on a modest,
yet informative, example: a transmon qubit coupled to a cavity mode
(see Fig.~\ref{fig:simple-example-transmon-cavity}). The transmon~\citep{Koch2007}
consists of a Josephson junction shunted by a capacitance. It is embedded
in the cavity, which we will consider as a black-box distributed structure.
The Hamiltonian of this system~$\hat{H}_{\mathrm{full}}$ can be
conceptually separated in two contributions (see Supplementary Section~\ref{app:Josephson-system}),
\begin{equation}
\hat{H}_{\mathrm{full}}=\hat{H}_{\mathrm{lin}}+\hat{H}_{\mathrm{nl}}\;,\label{ex:1jj:Hfull}
\end{equation}
where~$\hat{H}_{\mathrm{lin}}$ consists of \textit{\emph{all}} terms
associated with the linear response of the junction and the resonator
structure, and~$\hat{H}_{\mathrm{nl}}$ consists of terms associated
with the nonlinear response of the junction. Restricting our attention
to the cavity and qubit modes of the otherwise black-box structure,
the analytical form of the Hamiltonian follows from standard circuit
quantization \citep{Yurke1984,Devoret1995} (see Supplementary Section~\ref{app:diagonalize-eigenmodes}):
\begin{eqnarray}
\hat{H}_{\mathrm{lin}} & = & \hbar\omega_{c}\hat{a}_{c}^{\dagger}\hat{a}_{c}+\hbar\omega_{q}\hat{a}_{q}^{\dagger}\hat{a}_{q}\;,\label{eq:1jj:Hlin}\\
\hat{H}_{\mathrm{nl}} & = & -E_{J}\left[\cos\left(\hat{\varphi}_{J}\right)+\hat{\varphi}_{J}^{2}/2\right]\;,\label{eq:1jj:Hnl}\\
\hat{\varphi}_{J} & = & \varphi_{q}\left(\hat{a}_{q}+\hat{a}_{q}^{\dagger}\right)+\varphi_{c}\left(\hat{a}_{c}+\hat{a}_{c}^{\dagger}\right)\;,\label{eq:1jj:ZPF defn}
\end{eqnarray}
where~$\omega_{c}$ and~$\omega_{q}$ are the angular frequencies
of the cavity and qubit eigenmodes defined associated with~$\hat{H}_{\mathrm{lin}}$,
respectively, and where~$\hat{a}_{c}$ and~$\hat{a}_{q}$ are their
annihilation operators, respectively. The Josephson energy~$E_{J}$
can be computed from the Ambegaokar-Baratoff formula adapted to the
measured room-temperature resistance of the junction \citep{Gloos2000}.
The junction reduced generalized flux~$\hat{\varphi}_{J}$ corresponds
to the classical variable~$\varphi_{J}\left(t\right)\isdef\int_{-\infty}^{t}v_{J}\left(\tau\right)\,\mathrm{d}\tau/\phi_{0}$,
where~$v_{J}(\tau)$ is the instantaneous voltage across the junction~\citep{Yurke1984,Devoret1995},
and~$\phi_{0}\isdef\hbar/2e$ is the reduced flux quantum. The junction
flux operator {[}Eq.~\eqref{eq:1jj:ZPF defn}{]} is a linear, real-valued,
and non-negative combination of the mode operators (see Supplementary
Section~\ref{app:diagonalize-eigenmodes}), and in its expression,~$\varphi{}_{c}$
and~$\varphi_{q}$ are the quantum zero-point fluctuations of junction
flux in the cavity and qubit mode, respectively. It is worth stating
that the linear coupling between the cavity and qubit, commonly denoted~$g$
\citep{Koch2007}, is fully factored in our analysis, and is implicitly
handled in the extraction of the operators from the electromagnetic
simulation.

Our principal aim is to determine the unknown quantities:~$\omega_{c},\omega_{q},\varphi_{q}$
and~$\varphi_{c}$. As we will show, we extract and compute these
quantities from an eigenanalysis of the classical distributed circuit
corresponding to~$\hat{H}_{\mathrm{lin}}$. This includes the qubit-cavity
layout, materials, electromagnetic boundary conditions, and a model
of the junction as a lumped-element, linear inductor. The eigensolver
returns the requested set of eigenmodes and their frequencies, quality
factors, and field solutions. By running the eigensolver in the frequency
range of interest, we obtain the hybridized cavity and qubit modes,
whose eigenfrequencies~$\omega_{c}$ and~$\omega_{q}$ fully determine~$\hat{H}_{\text{lin}}$
(see Supplementary Section~\ref{app:FE-sims} for FE methodology).

To determine~$\hat{H}_{\mathrm{nl}}$, we need the quantum zero-point
fluctuations~$\varphi_{q}$ and~$\varphi_{c}$, which are calculated
from the participation of the junction in the eigenfield solutions.
The participation~$p_{m}$ of the junction in mode~$m\in\left\{ c,q\right\} $
is defined to be the fraction of inductive energy stored in the junction
relative to the total inductive energy stored in the entire circuit,
\begin{equation}
p_{m}\isdef\frac{\text{Inductive energy stored in the junction}}{\text{Total inductive energy stored in mode }m}\;,\label{eq:p_m defn}
\end{equation}
evaluated when only mode~$m$ is excited. Thus, $p_{m}$ can be computed
from the electric~$\vec{E}_{m}(\vec{r})$ and magnetic~$\vec{H}_{m}(\vec{r})$
eigenfields as detailed in Supplementary Section~\ref{app:FE-epr-single-junc};
$\vec{r}$ denotes spatial position. In the quantum setting, Eq.~\eqref{eq:p_m defn}
links~$p_{m}$, $\hat{\varphi}_{J}$, and the state of the circuit,
\begin{equation}
p_{m}=\frac{\langle\psi_{m}|\frac{1}{2}E_{J}\hat{\varphi}_{J}^{2}|\psi_{m}\rangle}{\langle\psi_{m}|\frac{1}{2}\hat{H}_{\mathrm{lin}}|\psi_{m}\rangle}\;,\label{eq:PJ-1JJ}
\end{equation}
where~$\ket{\psi_{m}}$ denotes a coherent state or a Fock excitation
of mode~$m$. Note that normal-ordering must be used in Eq.~\eqref{eq:PJ-1JJ};
this correct treatment of vacuum fluctuations is detailed in Supplementary
Section~\ref{app:EPR-defn}. Simplifying Eq.~\eqref{eq:PJ-1JJ},
one expresses the variance of the quantum zero-point fluctuations~$\varphi_{c}$
and~$\varphi_{q}$ as a function of the classical energy participations~$p_{m}$,
\begin{equation}
\varphi_{c}{}^{2}=p_{c}\frac{\hbar\omega_{c}}{2E_{J}}\quad\text{and}\quad\varphi_{q}{}^{2}=p_{q}\frac{\hbar\omega_{q}}{2E_{J}}\;,\label{eq:ZPF-pj-1jj}
\end{equation}
which completely determines~$\hat{H}_{\mathrm{nl}}$, and thus completes
the description of the system Hamiltonian~$\hat{H}_{\mathrm{full}}$.
Here,~$\varphi_{c}$ and~$\varphi_{q}$ can be taken as positive
numbers. As we will see in the next section, in the presence of multiple
junctions, this is not always true. 

Designing experiments with the EPR requires one to further extract
from~$\hat{H}_{\mathrm{full}}$ the transition frequencies and nonlinear
couplings between modes. Depending on the case, this can be done approximately
or exactly using numerical or analytical techniques \citep{DiPaolo2019}.
This task is easily achieved if~$\hat{H}_{\mathrm{nl}}$ is a perturbation
to~$\hat{H}_{\text{lin}}$~\citep{Koch2007}. In this limit, $\hat{H}_{\mathrm{full}}$
for our qubit-cavity example can be approximated by the effective,
excitation-number-conserving Hamiltonian, see Supplementary Eq.~\eqref{eq:app:mixing:energy-conserving:H4-bar},
\begin{multline}
\hat{H}_{\text{eff}}=\left(\omega_{q}-\Delta_{q}\right)\hat{n}_{q}+\left(\omega_{c}-\Delta_{c}\right)\hat{n}_{c}-\chi_{qc}\hat{n}_{q}\hat{n}_{c}\\
-\frac{1}{2}\alpha_{q}\hat{n}_{q}\left(\hat{n}_{q}-\hat{1}\right)-\frac{1}{2}\alpha_{c}\hat{n}_{c}\left(\hat{n}_{c}-\hat{1}\right)\;,\label{eq:qubitcavityHtot}
\end{multline}
where~$\hat{n}_{q}=\hat{a}_{q}^{\dagger}\hat{a}_{q}$ and~$\hat{n}_{c}=\hat{a}_{c}^{\dagger}\hat{a}_{c}$
denote the qubit and cavity excitation-number operators, respectively,
$\Delta_{q}$ denotes the `Lamb shift' of the qubit frequency due
to the dressing of this nonlinear mode by quantum fluctuations of
the fields, $\alpha_{q}$ ($\alpha_{c}$) is the qubit (cavity) anharmonicity,
and~$\chi_{qc}$ is the qubit-cavity dispersive shift (cross-Kerr
coupling). The Hamiltonian parameters can be calculated directly from
the EPR, see Supplementary Section~\ref{app:nonlinear-interactions},
\begin{align}
\alpha_{q} & =\frac{1}{2}\chi_{qq}=p_{q}^{2}\frac{\hbar\omega_{q}^{2}}{8E_{J}}\;,\label{eq:alpha-1jj-old}\\
\alpha_{c} & =\frac{1}{2}\chi_{cc}=p_{c}^{2}\frac{\hbar\omega_{c}^{2}}{8E_{J}}\;,\\
\chi_{qc} & =p_{q}p_{c}\frac{\hbar\omega_{q}\omega_{c}}{4E_{J}}\;.
\end{align}
Experimentally, the qubit Lamb shift can be obtained as~$\Delta_{q}=\alpha_{q}-\chi_{qc}/2$.
Since a single EPR~$p_{m}$ determines the nonlinear interaction
for each mode, the parameters~$\chi_{qc}$ and~$\alpha_{q}$ are
interdependent, 
\begin{equation}
\chi_{qc}=\sqrt{\chi_{qq}\chi_{cc}}=2\sqrt{\alpha_{q}\alpha_{c}}\;.\label{eq:1jj_chi_qc_const}
\end{equation}

As shown in Supplementary Section\textbf{~}\ref{app:universal-constraints-epr},
the EPRs~$p_{c}$ and~$p_{q}$ obey the constraints
\begin{equation}
0\leq p_{q},p_{c}\leq1\quad\text{and}\quad p_{q}+p_{c}=1\;.\label{eq:1jj constrain}
\end{equation}
These relations together with Eqs.~\eqref{eq:alpha-1jj-old} and~\eqref{eq:1jj_chi_qc_const}
are useful to budget the dilution of the nonlinearity of the junction
(see Supplementary Section~\ref{app:EPR-matrix}) and to provide
insight on the limits of accessible parameters (see Methods). Further,
Eq.~\eqref{eq:1jj constrain} is used to validate the convergence
of the FE simulation.

\begin{figure}[t]
\centering{}\includegraphics[width=3.375in]{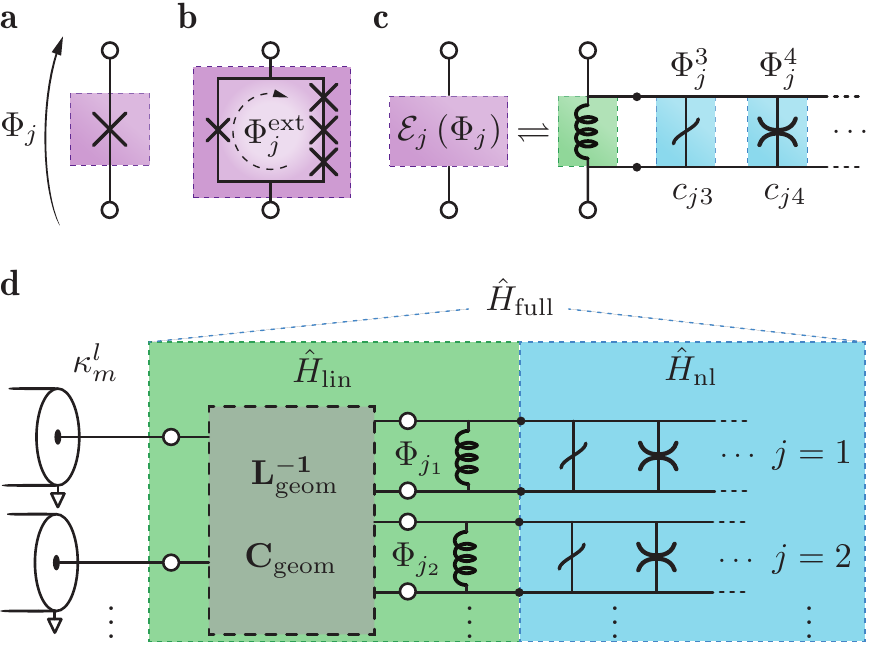}
\caption{\label{fig:Jos-system-conceptual-rep} \textbf{Schematic representation
of the Josephson circuit and its nonlinear elements.} \textbf{a}~A
simple example of a Josephson dipole\textemdash a Josephson tunnel
junction. \textbf{b}~An example of a composite junction, comprising
four Josephson junctions in a ring, frustrated by an external magnetic
flux~$\Phi_{j}^{\mathrm{ext}}$ threading the loop. \textbf{c} Conceptual
decomposition of a general Josephson dipole, denoted~$j$. For convenience,
its potential energy function~$\mathcal{E}_{j}(\Phi_{j};\Phi_{j}^{\mathrm{ext}})$
can be Taylor expanded in a sum of nonlinear inductive contributions
of increasing order~$\Phi_{j}^{p}$, with relative amplitude~$c_{jp}$,
where~$p$ denotes the index in the series. The energy function can
be subjected to external bias parameters~$\Phi_{j}^{\mathrm{ext}}$,
such as flux or voltages. \textbf{d}~Schematic diagram of a general
Josephson circuit conceptually resolved into a purely dissipative
(left,~$\kappa_{m}^{l}$), linear (middle,~$\hat{H}_{\mathrm{lin}}$),
and nonlinear (right,~$\hat{H}_{\mathrm{nl}}$) constitutions.}
\end{figure}

\textbf{
\subsection{Quantizing the general Josephson system}}\label{sec:general-system} The simple results obtained in the preceding
section will now be generalized to arbitrary nonlinear devices enclosed
in a black-box, distributed, electromagnetic structure. While such
structures are frequently classified as planar \citep{Blais2004,Wallraff2004,Barends2013,FYan2016}~(2D),
quasi-planar \citep{Minev2013,Minev2016,Brecht2016,Rosenberg2017}~(2.5D),
or three-dimensional \citep{Paik2011,Rigetti2012,Reagor2016-cavity,Axline2016}~(3D),
we will treat all classes on equal footing. The electromagnetic structure
is assumed to be linear in the absence of the enclosed nonlinear devices.
For simplicity of discussion, we can consider these devices to be
inductive and lumped; distributed nonlinear devices, such as kinetic-inductance
transmission lines \citep{Yurke2006,HoEom2012,Vissers2015,Mortensen2016},
can be thought of as a series of lumped ones.

The simplest nonlinear device comprises a single element, such as
a Josephson tunnel junction {[}see Fig.~\ref{fig:Jos-system-conceptual-rep}(a){]},
an atomic-point contact \citep{Koops1996,Bretheau2013}, a nanobridge
\citep{Vijay2010,Peltonen2016}, a semiconducting nanowire \citep{Mooij2006,Ku2010,Abay2014,Larsen2015,DeLange2015-nanowire,Casparis2016},
or another hybrid structure \citep{Shim2014}. A \emph{multi-element
device}, such as a SQUID \citep{Zimmerman1966,Clarke2004}, a SNAIL
\citep{Frattini2017} {[}see Fig.~\ref{fig:Jos-system-conceptual-rep}(b){]},
a superinductance \citep{Manucharyan2012,Pop2014,Muppalla2017}, or
a junction array \citep{Corlevi2006,Hutter2011,Bourassa2012,Masluk2012,Bell2012,Weibl2015,Macklin2015}
refers to a subcircuit composed of purely inductive lumped elements.
This subcircuit can also be subjected to external controls, such as
voltage or flux biases.

The general nonlinear device that we now consider, referred to as
a \emph{Josephson dipole,} is any lumped, purely-inductive, nonlinear
subcircuit with two terminals. The key characteristic of the Josephson
dipole is that it possesses a characteristic energy function, which
encapsulates all details of its constitution. For example, the two-terminal
nonlinear device known as the symmetric SQUID \citep{Clarke2004}
is described by the energy function~$\mathcal{E}_{j}\left(\Phi_{j};\Phi_{j}^{\mathrm{ext}}\right)=-E_{j}\left(\Phi_{j}^{\mathrm{ext}}\right)\cos\left(\Phi_{j}/\phi_{0}\right)$,
where~$\Phi_{j}$ is the generalized flux across the device terminals
\citep{Yurke1984,Devoret1995},~$E_{j}$ is the effective Josephson
energy,~$\Phi_{j}^{\mathrm{ext}}$ is the external flux bias, and
the subscript~$j$ denotes the~$j$-th Josephson dipole in the circuit.
The flux~$\Phi_{j}$ is defined as the deviation away from the value
in equilibrium, as discussed below. To ease the notation, parameters
such as~$\Phi_{j}^{\mathrm{ext}}$ will be implicit hereafter. Similarly
to the example of the single-junction transmon , the energy of a
Josephson dipole can be separated in two parts. One part~$\ejlin$
accounts for the linear response of the dipole, while the other~$\ejnl$
accounts for the nonlinear response, 
\begin{equation}
\mathcal{E}_{j}\left(\Phi_{j}\right)=\ejlin\left(\Phi_{j}\right)+\ejnl\left(\Phi_{j}\right)\;,\label{eq:circuit: defn of junc energy}
\end{equation}
where 
\begin{eqnarray}
\ejlin\left(\Phi_{j}\right) & \isdef & \frac{1}{2}E_{j}\left(\frac{\Phi_{j}}{\phi_{0}}\right)^{2}\;,\label{eq:circuit:Uj-lin}
\end{eqnarray}
and where the constant~$E_{j}$ sets the scale of the junction energy.
This energy scale can be represented by the linear inductance~$L_{j}\isdef\phi_{0}/E_{j}$
presented by the Josephson dipole when submitted to a small excitation
about its equilibrium.

\paragraph{Frustrated equilibrium.}

External biases can set up persistent currents in the circuit. These
can alter the static {[}direct-current (dc){]} equilibrium of the
Josephson system. For example, frustrating a superconducting ring
with a magnetic flux sets up a persistent circulating current in the
ring. For a Josephson dipole in such a loop, the definition of the
flux~$\Phi_{j}$ will differ in Eqs.~\eqref{eq:circuit: defn of junc energy}
and~\eqref{eq:circuit:Uj-lin} as a function of the equilibrium.
Due to this adjustment, terms linear in~$\Phi_{j}$ are absent from
Eq.~\eqref{eq:circuit:Uj-lin} by construction. Supplementary Sections~\ref{app:equilibrium-point-of-circuit} and \ref{subsec:The-biased-Josephson-simplications}
discuss these equilibrium considerations in detail.

\subsubsection{Quantum Hamiltonian.}

Having conceptually carved out nonlinear contributions from the system
Hamiltonian~$\hat{H}_{\mathrm{full}}$, and collected them in the
set of~$\ejnl$ functions, we define the \textit{linearized Josephson
circuit} to correspond to everything left over in the system. This
linear circuit consists of the electromagnetic circuit external to
the Josephson dipoles, combined with their linear inductances~$L_{j}$.
We will use the eigenmodes of the linearized circuit to explicitly
construct~$\hat{H}_{\mathrm{full}}$. The eigenmode frequencies and
field distributions are readily obtained using a conventional finite-element
solver (see Supplementary Section~\ref{app:FE-sims}). The Hamiltonian
of the linearized Josephson circuit can thus be expressed as (see
Supplementary Section~\ref{app:quantize-modes}) 
\begin{align}
\hat{H}_{\mathrm{lin}} & =\sum_{m=1}^{M}\hbar\omega_{m}\hat{a}_{m}^{\dagger}\hat{a}_{m}\;,\label{eq:Hlin-multi}
\end{align}
where~$M$ is the number of modes addressed by the numerical simulation,~$\omega_{m}$
is the solution eigenfrequency of mode~$m$, and~$\hat{a}_{m}$
the corresponding mode amplitude (annihilation operator), defined
by the mode eigenvector. We emphasize that the frequencies~$\omega_{m}$
will be significantly perturbed by the Lamb shifts~$\Delta_{m}$,
and should be seen as an intermediate parameter entering in the calculation
of the rest of the nonlinear Hamiltonian, 
\begin{align}
\hat{H}_{\mathrm{nl}} & =\sum_{j=1}^{J}\ejnl=\sum_{j=1}^{J}E_{j}\left(c_{j3}\hat{\varphi}_{j}^{3}+c_{j4}\hat{\varphi}_{j}^{4}+\cdots\right)\label{eq:Hnl-multi-old}\\
 & =\sum_{j=1}^{J}E_{j}\sum_{p=3}^{\infty}c_{jp}\hat{\varphi}_{j}^{p}\;,\label{eq:Hnl-series}\\
\hat{\varphi}_{j} & =\sum_{m=1}^{M}\varphi_{mj}\left(\hat{a}_{m}^{\dagger}+\hat{a}_{m}\right)\;,\label{eq:multijj:ZPF defn}
\end{align}
where~$J$ is the total number of junctions and~$\hat{\varphi}_{j}\isdef\hat{\Phi}_{j}/\phi_{0}$.
In Eq.~\ref{eq:Hnl-multi-old}, we have introduced a Taylor expansion
of~$\ejnl$, where the energy~$E_{j}$ and expansion coefficients~$c_{jp}$
are known from the fabrication of the Josephson circuit, see Fig.~\ref{fig:Jos-system-conceptual-rep}(c).
For example, for a Josephson junction, the constant~$E_{j}$ is just
the Josephson energy, while~$c_{jp}$ are the coefficients of the
cosine expansion; i.e.,~$c_{jp}$ is~$0$ for odd~$p$ and~$\left(-1\right)^{p/2+1}/p!$
for even~$p$. The expansion is helpful for analytics but does not
need to be used in the numerical analysis of~$\hat{H}_{\mathrm{nl}}$,
see Supplementary Section~\ref{app:main-derivation}.

The Hamiltonian~$\hat{H}_{\mathrm{full}}$ is specified since the
operators~$\hat{\varphi}_{j}$ are expressed in terms of the mode
amplitudes as a linear combination (see Supplementary Section~\ref{app:quantize-modes}).
Here,~$\varphi_{mj}$ are the dimensionless, \textit{real}-valued,
quantum zero-point fluctuations of the reduced flux of junction~$j$
in mode~$m$. Determination of~$\hat{H}_{\mathrm{full}}$ is now
reduced to computing~$\varphi_{mj}$. We achieve this by employing
a generalization of the energy-participation ratio.

The \emph{energy-participation ratio}~$p_{mj}$ of junction~$j$
in eigenmode~$m$ is defined to be the fraction of inductive energy
stored in the junction when only that mode is excited, 
\begin{align}
p_{mj} & \isdef\frac{\text{Inductive energy stored in junction }j}{\text{Inductive energy stored in mode }m}\nonumber \\
 & =\frac{\langle\psi_{m}|\frac{1}{2}E_{j}\hat{\varphi}_{j}^{2}|\psi_{m}\rangle}{\langle\psi_{m}|\frac{1}{2}\hat{H}_{\mathrm{lin}}|\psi_{m}\rangle}\;,\label{eq:Pmj}
\end{align}
which is a straightforward extension of Eq.~\eqref{eq:p_m defn},
and is similarly computed using normal ordering (see Supplementary
Section~\ref{app:EPR-defn}). The EPR~$p_{mj}$ is computed from
the eigenfield solutions~$\vec{E}_{m}(\vec{r})$ and~$\vec{H}_{m}(\vec{r})$
as explained in Supplementary Section~\ref{app:FE-multiple-JJ}.
It is a bounded, non-negative, real number, $0\leq p_{mj}\leq1$.
A zero EPR~$p_{mj}=0$ means that junction~$j$ is not excited in
mode~$m$. A unity EPR~$p_{mj}=1$ means that junction~$j$ is
the only inductive element excited in the mode.

From the EPR~$p_{mj}$, one directly computes the variance of the
quantum zero-point fluctuations (see Appendix~\ref{app:EPR-defn}),
\begin{equation}
\boxed{\varphi_{mj}^{2}=p_{mj}\frac{\hbar\omega_{m}}{2E_{j}}\;.}\label{eq:pmj_multi_zpf}
\end{equation}
Equation~\eqref{eq:pmj_multi_zpf} constitutes the bridge between
the classical solution of the linearized Josephson circuit and the
quantum Hamiltonian~$\hat{H}_{\mathrm{full}}$ of the full Josephson
system, up to the sign of~$\varphi_{mj}$.

\subsubsection{Universal EPR properties}

The quantum fluctuations~$\varphi_{mj}$ are not independent of each
other, since the EPRs are submitted to three types of universal constraints\textemdash valid
regardless of the circuit topology and nature of the Josephson dipoles.
These are of practical importance, as they are useful guides in evaluating
the performance of possible designs and assessing their limitations.
As shown in Supplementary Section~\ref{app:universal-constraints-epr},
the EPRs obey one sum rule per junction~$j$ and one set of inequalities
per mode~$m$, 
\begin{equation}
\sum_{m=1}^{M}p_{mj}=1\quad\text{and}\quad0\leq\sum_{j=1}^{J}p_{mj}\leq1\;.\label{eq:constrains-on-p-for-m-and-j}
\end{equation}
The total EPR of a Josephson dipole is a quantity that is independent
of the number of modes\textemdash it is precisely unity for all circuits
in which the dipole is embedded. It can only be diluted among the
modes. On the other hand, a given mode can accept at most a total
EPR of unity from all the dipoles. In practice, this sum rule can
be fully exploited only if the bound~$M$ reaches the total number
of relevant modes of the system.

The next fundamental property concerns the orthogonality of the EPRs.
Rewriting Eq.~\eqref{eq:pmj_multi_zpf} in terms of the amplitude
of the zero-point fluctuation we have
\begin{equation}
\varphi_{mj}=s_{mj}\sqrt{p_{mj}\hbar\omega_{m}/2E_{j}}\;,\label{eq:zpf-general-Smj-Pmj}
\end{equation}
where the \textit{\emph{EPR}}\textit{ }sign~$s_{mj}$ of junction~$j$
in mode~$m$ is either~$+1$ or~$-1$. The EPR sign encodes the
\emph{relative }direction of current flowing across the junction.
Only the \textit{\emph{relative}} value between~$s_{mj}$ and~$s_{mj'}$
for~$j\neq j'$ has physical significance (see Supplementary Figure~\ref{fig:port-signs}).
The EPR sign~$s_{mj}$ is calculated in parallel with the process
of calculating~$p_{mj}$, from the field solution~$\vec{H}(\vec{r})$,
see Supplementary Section~\eqref{eq:Smj-calc}. We now obtain the
EPR orthogonality relationship 
\begin{align}
\sum_{m=1}^{M}s_{mj}s_{mj'}\sqrt{p_{mj}p_{mj'}}\, & =0\;,\label{eq:pmjpmj' orthogonality}
\end{align}
valid when the sum from~$1$ to~$M$ covers all the relevant modes,
see Supplementary  Eq.~\eqref{eq:epr-orthogonality}.

\subsubsection{Excitation-number-conserving interactions\label{sec:multi-junction:energy-conserving}}

Thus, as announced, knowledge of the energy-participation ratios completely
specifies~$\hat{H}_{\mathrm{nl}}$, through Eqs.~\eqref{eq:Hnl-multi-old},~\eqref{eq:multijj:ZPF defn},
and~\eqref{eq:pmj_multi_zpf}. The Hamiltonian can now be analytically
or numerically diagonalized using various computational techniques
\citep{DiPaolo2019}. In this section, our focus will be now to explicitly
handle the effect of the nonlinear interactions~$\hat{H}_{\mathrm{nl}}$
on the eigenmodes. Before treating the case of a general nonlinear
interaction, we focus on the leading-order effect of~$\hat{H}_{\mathrm{nl}}$
in the case of the `canonical' Josephson system. In this case, the~$J$
Josephson dipoles are all Josephson tunnel junctions, characterized by Eq.~\eqref{eq:Ej-jj-dipole-nl},
and the dispersive regime is satisfied for all pairs of modes~$k$
and~$m$; i.e., $\omega_{k}-\omega_{m}\gg E_{j}c_{jp}\left<\hat{\varphi}_{j}^{p}\right>$
for~$p\geq3$ and in the absence of strong drives. The leading-order
nonlinear terms are the subset of~$p=4$ terms that conserve excitation
number. After normal ordering, see Supplementary Section~\ref{app:excitation-conserving-mixing},
one finds the effective Hamiltonian 
\begin{equation}
\hat{\overline{H}}_{4}=-\hbar\sum_{m=1}^{M}\Delta_{m}\hat{a}_{m}^{\dagger}\hat{a}_{m}+\frac{\alpha_{m}}{2}\hat{a}_{m}^{\dagger2}\hat{a}_{m}^{2}+\sum_{n<m}\chi_{mn}\hat{a}_{m}^{\dagger}\hat{a}_{n}^{\dagger}\hat{a}_{m}\hat{a}_{n}\;,\label{eq:H4 - RWA main}
\end{equation}
which is a generalization of the one found in Eq.~\eqref{eq:qubitcavityHtot}.
In Eq.~\eqref{eq:H4 - RWA main}, we have introduced the Lamb shift~$\Delta_{m}$
of mode~$m$, the anharmonicity~$\alpha_{m}$ of the mode, and its
total dispersive shift~$\chi_{mn}$ (so-called cross-Kerr term) with
a different mode, labeled~$n$. Each of these parameters is directly
calculated from the EPRs. As shown in Supplementary Section~\eqref{eq:app:conserve-E:chimn_zpf},
for arbitrary~$m$ and~$n$,
\begin{equation}
\chi_{mn}=\sum_{j=1}^{J}\frac{\hbar\omega_{m}\omega_{n}}{4E_{j}}p_{mj}p_{nj}\;,\label{eq:chi_mj}
\end{equation}
while~$\alpha_{m}=\chi_{mm}/2$ and~$\Delta_{m}=\sum_{n=1}^{M}\chi_{mn}/2$.
Equation~\eqref{eq:chi_mj} implements, mathematically, the idea
that the amplitude of these nonlinear couplings is the result of a
spatial-mode scalar product of the EPRs. Remarkably, from Eq.~\eqref{eq:chi_mj}
it is seen that the EPRs are essentially the only free parameters
subject to design, when determining the nonlinear couplings, since~$\omega_{m}$,
$\omega_{n}$, and~$E_{j}$ are generally tightly constrained by
experimental considerations.

Equation~\eqref{eq:chi_mj} can be cast in matrix form by introducing
the EPR matrix 
\begin{equation}
\m P\isdef\left(\begin{array}{ccc}
p_{11} & \cdots & p_{1J}\\
\vdots & \ddots & \vdots\\
p_{M1} & \cdots & p_{MJ}
\end{array}\right)\;,\label{ex:P matrix}
\end{equation}
which we have found useful in handling large circuits, especially
for those in excess of 100~modes. We also introduce the diagonal
matrices of eigenfrequencies~$\m{\Omega}\isdef\diag\left(\omega_{1},\ldots,\omega_{M}\right)$
and junction energies~$\m{E_{J}}\isdef\diag\left(E_{1},\ldots,E_{J}\right)$,
which lead to the matrix form of Eq.~\eqref{eq:chi_mj}, 
\begin{equation}
\begin{array}{rcclc}
\text{Kerr matrix:} &  & \bm{\chi} & = & \frac{\hbar}{4}\m{\Omega PE_{J}^{-1}P^{\intercal}}\boldsymbol{\Omega}\;,\\
\text{Anharmonicity:} &  & \alpha_{m} & = & \frac{1}{2}\left[\bm{\chi}\right]_{mm}\;,\\
\text{Lamb shift:} &  & \Delta_{m} & = & \frac{1}{2}\sum_{m'=1}^{M}\left[\bm{\chi}\right]_{mm'}\;.
\end{array}\label{eq:quantize arb circ: energy-conserv: Chi matrix}
\end{equation}
We have defined the symmetric matrix of dispersive shifts~$\boldsymbol{\chi}$,
with elements~$\left[\boldsymbol{\chi}\right]_{mm'}=\chi_{mm'}$.
Further discussion of the matrix approach and applications to~$p$th-order
corrections is deferred to Supplementary Section~\ref{app:EPR-matrix},
and the amplitude of an arbitrary multi-photon interaction stemming
from the full~$\hat{H}_{\text{nl}}$ is calculated in Supplementary
Section~\ref{app:general-interactions}.

\subsubsection{EPR for dissipation in the circuit\label{sec:Dissipation}}

The EPR method treats the calculation of Hamiltonian and dissipation
parameters on equal footing. Unlike in the impedance method \citep{Nigg2012},
one can completely characterize both~$\hat{H}_{\mathrm{full}}$ and
the effect of dissipative elements in the circuit from the eigenfield
solutions,~$\vec{E}_{m}(\vec{r})$ and~$\vec{H}_{m}(\vec{r})$.
The list of dissipative elements include bulk and surface dielectrics~\citep{Martinis2005,Patel2013,Dial2016},
thin-film metals~\citep{Vissers2012,Minev2013}, surface interfaces~\citep{Wenner2011,Geerlings2012,Sandberg2013,Wang2015,Bruno2015-loss},
and metal seams~\citep{Brecht2015}. The\emph{ }energy-participation
ratio of a dissipative element~$l$ in mode~$m$ will be denoted~$p_{ml}$.
It is calculated similarly to~$p_{mj}$, as summarized in Supplementary
Section~\ref{app:dissipation}. The participation~$p_{ml}$ and
the quality factor~$Q_{l}$ of the material of this element are used
to estimate the total quality factor of mode~$m$ in the standard
way when the fields are not greatly altered by the dissipation ($Q_{m}\gg1$)~\citep{Gao2008,Geerlings2013,Reagor2016-cavity,Brecht2017-micromachined},
\begin{equation}
Q_{m}^{-1}=\sum_{l}{p_{ml}Q_{l}^{-1}}\;.\label{eq:Qm total}
\end{equation}
Experimental values of~$Q_{l}$ are found in the literature, and some
are provided in Supplementary Section~\ref{app:dissipation}. Equation~\eqref{eq:Qm total}
and the dissipative EPR~$p_{ml}$ provide a dissipation budget for
the individual influence of each dissipation mechanism in the system,
providing a useful tool to optimize design layout for quantum coherence~\citep{Martinis2014}.

\begin{figure}[t]
\centering{}\includegraphics[width=3.375in]{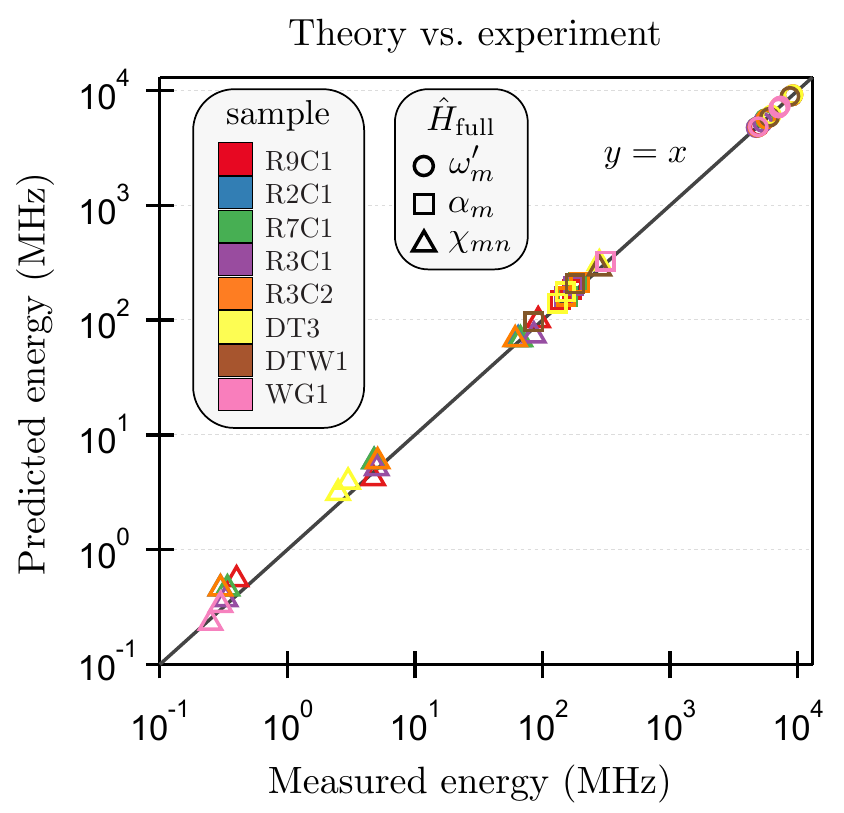}
\caption{\label{fig:graph_DiTr}\textbf{Comparison between theory and experiment
}over five-orders of magnitude in energy scale of the system Hamiltonian~$\hat{H}_{\mathrm{full}}$
for eight distinct, multi-mode device samples, described in detail
in the Methods, including 3D, flip-chip (2.5D), and~3D waveguide
architectures incorporating readout and storage resonators and qubit
modes. For each device, the dominant parameters in~$\hat{H}_{\mathrm{full}}$,
dressed frequencies~$\omega_{m}^{\prime}$, bare anharmonicities~$\alpha_{m}$,
and cross-Kerr interactions~$\chi_{mn}$, were measured and calculated
using the EPR method with our open-source \textsc{pyEPR} package \citep{Note1}.
Gray line is of slope one, representing ideal agreement between theory
and experiment.}
\end{figure}

\subsection{Comparison between theory and experiment}

\textbf{}\label{sec:Main-Text-Experiment-Discussion} Applying the
EPR method, we designed 8 superconducting samples to test the agreement
between the EPR theory and experimental results. We tested several
sample configurations, comprising 15 qubits, 8 cavity modes, and one
waveguide in three different circuit-QED architectures. The samples
were measured in a standard cQED setup, see Methods, at the 15\,mK
stage of a dilution unit, over multiple cool downs.

Six of the samples were each composed of 2 qubits and one 3D cavity,
one sample was composed of 2 qubits and a waveguide, and one sample
was a flip-chip, 2.5D system~\citep{Minev2016} consisting of a flip-chip
qubit embedded in a two-mode whispering gallery mode resonator \citep{Minev2013}
(WGMR). The specifics of each sample are discussed in the Methods.

For each sample, we measured the circuit parameters of interest: dressed
mode frequencies~$\omega_{m}-\Delta_{m}$, anharmonicities of qubits
and high-Q cavities~$\alpha_{m}$, cross-Kerr frequencies~$\chi_{mn}$,
and input-output quality factors~$Q_{C}$ for any readout modes.
Our measurement methodology is detailed in the Methods.

The measured parameters were compared to those calculated using the
energy-participation method. The linearized Josephson circuit of each
sample was modeled in \emph{Ansys High-Frequency Electromagnetic-Field
Simulator} (HFSS). Junctions were modeled as lumped inductors, whose
nominal energy~$E_{j}$ was inferred from room-temperature resistance
measurements~\citep{Gloos2000}. To account for the error bars of
the measurement and the drift in resistance over time,~$E_{j}$ was
adjusted by no more than~10\% to fit the measured qubit frequency.
To minimize the number of free parameters, we neglect the small junction
intrinsic capacitance~$C_{J}$ in our modeling. The tradeoff is a
small and estimable systematic offset of the bare simulated mode anharmonicities.
We estimate this correction to be on the order of~4\% for a~$C_{J}=4\,\mathrm{fF}$.
From the eigenfield solutions, we calculated the EPRs~$p_{mj}$ and
the sign~$s_{mj}$ to construct~$\hat{H}$ and extract its parameters.
Detailed steps of the procedure can be found in Supplementary Section~\ref{app:FE-sims}.
The results are presented in Tables~\ref{tab:ditransmon}\textendash \ref{tab:DiTransmon in waveguide}.

Figure~\ref{fig:graph_DiTr} summarizes the agreement of the measured
and calculated sample parameters, which span five orders of magnitude
in frequency. Accounting for~$C_{J}$, we find that mode frequencies
are calculated to one percent accuracy, large nonlinear interaction
energies (namely, anharmonicity and cross-Kerr frequencies greater
than 10~MHz) are calculated at the~5\% level, and small nonlinear
interaction energies agree at the~10\% level. We highlight that we
have used minimal, coarse adjustment to account for shifts in~$E_{j}$,
and otherwise, by neglecting~$C_{j}$, the calculation is free from
adjustable parameters.

The results of Fig.~\ref{fig:graph_DiTr} demonstrate the accuracy
and applicability of the EPR method. For each device, the EPR results
are obtained from a single eigenmode simulation, using full automation
of the analysis, provided by our open-source package \textsc{pyEPR}
\citep{Note1}. For current standard applications, we find the agreement
sufficient. Further improvements in accuracy would require improved
ability to estimate the Josephson dipole energy~$E_{j}$ and its
intrinsic capacitance~$C_{J}$. At the same level of accuracy, improvements
in the precision and reproducibility of the implementation and assembly
of the Josephson circuit design are needed, such as in chip-clamping
techniques, precision machining of the device sample holder and \textit{\emph{input-output}}
couplers.

\paragraph{Conclusion.}

An intuitive, easy-to-use and efficient method is needed to design
and analyze Josephson microwave quantum circuits. We have described
in this article such a method, based on the distribution of the electromagnetic
energy in the circuit and its participation in nonlinear and dissipative
elements. This so-called EPR method offers physical insight helping
the design process, and provides a simple link between the classical
circuit and its quantum properties. By comparing our theory to 8 experimental
devices incorporating Josephson junctions, we have shown that our
method is accurate and applicable to a large range of quantum circuit
architectures. It is directly applicable to a broader class of nonlinear
inductive elements, such as weak-link nanobridges \citep{Vijay2010,Peltonen2016},
nanowires \citep{Mooij2006,Abay2014,Larsen2015,DeLange2015-nanowire,Casparis2016},
and kinetic-inductance thin-films \citep{HoEom2012,Vissers2015,Maleeva2018}.
While best suited for weakly nonlinear systems, the EPR method is
derived within circuit theory without approximations. It can be seen
as arising from a change of basis adapted to nonlinear elements, as
detailed in Supplementary Section~\ref{app:main-derivation}. In
practice, the useful reach of the method is set by the numerical ability
to include all relevant electromagnetic modes and to compute the spectrum
of the extracted Hamiltonian \citep{DiPaolo2019}. We contribute an
open-source package \textsc{pyEPR} \footnote{See the \textsc{pyEPR} \citep{pyEPR} code repository at \href{http://github.com/zlatko-minev/pyEPR}{http://github.com/zlatko-minev/pyEPR}.},
which automates the EPR method, and was tested in the design of several
further experiments~\citep{Leghtas2015,Minev2016,Mundhada2017,Touzard2017,Campagne-Ibarcq2018-RemoteEnt,Muppalla2017,Wang2019-cav-atten,Grimm2019,Minev2019Nature,Campagne-Ibarcq2019,Winkel2019,Winkel2019a}.

\section{METHODS}

{
\small

\subsection{Methods of the experiment}

\label{sec:Experimental-methods}\label{app:Measured Devices:Setup}

\begin{table*}
\begin{centering}
{
\zkmDataTablePrep
\begin{tabular}{c|ccc|>{\centering}p{4em}>{\centering}p{4em}|r@{\extracolsep{0pt}.}lr@{\extracolsep{0pt}.}lr@{\extracolsep{0pt}.}l|c}
\multicolumn{1}{c}{Device} & \multicolumn{3}{c}{Frequency (MHz)} & \multicolumn{2}{c}{Anharmonicity (MHz)} & \multicolumn{6}{c}{Cross-Kerr (MHz)} & \emph{I-O} coupling\tabularnewline[1pt]
 & $\omega_{\mathrm{D}}/2\pi$ & $\omega_{\mathrm{B}}/2\pi$ & $\omega_{\mathrm{C}}/2\pi$ & $\alpha_{\mathrm{D}}/2\pi$ & $\alpha_{\mathrm{B}}/2\pi$ & \multicolumn{2}{c}{$\chi_{\mathrm{DB}}/2\pi$} & \multicolumn{2}{c}{$\chi_{\mathrm{BC}}/2\pi$} & \multicolumn{2}{c|}{$\chi_{\mathrm{DC}}/2\pi$} & $Q_{\mathrm{C}}$\tabularnewline
\hline 
\noalign{\vskip\doublerulesep}
R9C1 & 4951 & 5664 & 9158 & 138 & 170 & 92& & 4&7 & 0&4 & $5.20\times10^{3}$\tabularnewline
 & 4866 & 5691 & 9154 & 150 & 185 & 99& & 4&2 & 0&55 & $7.40\times10^{3}$\tabularnewline
 & -1.7\% & 0.5\% & -0.04\% & 8\% & 8\% & \multicolumn{2}{c}{7\%} & \multicolumn{2}{c}{-12\%} & \multicolumn{2}{c|}{27\%} & 29\%\tabularnewline[\doublerulesep]
\hline 
R2C1 & 4823 & 5567 & 8947 & 150 & 192 & 64&5 & 4&8 & 0&3 & $4.97\times10^{3}$\tabularnewline
 & 4770 & 5640 & 8950 & 161 & 211 & 67&7 & 5&88 & 0&46 & $5.44\times10^{3}$\tabularnewline
 & -1.1\% & 1.3\% & 0.03\% & 6.8\% & 9\% & 4&7\% & \multicolumn{2}{c}{18\%} & \multicolumn{2}{c|}{35\%} & 9 \%\tabularnewline[\doublerulesep]
\hline 
R7C1 & 4726 & 5475 & 8999 & 156 & 189 & 67& & 4&8 & 0&34 & $2.68\times10^{3}$\tabularnewline
 & 4770 & 5640 & 8950 & 161 & 211 & 67&7 & 5&88 & 0&46 & $3.07\times10^{3}$\tabularnewline
 & 0.9\% & 2.9\% & -0.55\% & 3.1\% & 10\% & \multicolumn{2}{c}{1\%} & \multicolumn{2}{c}{18\%} & \multicolumn{2}{c|}{26\%} & 13\%\tabularnewline[\doublerulesep]
\hline 
R3C2 & 4845 & 5620 & 8979 & 152 & 195 & \multicolumn{2}{c}{61} & 5&1 & 0&3 & $2.11\times10^{3}$\tabularnewline
 & 4770 & 5640 & 8950 & 161 & 211 & 67&7 & 5&88 & 0&46 & $1.78\times10^{3}$\tabularnewline
 & -1.5\% & 0.4\% & -0.3\% & 5.6\% & 7.6\% & 9&9\% & \multicolumn{2}{c}{13\%} & \multicolumn{2}{c|}{35\%} & -19\%\tabularnewline[\doublerulesep]
\hline 
R3C1 & 4688 & 5300 & 9003 & 148 & 174 & \multicolumn{2}{c}{85} & 5& & 0&33 & $2.43\times10^{3}$\tabularnewline
 & 4745 & 5265 & 8922 & 159 & 198 & \multicolumn{2}{c}{73} & 5&1 & 0&37 & $5.65\times10^{3}$\tabularnewline
 & 1.2\% & -0.7\% & -0.9\% & 6.9\% & 12.1\% & \multicolumn{2}{c}{-16\%} & \multicolumn{2}{c}{2\%} & \multicolumn{2}{c|}{9\%} & 57\%\tabularnewline[\doublerulesep]
\hline 
DT3 & 6160 & 7110 & 9170 & 130 & 150 & \multicolumn{2}{c}{278} & 3& & 2&5 & $9.17\times10^{3}$\tabularnewline
 & 6100 & 7141 & 9155 & 140 & 177 & \multicolumn{2}{c}{312} & 3&9 & 3&1 & $7.33\times10^{3}$\tabularnewline
 & -1.0\% & 0.4\% & -0.15\% & 7\% & 15\% & \multicolumn{2}{c}{11\%} & \multicolumn{2}{c}{23\%} & \multicolumn{2}{c|}{19\%} & -25\%\tabularnewline
\end{tabular}} 
\par\end{centering}
\caption{\textbf{Two-qubit, one-cavity devices.} Summary of measured and calculated
Hamiltonian and input-output (\emph{I-O}) coupling parameters for
the six devices described in Methods. Indices D, B, C denote the
dark, bright, and cavity modes respectively. The input-output quality
factor to the readout cavity is denoted~$Q_{\mathrm{C}}$. For each
device, the first (second) row quantifies the measured, $m$, (bare
calculated, $c$) values. The third row quantifies the bare agreement,
i.e., $\left(c-m\right)/c$. In the anharmonicity column, the bare
agreement should be corrected by the systematic shift due to our choice
to neglect the junction intrinsic capacitance in our modeling (see
Methods). We evaluate the correction to be of order 4\%, estimated
by taking a nominal junction~$C_{J}=4\,\mathrm{fF}$; hence, an overall
corrected agreement of~4.3\% for this column.\label{tab:ditransmon}}
\end{table*}

\begin{table*}
\begin{centering}
{
\zkmDataTablePrep
\begin{tabular}{c|ccc|c|cc|c}
\multicolumn{1}{c}{Device} & \multicolumn{3}{c}{Frequency (MHz)} & \multicolumn{1}{c}{Anharmonicity (MHz)} & \multicolumn{2}{c}{Cross-Kerr (MHz)} & \emph{I-O} coupling\tabularnewline[1pt]
 & $\omega_{\mathrm{Q}}/2\pi$ & $\omega_{\mathrm{S}}/2\pi$ & $\omega_{\mathrm{C}}/2\pi$ & $\alpha_{\mathrm{D}}/2\pi$ & $\chi_{\mathrm{QS}}/2\pi$ & $\chi_{\mathrm{QC}}/2\pi$ & $Q_{\mathrm{C}}$\tabularnewline
\hline 
\noalign{\vskip\doublerulesep}
WG1 & 4890 & 7070 & 7267 & 310 & 0.25 & 0.30 & $20\times10^{3}$\tabularnewline
 & 4820 & 7020 & 7340 & 325 & 0.29 & 0.33 & $16\times10^{3}$\tabularnewline
 & -1.4\% & -0.7\% & 1.0\% & 4.6\% & 13\% & 9\% & -22\%\tabularnewline[\doublerulesep]
\end{tabular}} 
\par\end{centering}
\caption{\label{tab:wgmr}\textbf{Flip-chip (2.5D), one-qubit, one-storage-cavity,
one-readout-cavity devices. }Summary of measured and calculated Hamiltonian
and input-output (\emph{I-O}) coupling parameters for the device described
in Methods. Indices Q, S, C denote the qubit, storage, and readout
cavity modes respectively. The input-output quality factor to the
readout cavity is denoted~$Q_{\mathrm{C}}$. For each device, the
first (second) row quantifies the measured, $m$, (bare calculated,
$c$) values. The third row quantifies the bare agreement, i.e., $\left(c-m\right)/c$.
}
\end{table*}

\emph{Device fabrication. }Unless otherwise noted, samples were fabricated
according to the following methodology. Sample patterns, both large
and fine features, were defined by a 100~kV electron-beam pattern
generator (\emph{Raith EBPG~5000+}) in a single step on a PMAA/MAA
(\emph{Microchem~A-4/Microchem~EL-13)} resist bilayer coated on
a 430\,$\mu$m thick, double-side-polished, c-plane sapphire wafer,
grown with the edge-defined film-fed growth (EFG) technique. Using
the bridge-free fabrication technique \citep{Rigetti2009,Lecocq2011-bridge-free,Pop2012-Junction}
the Al/Al$\text{O}_{\text{x}}$/Al Josephson tunnel junctions were
formed by a double-angle aluminum evaporation under ultra-high vacuum
in a multi-chamber\emph{ Plassys UMS300~UHV.} The two depositions
were interrupted by a thermal oxidation step, static 100~Torr environment
of 85\%~argon and 15\%~oxygen, to form the thin AlO$_{x}$ barrier
of the tunnel junction. Prior to the first deposition, to reduce junction
aging \citep{Pop2012-Junction}, the exposed wafer surfaces were exposed
to 1~minute oxygen-argon plasma cleaning, under a pressure of~$3\times10^{-3}$~mbar.
After wafer dicing (\emph{ADT ProVecturs~7100}) and chip cleaning,
the normal-state resistance~$R_{N}$ of the Josephson junctions was
measured to provide an estimate of the Josephson energy, $E_{J}$,
of the device junctions. The junction energy was to first order estimated
by an extrapolation of~$R_{N}$ from room temperature to the operating
sample temperature, at approximately~$15$~mK, using the Ambegaokar-Baratoff
relation \citep{Ambegaokar1963},
\begin{equation}
E_{J}=\frac{1}{2}\frac{h\Delta}{\left(2e\right)^{2}}R_{N}^{-1}\;,
\end{equation}
where~$\Delta$ is the superconducting gap of aluminum, $e$ is the
elementary charge, and~$h$ is Planck's constant.

\begin{table*}
\begin{centering}
{
\zkmDataTablePrep
\begin{tabular}{cc|>{\centering}p{4em}>{\centering}p{4em}|c}
\multicolumn{2}{c}{Frequency (MHz)} & \multicolumn{2}{c}{Anharmonicity (MHz)} & \multicolumn{1}{c}{Cross-Kerr (MHz)}\tabularnewline[1pt]
$\omega_{\mathrm{D}}/2\pi$ & $\omega_{\mathrm{B}}/2\pi$ & $\alpha_{\mathrm{D}}/2\pi$ & $\alpha_{\mathrm{B}}/2\pi$ & $\chi_{\mathrm{DB}}/2\pi$\tabularnewline
\hline 
\noalign{\vskip\doublerulesep}
6010 & 8670 & 85 & 180 & 278\tabularnewline
5824 & 8878 & 97 & 206 & 281\tabularnewline
-3.2\% & 2.3\% & 12\% & 13\% & 1.1\%\tabularnewline[\doublerulesep]
\end{tabular}} 
\par\end{centering}
\caption{\label{tab:DiTransmon in waveguide}\textbf{Two-qubit, one-waveguide
devices. }Summary of measured and calculated Hamiltonian parameters
for the device described in Methods . Indices D and B denote the
dark and bright modes, respectively. For each device, the first (second)
row summarizes the measured,~$m$, (bare calculated,~$c$) values.
The third row quantifies the bare agreement, i.e., $\left(c-m\right)/c$.}
\end{table*}

\emph{Sample holder.} Sample holders were machined in aluminum alloy
6061, seams were formed using thin indium gaskets placed in machined
grooves in one of the mating surfaces. Only non-magnetic components
were used in proximity to the samples, molybdenum washers, aircraft-alloy
7075 screws (\emph{McMaster}/\emph{Fastener Express}) with less than
1\% iron impurities, and non-magnetic \emph{SubMiniature version A}
(SMA) connectors.

\emph{Cryogenic setup. }Samples were thermally anchored to the 15\,mK
stage of a cryogen-free dilution refrigerator (\textit{Oxford Triton
200}) and were measured using a standard cQED measurement setup \citep{Wallraff2004,Leghtas2015,Minev2019-Thesis}.
High-magnetic-permeability, $\mu$-metal (\emph{Amumetal A4K}) shields
together with aluminum superconducting shields enclosed all samples.
Microwave input and output lines were filtered with \emph{Eccosorb
CR-110 }infrared-frequency filters \citep{Rigetti2012,Geerlings2013},
thermally anchored at the 15~mK stage. Output lines were additionally
filtered with cryogenic isolators\emph{ }(\emph{Quinstar CWJ1019-K414})
and \emph{12\,GHz K\&L }multi-section lowpass filters. Output lines
leading up to the high-electron-mobility transistor (HEMT) amplifier\emph{
(Low Noise Factory)},\emph{ }anchored at 4~K, were superconducting
(\emph{CoaxCo Ltd. SC-086/50-NbTi-NbTi PTFE}).

\emph{Quantum amplifier. }The output signal of a sample was processed
by a \emph{Josephson parametric converter} (JPC) anchored at the~15~mK
stage and operated in amplification mode \citep{Bergeal2010,Abdo2013},
before routing to the HEMT. The JPC provide a typical gain of~21\,dB
with a typical noise-visibility ratio of 6\,dB.  See Ref.~\citep{Roy2017-Review}
for a review of the parametric amplification.

\emph{Frequency and input-output (I-O) coupling measurements. }Spectroscopic
measurements were used to determine the frequencies of the resonator
modes. Anharmonicities were determined in two-tone spectroscopy \citep{Geerlings2013,Reagor2016}.
Cross-Kerr energies were determined from dressed dephasing measurements
\citep{Gambetta2006-dephasing,Gambetta2008-qm-traj}. In particular,
the dressed-dephasing measurement sequence consisted of first preparing
the qubit in the ground state, then exciting it to the equator by
a~$\pi/2$ pulse. Subsequently, a weak readout tone excited the readout
cavity of the qubit for a fixed duration, 10~times the readout cavity
lifetime~$\kappa_{r}$, after which we measure the qubit~X and~Y
Bloch vectors, after waiting for a time~$5/\kappa_{r}$ for any photons
in the cavity to leak out. By varying the amplitude and frequency
of the applied weak-readout tone, we could calibrate both the strength
of our readout, in steady-state photon number in the readout cavity,
and the value of the cross-Kerr frequency shift between the qubit
and readout resonator. The values could be obtained from fits of the~X
and~Y quadratures. For each sample, the coupling quality factor of
the readout-cavity mode, denoted~$Q_{C}$, was extracted from the
spectroscopic response of the readout cavity at low photon numbers~\citep{Geerlings2013,Reagor2016},
by measuring the scattering parameters, $S_{21}$ or~$S_{11}$.

To test EPR's robustness to experimental variability and its applicability
over wide range of experimental conditions, the presented samples
were fabricated in multiple runs and measured in different cooldowns.
Some devices were subjected to as many as 6 thermal cycles.

The Hamiltonian parameters and coupling energies for each sample were
also calculated, following the EPR method presented in section on
the general approach. In particular, we modeled the sample geometry
and materials in a FE electromagnetic simulation, as explicated in
Supplementary Section~\ref{app:FE-sims}. Our aim in writing this
supplementary section has been to provide an easy access point to
the practical use of the EPR method, which we hope will benefit the
reader, and allow them to adopt it easily. Our choice of simulation
software was the \emph{Ansys High Frequency Electromagnetic Field
Simulation} (HFSS), although we emphasize that the EPR ideas translate
to any standard EM eigenmode simulation package. Further, we modeled
the loss due to the input-output couplers in the simulation as~$50\,\Omega$
resistive sheets, see Supplementary Section~\ref{app:dissipation:radiative}.
The eigenmode analysis provided the calculated \emph{I-O} quality
factors and Purcell limits. All electromagnetic and quantum analyses,
including the extraction of participations form the eigenfields and
the numerical diagonalization of the Hamiltonian to extract its quantum
spectrum, were performed in a fully automated manner using the freely
available \textsc{pyEPR} package \citep{Note1}.

The mode quality due to the input-output coupling, $Q_{C}$, was set
by the length of the \emph{I-O} SMA-coupler pin. Its length inside
the sample-holder box was measured at room temperature using calipers.
This nominal length was used then used in the HFSS model to create
a 3D model of the pin inside the sample holder. The quality factor~$Q_{C}$ was then obtained from the eigenmode eigenvalue. We remark
that the measurement of the pin-length is accurate to no more than
20\%; further, it can be affected by various idiosyncrasies, such
as bending of the thin SMA center pin. Nonetheless, the predictions
of the quality factors for low-Q modes were observed to be very reasonable
estimates, and, similarly, the predicted Purcell limits for qubit
and high-Q cavity modes were consistent with estimates from measurements.

\begin{figure}
\centering{}\includegraphics[width=3.375in]{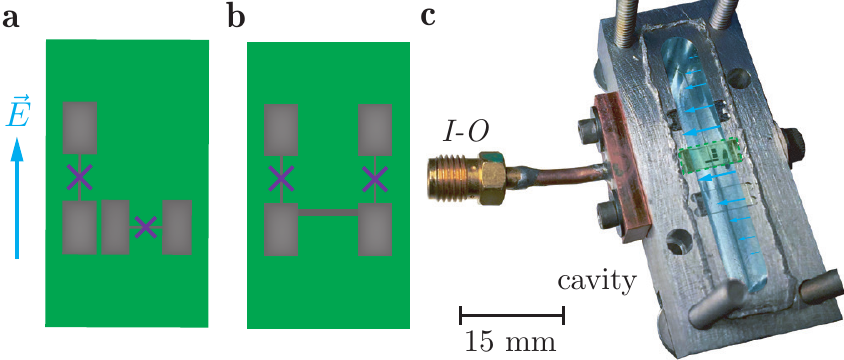}
\caption{\label{fig:devices:2q1c-ditransmon}\textbf{Two\textendash qubit,
one-cavity devices.} \textbf{a} and \textbf{b} Not-to-scale diagram
illustrating chip configurations~A and~B, respectively. Vertical
blue arrow indicates cavity electric field orientation. Crosses mark
the location of Josephson tunnel junctions. \textbf{c} Optical photograph
of sample R2C1. Bottom half of aluminum sample holder is visible;
top half is removed. The two-qubit chip (outlined by the dashed green
box) is housed in the middle of the readout cavity (highlighted in
blue). Cavity fundamental mode electric field profile~$\vec{E}$ depicted
by arrows. \emph{Input-output }SMA pin coupler labeled\emph{~I-O}.}
\end{figure}

\subsection{Devices}

\subsubsection{Two-qubit, one-cavity devices}

\textbf{\label{app:devices:2q1c}}

\paragraph{Device description.}

We measured 6 samples that were each comprised of two qubits and one
cavity. The cavity was a standard, machined aluminum cavity \citep{Paik2011}.
It housed either one or two sapphire chips, which were either patterned
with transmon qubits or simply blank. Each transmon consisted of two
thin-film aluminum pads connected by a Josephson junction. We tested
two configurations of chips and patterns. Configuration A consisted
of one chip with two \textit{orthogonal} qubits, as depicted in Fig.~\ref{fig:devices:2q1c-ditransmon}(a).
Similarly, configuration B consisted of one chip with two \textit{parallel}
qubits, depicted in Fig.~\ref{fig:devices:2q1c-ditransmon}(b). The
two qubits were aligned parallel to each other; however, unlike configuration
B, there was no galvanic connection between them. The results of the
measurements are presented in Table~\ref{tab:ditransmon}.

Samples R1C9, R2C1, R7C1, and R2C1, R3C2 were fabricated in configuration
A, sample DT3 was fabricated in configuration B. Three of the sample
(R2C1, R7C1, R3C1) were fabricated simultaneously on the same sapphire
wafer, all with nominally identical dimensions. Additionally, R2C1
and R7C1 were designed to have nominally the same Josephson junctions
energy, $E_{j}$. The rest of the samples (R1C9, DT3, and R3C2) 
were fabricated at different times and on different wafers. The dimensions
of their transmons and the inductance of the junctions were designed
to be different. Only sample R3C2 was designed to be very similar
to the nominally identical sample R2C1 and R7C1, but with adjusted~$E_{j}$. For samples R2C1, R7C1, R3C1, and R3C2 a second, un-processed,
un-patterned, blank sapphire chip was placed in parallel with the
qubit carrying chip {[}see Fig.~\ref{fig:devices:2q1c-ditransmon}(c){]}
to purposefully lower the readout cavity frequency, thus bringing
it within the JPC amplification band.

Configurations A and B were designed to test the ability of the EPR
method to calculate the mixing between strongly coupled modes. The
strong coupling was achieved in two distinct ways. First, configuration~A
used the spatial proximity of the two qubits to yielded a strong capacitive
coupling between them, which resulted in large qubit-qubit mixing.
Second, instead of spatial proximity, configuration B used a galvanic
connection between the qubits to yield strong hybridization. Our two-qubit
designs share some similar-in-spirit characteristics with the promising
recent developments reported in Refs.~\onlinecite{Gambetta2011-Purcell, Srinivasan2011, Diniz2013, Dumur2015, Zhang2017,
Roy2017-3qubits}, but our implementation is distinct and is designed to provide several
unique advantages.

\paragraph{Mode structure and interesting physical insights.}

\noindent \textit{Configuration A} is characterized by strong capacitive
coupling between the two transmons, which have different pad sizes,
see Fig.~\ref{fig:devices:2q1c-ditransmon}(c), and hence different
normal-mode frequencies. Due to the strong hybridization, each qubit
normal mode consists of some excitation in the vertical and some in
the horizontal transmon. With some foresight, we will label the vertical
mode \textit{bright} (B), and the horizontal \textit{dark} (D). The
bright-mode resonance is higher in frequency, and thus is closer to
the resonance of the readout cavity mode (C). This smaller detuning
made it a natural choice for designing stronger coupling between it,
(B), and the readout mode (C). This was implemented by orienting the
transmon design that participates in mode (B) vertical.

To understand this design choice, let us first consider the popular
analogy \citep{Koch2007,Devoret2007-HowStrongCanCouplingBe} between
circuit-QED and cavity-QED, often used to discuss mode couplings.
In this atomic analogy, the transmon qubit is analogous to a real
atom inside the cavity. Thus, it can described by an electric dipole
moment~$\vec{\mathrm{d}}_{B}$. Meanwhile, its coupling, cross-Kerr,
etc.~to the cavity mode are derived from the electric-dipole coupling
interaction. In particular, the coupling amplitude is proportional
to~$\vec{\mathrm{d}}_{B}\cdot\vec{E}$, where~$\vec{E}$ is the
cavity electric field at the transmon junction. From this analogy,
one can infer that the coupling is maximized when the two are parallel,
$\vec{\mathrm{d}}_{B}\parallel\vec{E}$, and one could hope to measure
a strong cross-Kerr between the bright qubit and the cavity. This
successful conclusion is true, but a coincidence. We will shortly
discuss how this popular analogy \textit{fails} spectacularly for
the dark mode in configuration~B. Instead, we will argue that a correct
way to understand the nonlinear coupling between the two modes is
through the participation ratio, which will provide the correct coupling
for both configuration~A and~B.

Before proceeding to configuration B, we note one further useful features
that configuration~A exhibits. In particular, while the bright qubit
mode can be Purcell limited \citep{Houck2008-Purcell,Gambetta2011-Purcell},
the dark mode is simultaneously Purcell protected. Thus, one can potentially
achieve a high ratio in the \emph{I-O} bath coupling of the two qubits.

\textit{Configuration B} has two qubit modes, which we will also label
\textit{dark}~(D) and \textit{bright} (B). Since both transmons are
designed with the exact same transmon pad geometry and junction energy~$E_{J}$,
see Fig.~\ref{fig:devices:2q1c-ditransmon}, we can expect that no
single junction is preferred, due to the symmetry of the sample. This
is in sharp contrast to the asymmetric energy distribution in configuration
A. Returning to configuration B, we can estimate that in each qubit
mode, both junctions participate equally and with near maximal allowed
participation, 
\begin{equation}
p_{\mathrm{D}1}=p_{\mathrm{D}2}=p_{\mathrm{B}1}=p_{\mathrm{B}2}\approx\frac{1}{2}\;.\label{eq:ditrans-p=00003D0.5}
\end{equation}
If the two transmons were well-separated spatially and not connected,
they would be uncoupled. However, the galvanic connection between
the two lower pads, see Fig.~\ref{fig:devices:2q1c-ditransmon}(a),
results in a very strong hybridization and splitting between the nominally
identical transmons. The result of the strong hybridization is a symmetric
and antisymmetric combination of the two bare transmons. In other
words, the hybridization results in a common mode, namely~(B), where
both junctions oscillate in-phase, and a differential mode namely~(D),
where both junctions oscillate out-of-phase. These phase relationships
are captured by the signs:
\begin{align}
s_{\mathrm{D}1} & =1\;, & s_{\mathrm{D}2} & =-1\;,\label{eq:ditransmon-ex-1}\\
s_{\mathrm{B}1} & =1\;, & s_{\mathrm{B}2} & =+1\;.\label{eq:ditransmon-ex-2}
\end{align}

In an attempt to understand how these hybridized qubit modes will
couple to the cavity mode (C), let us first consider the atomic analogy
again. When the two junctions oscillate in phase, in the (B) mode,
the net dipole moment of the bright mode, $\vec{d}_{\mathrm{B}}$,
must be large, since it is the sum of the two junction dipole contributions.
Secondly, $\vec{d}_{\mathrm{B}}$ must be oriented in the vertical
direction, parallel to the cavity electric field~$\vec{E}$. Hence,
we would conclude that the bright mode coupling is large, $\vec{d}_{\mathrm{B}}\cdot\vec{E}\gg0$,
and there should be a strong cross-Kerr interaction between the cavity
and bright qubit. Continuing the analogy in the case of the dark mode,
we would deduce that the net dipole moment of the dark mode is zero,
since the two junctions oscillate out of phase, and cancel each other's
contribution, $\vec{d}_{\mathrm{D}}=0$. Thus, we should not expect
any coupling between the dark qubit and the cavity mode, $\vec{d}_{\mathrm{D}}\cdot\vec{E}=0$.
To the contrary of this conclusion, as can be seen in the measured
results in Table~\ref{tab:ditransmon}, the nonlinear coupling of
the dark and bright qubit to the cavity is nearly equal. The atomic
analogy and the dipole argument have failed completely. We can understand
the origin of this failure and how to arrive at the correct conclusion
by using the energy-participation ratio. As embodied in Eq.~\eqref{eq:chi_mj},
in the dispersive regime, the nonlinear coupling between two modes,
in this case a qubit and cavity, is given by the overlap of the EPR
distribution. In particular, the cross-Kerr amplitude between the
dark qubit and the readout cavity mode is given by 
\begin{align}
\chi_{\mathrm{DC}} & =\frac{\hbar\omega_{D}\omega_{C}}{4E_{J}}\left(p_{\mathrm{D}1}p_{\mathrm{C}1}+p_{\mathrm{D}2}p_{\mathrm{C}2}\right)\;,\label{eq:ditrans chi }
\end{align}
where both junctions have the same junction energy~$E_{J}$, and~$\omega_{B}$
(resp: $\omega_{C}$) denotes the dark qubit (resp: cavity) mode frequency.
The signs, used in the atomic dipole logic do not factor into the
coupling, because the Josephson mechanics is fundamentally different.
To obtain~$\chi_{\mathrm{BC}}$, one can replace the label `D' with
`B' in Eq.~\eqref{eq:ditrans chi }. Then, it is easy to use Eq.~\eqref{eq:ditrans-p=00003D0.5}
to show that the ratio of two Kerr couplings is not zero, but rather
of order unity, 
\begin{equation}
\chi_{\mathrm{BC}}/\chi_{\mathrm{DC}}=\omega_{B}/\omega_{D}\;.
\end{equation}

\paragraph{Failure of the conventional dipole approach.}

We showed that although the heuristic atomic analogy seems seductively
accurate, it fails completely in some cases to predict the nonlinear
couplings. Instead, one can use the intuition and calculation method
provided by the EPRs.

As an added note, we observe that Eqs.~\eqref{eq:ditransmon-ex-1}
and~\eqref{eq:ditransmon-ex-1} embodies the orthogonality of the
participations, see Eq.~\eqref{eq:pmjpmj' orthogonality}. We also
remark that although the atomic analogy fails in the case of the nonlinear
couplings, it can yield some guidance when considering the \textit{linear}
mixing of the modes, useful for discussing the Purcell effect. To
illustrate, let us briefly extend the atomic analogy. The dipole-like
coupling between the bright mode and the cavity suggests that the
bright mode will inherit some coupling to the environment, mediated
by the cavity. Thus, since~$\vec{d}_{\mathrm{B}}\cdot\vec{E}\gg0$,
we can expect the bright qubit to potentially be Purcell limited.
In contrast, since~$\vec{d}_{\mathrm{D}}\cdot\vec{E}=0$, we could
expect the dark qubit to be Purcell protected. Both of these qualitative
Purcell predictions are valid, but to quantify them, we will use the
EPR method and FE eigenmode simulation of the sample, as will be discussed
shortly.

\paragraph{Experimental results.}

\noindent Table~\ref{tab:ditransmon} summarizes the results of the
agreement between the measured and calculated Hamiltonian parameters
for all two-qubit, one-cavity samples. The three modes in each sample
are labeled dark~(D), bright~(B), and cavity~(C); the reason for
this convention is described above. In all samples, the qubits were
designed to be in the dispersive regime with respect to the cavity,
which was detuned by 2\textendash 4\,GHz. However, in a large number
of the samples, the two qubits were strongly hybridized, often necessitating
higher-order nonlinear corrections to be included in the calculation.
This strong hybridization was used as a test of the theory in this
more challenging and fickle regime.

In total, for each sample we measured and calculated 8 frequency parameters
and one dimensionless, coupling quality factor,~$Q_{C}$, of the
readout cavity mode. In particular, in the low-excitation limit, the
nonlinear interactions among the modes were characterized by the effective
dispersive Hamiltonian
\begin{multline}
\hat{\overline{H}}/\hbar=\omega_{D}\hat{n}_{D}+\omega_{B}\hat{n}_{B}+\omega_{C}\hat{n}_{C}\\
-\frac{1}{2}\alpha_{D}\hat{n}_{D}\left(\hat{n}_{D}-\hat{1}\right)-\frac{1}{2}\alpha_{B}\hat{n}_{B}\left(\hat{n}_{B}-\hat{1}\right)\\
-\chi_{DB}\hat{n}_{D}\hat{n}_{B}-\chi_{DC}\hat{n}_{D}\hat{n}_{C}-\chi_{BC}\hat{n}_{B}\hat{n}_{C}\;,\label{eq:del}
\end{multline}
where~$\hat{n}_{D},\hat{n}_{B}$, and~$\hat{n}_{C}$ denote the dark,
bright, and cavity photon-number operator, respectively. The coupling
of the resonator mode to the bath is given by the Lindblad superoperator
term~$\kappa_{C}\ensuremath{\mathcal{D}[\hat{a}_{C}]}\rho$, where~$\kappa_{C}=\omega_{C}/Q_{C}$,
and~$\rho$ is the density operator.

We remark that all samples in configuration A demonstrated a large
asymmetry in the Kerr coupling between the bright-to-cavity and dark-to-cavity
coupling, $\chi_{BC}\gg\chi_{DC}$. In contrast, samples in configuration
B demonstrated near equal coupling, $\chi_{BC}\approx\chi_{DC}$.
In both configurations A and B, the dark mode was Purcell protected,
we calculated a Purcell coupling factor of~$Q_{\text{Purcell}}^{D}\gg10^{7}$,
using the eigenmode method described in Supplementary Section~\ref{app:dissipation:radiative}.
On the other hand, the bright mode was somewhat Purcell limited, $Q_{\text{Purcell}}^{B}\approx10^{6}$.
From the relative Rabi amplitudes of the dark and bright qubits, we
could verify the order of magnitude scaling calculated for the Purcell
effect.

\begin{figure}
\centering{}\includegraphics[width=3.375in]{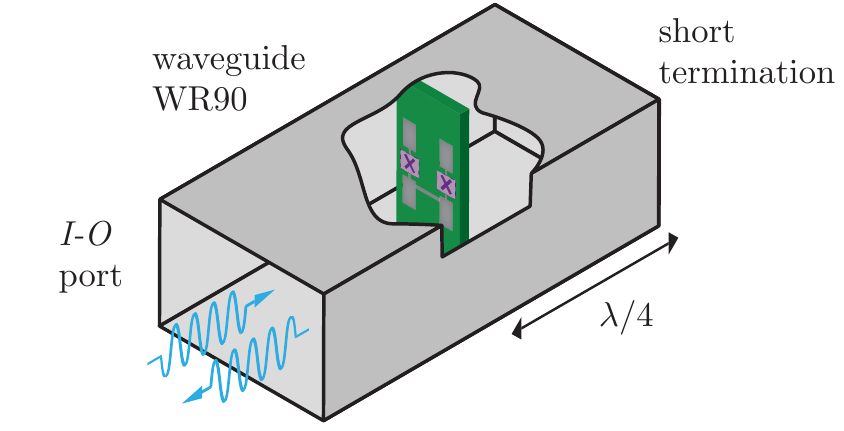} \caption{\label{fig:appdx:devices:2q1w} \textbf{Schematic representation (not
to scale) of superconducting waveguide device DTW1}. Two-qubit chip
is placed~$\lambda/4$ from a short termination in the waveguide.
Guided input-output waves are launched and monitored through the \emph{I-O}
port connecting to a standard SMA adapter (not shown).}
\end{figure}

\subsubsection{Two-qubit, single-waveguide devices}

\textbf{\label{app:2q1w}}

\noindent We measured a two qubit sample inside of a waveguide. Figure~\ref{fig:appdx:devices:2q1w}
presents the setup, and depicts the sample, which was of the configuration
B type presented in Fig.~\ref{fig:devices:2q1c-ditransmon}. The
sample chip was positioned inside an aluminum WR90 waveguide. The
waveguide was terminated in a short at one side, and attached to an
impedance-matched SMA coupler port on the input-launcher side, which
was used to drive and measure the waveguide. The chip was centered
inside the cross-section of the waveguide, and placed~$\lambda/4$
away from the termination wall, at the measurement frequency. The
rest of the experimental setup was identical to that described in
section `Methods of the experiment'. The two qubit modes were labeled
dark and bright, similarly to the samples discussed in the section
`Two-qubit, one-cavity devices.'.

Table~\ref{tab:DiTransmon in waveguide} presents the agreement between
the measured and calculated key Hamiltonian parameters of the sample.
These consist of the two mode frequencies, two qubit anharmonicities,
and the strong cross-Kerr interaction between the two qubits.

\begin{figure}
\centering{}\vspace{0.5em}
 \includegraphics[width=3.375in]{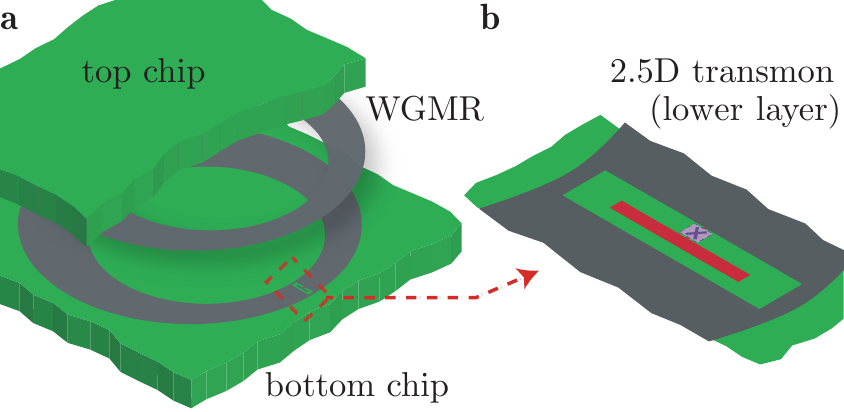} \caption{\label{fig:WGMR}\textbf{Illustration of flip-chip (2.5D) device WG1.}
\textbf{a} Depiction (not to scale) of chip stack consisting of two
chips separated by a 100~$\upmu\text{m}$ vacuum gap. The inner face
of each chip supports part of the pattern of a multi-layer whispering
gallery-mode resonator (WGMR) resonator \citep{Minev2013}. The lower
layer contains a 2.5D, aperture transmon qubit \citep{Minev2016}
embedded in the WGMR. \textbf{b} Zoomed-in view of the lower layer
of the aperture transmon. The cross marks the location of the Josephson
tunnel junction device, which connects the lower center trace to an
island embedded in the lower aperture. The capacitance to the top
layer significantly participates in determining qubit parameters.}
\end{figure}

\subsubsection{Flip-chip (2.5D), one-qubit, one-storage-cavity, one-readout-cavity
devices}

\textbf{\label{app:multilayer planar cQED}}

\noindent We also designed a multilayer planar \citep{Minev2016}
(2.5D) circuit-QED sample, depicted in Fig.~\ref{fig:WGMR}, with
the EPR method. It consisted of high-Q storage mode (S), one low-Q
readout cavity (C), and one control transmon qubit (Q). The two cavity
modes were formed in the footprint of a single whispering gallery
mode resonator \citep{Minev2013}. The three modes were in the dispersive
regime, and the storage mode was used to encode and decode quantum
information, as well as to observe parity revivals. Details of the
sample design have been reported in Ref.~\onlinecite{Minev2016}.
The agreement between the measured parameters of the sample and those
obtained by the EPR calculation methods are presented in Table~\ref{tab:wgmr}.\bigskip{}

}
{
\titleformat*{\paragraph}{\bfseries\rmfamily}
\setlength{\parskip}{1em}

\paragraph{Data availability.}

\noindent Data are available from the authors on reasonable request.

\paragraph{Code availability.}

\noindent The source code for the EPR method is open-sourced and can
be found at \href{http://github.com/zlatko-minev/pyEPR}{http://github.com/zlatko-minev/pyEPR}.

\paragraph{Acknowledgments.}

\noindent We thank S.M.~Girvin, R.J.~Schoelkopf, S.~Nigg, H.~Paik,
A.~Blais, H.E.~T\"ureci, F.~Solgun, V.~Sivak, S.~Touzard, N.~Frattini,
S.~Shankar, C.~Axline, V.V.~Albert, K.~Chou, A.~Petrescu, D.~Cody,
A.~Eickbusch, and E.~Flurin for valuable discussions, and the I.~Siddiqi
and B.~Huard groups for using the early versions of \textsc{pyEPR}.
This research was supported by the US Army Research Office (ARO) Grant
No. W911NF-18-1-0212. Z.K.M. acknowledges partial support from the
ARO (W911NF-16-1-0349). M.H.D. acknowledges partial support from the
ARO (W911NF-18-1-0020) and the Air Force Office of Scientific Research
(FA9550-19-1-0399). The view and conclusions contained in this document
are those of the authors and should not be interpreted as representing
the official policies, either expressed or implied, of the Army Research
Office or the US Government. The US Government is authorized to reproduce
and distribute reprints for Government purposes notwithstanding any
copyright notation herein.

\paragraph*{Author contributions.}

\noindent Z.K.M. conceived and developed the method and the theory,
performed the experiments, and analyzed the data. Z.L. and M.H.D.
contributed to the theory. I.M.P. and S.O.M. contributed to the measurement
and fabrication of the devices L.C. assisted with the simulations.
Z.K.M. and M.H.D. wrote the manuscript. All authors provided suggestions,
discussed the results and contributed to the manuscript.

\paragraph{Competing interests.}

\noindent The authors declare no competing interests.

} 
} 

\clearpage{}

\appendix
\global\long\def\thefigure{S\arabic{figure}}%
\global\long\def\ket#1{\left|#1\right\rangle }%

\global\long\def\bra#1{\left\langle #1\right|}%

\global\long\def\braket#1#2{\left\langle #1\middle|#2\right\rangle }%

\global\long\def\ketbra#1#2{\left|#1\vphantom{#2}\right\rangle \left\langle \vphantom{#1}#2\right|}%

\global\long\def\kb#1#2{\left|#1\vphantom{#2}\right\rangle \left\langle \vphantom{#1}#2\right|}%

\global\long\def\braOket#1#2#3{\left\langle #1\middle|#2\middle|#3\right\rangle }%

\global\long\def\isdef{\coloneqq}%

\global\long\def\m#1{\mathrm{\mathbf{#1}}}%

\global\long\def\mh#1{\boldsymbol{\hat{\mathrm{#1}}}}%

\global\long\def\abs#1{\left|#1\right|}%

\global\long\def\normord#1{{:\mathrel{\mspace{1mu}#1\mspace{1mu}}:}}%

\global\long\def\diag{\operatorname{diag}}%

\global\long\def\Re{\operatorname{Re}}%

\global\long\def\ddt{\frac{\mathrm{d}}{\mathrm{d}t}}%

\global\long\def\phit{\m{\Phi}_{\mathrm{t}}}%

\global\long\def\phithat{\m{\hat{\Phi}}_{\mathrm{t}}}%

\global\long\def\phitd{\dot{\m{\Phi}}_{\mathrm{t}}}%

\global\long\def\phim{\m{\Phi}_{\mathrm{m}}}%

\global\long\def\phimhat{\m{\hat{\Phi}}_{\mathrm{m}}}%

\global\long\def\phimd{\dot{\m{\Phi}}_{\mathrm{m}}}%

\global\long\def\dashedph{s}%

\global\long\def\ejlin{\mathcal{E}_{j}^{\mathrm{lin}}}%

\global\long\def\ejnl{\mathcal{E}_{j}^{\mathrm{nl}}}%

\global\long\def\pml{p_{ml}}%

\onecolumngrid 

\makeatletter

\def\p@subsection{}  
\def\thesection{\Alph{section}}
\def\thesubsection{\Alph{section}\arabic{subsection}}
\def\thesubsubsection{\Alph{section}\arabic{subsection}.\arabic{subsubsection}} 

\renewcommand{\figurename}{  \textbf{Supplementary Figure}}
\renewcommand{\tablename}{   \textbf{Supplementary Table}}
\renewcommand{\fnum@figure}{ \textbf{Supplementary Figure~\thefigure}}
\renewcommand{\fnum@table}{  \textbf{Supplementary Table~\thetable}}

\numberwithin{equation}{section} 
\renewcommand{\theequation}{\Alph{section}.\arabic{equation}}

\renewcommand{\refname}{Supplementary References}
\renewcommand{\bibname}{Supplementary References}

\makeatother

\renewcommand{\refname}{References}
\renewcommand{\bibname}{References}

\setcounter{figure}{0}
\setcounter{table}{0}

\renewcommand{\appendixname}{Supplementary Section}

\part*{Supplementary Information:\protect \\
Energy-participation quantization of Josephson circuits}

\tableofcontents{}

\clearpage{}

\section{Theoretical foundation of the energy-participation method\label{app:main-derivation}}

{
\setlength{\extrarowheight}{8pt}
\setlength{\doublerulesep}{4pt}
\addtolength{\tabcolsep}{4pt}

\definecolor{Column1}{gray}{0.93}
\definecolor{Row1}{gray}{0.93}

\newcolumntype{z}{>{\raggedleft}p{4.5em}}

\rowcolors{1}{}{gray!4}%
\begin{longtable}[c]{z|lp{0.7\textwidth}}
\caption{\label{tab:symbols-and-key-results}\textbf{Table of key symbols and
relationships} used in the derivation of the energy-participation-ratio
method.}
\tabularnewline
\endfirsthead
\noalign{\vskip\doublerulesep}
\multicolumn{3}{c}{\textbf{\cellcolor{HeaderColor}Basic Josephson circuit variables
and parameters}}\tabularnewline[4pt]
\hline 
\noalign{\vskip\doublerulesep}
\textbf{Symbol} & \textbf{Value} & \centering{}\textbf{Description}\tabularnewline[4pt]
$\Phi_{0}$ & $h/(2e)$ & Superconducting magnetic-flux quantum\textemdash the ratio of Planck's
quantum of electromagnetic action~$h$ and the charge of a Cooper
pair~$2e$.\tabularnewline~$\phi_{0}$ & $\Phi_{0}/\left(2\pi\right)$ & Reduced flux quantum.\tabularnewline~$\Phi_{b}\left(t\right)$ & $\int_{-\infty}^{t}v_{b}\left(\tau\right)\mathrm{d}\tau$ & Generalized magnetic flux of circuit branch~$b$, at time~$t$. The
instantaneous voltage~$v_{b}\left(\tau\right)$ across the terminals
of branch~$b$ at time~$\tau$ is equivalently denoted~$\dot{\Phi}_{b}\left(\tau\right)$.
In general, $\Phi_{b}\left(t\right)=\int_{-\infty}^{t}v_{b}\left(\tau\right)\mathrm{d}\tau+\Phi_{b}\left(-\infty\right)$,
but we take the initial flux~$\Phi_{b}\left(-\infty\right)$ to be
zero, corresponding to the circuit in an equilibrium state (see Sec.~\ref{app:equilibrium-point-of-circuit}).
With this convention,~$\Phi_{b}$ is a \emph{deviation }away from
equilibrium. This variable is analogous to the elongation of a mechanical
spring.\tabularnewline~$j,\,J$ &  & The subscript or index~$j$ denotes the~$j$-th Josephson dipole,
where~$j\in\left\{ 1,\,\ldots,\,J\right\} $, and~$J$ denotes the
total number of Josephson dipoles.\tabularnewline~$\Phi_{j}\left(t\right)$ &  & Generalized magnetic-flux \emph{deviation} of a non-linear device,
the~$j$-th Josephson dipole.\tabularnewline~$\varphi_{j}\left(t\right)$ & $\Phi_{j}\left(t\right)/\phi_{0}$ & Reduced magnetic-flux variable of Josephson dipole~$j$.\tabularnewline~$\mathcal{E}_{j}\left(\Phi_{j}\right)$ & $\ejlin\left(\Phi_{j}\right)+\ejnl\left(\Phi_{j}\right)$ & Energy function of Josephson dipole~$j$. Typically split into two
component:~$\ejlin$ associated with linear interactions and~$\ejnl$
associated with nonlinear ones.\tabularnewline~$\ejlin\left(\Phi_{j}\right)$ & $\frac{1}{2}E_{j}\left(\Phi_{j}/\phi_{0}\right)^{2}$ & Linear component of the energy function of junction~$j$, defined
by the energy scale~$E_{j}$. Terms linear in~$\Phi_{j}$ are absent
since we selected an equilibrium operating point of the circuit.
The energy function~$\ejlin$ may never contain non-linear terms.\tabularnewline~$\ejnl\left(\Phi_{j}\right)$ & $\mathcal{E}_{j}\left(\Phi_{j}\right)-\ejlin\left(\Phi_{j}\right)$ & The nonlinear component of the energy function of junction~$j$.
For certain situations,~it may be favorable to select the partition
of~$\mathcal{E}_{j}$ such that~$\ejnl$ contains linear interactions.
If possible, it is often convenient to expand~$\ejnl$ in a Taylor
series around the circuit operating point:~$\ejnl\left(\Phi_{j}\right)=E_{j}\sum_{p=3}^{\infty}c_{jp}\left(\frac{\Phi_{j}}{\phi_{0}}\right)^{p}$,
where~$c_{jp}$ are the dimensionless coefficients of the expansion.\tabularnewline~$\phit\left(t\right)$ & $\left(\Phi_{\mathrm{t}_{1}},\,\Phi_{\mathrm{t}_{2}},\,\ldots\right)^{\intercal}$ & Column vector consisting of the flux \emph{deviations} of all circuit
branches enclosed in the minimum spanning tree. The roman subscript~$\mathrm{t}$
denotes the tree. The flux of individual tree-branches are denoted~$\Phi_{\mathrm{t}_{1}},\,\Phi_{\mathrm{t}_{2}},\,\ldots$\tabularnewline
\noalign{\vskip8pt}
\multicolumn{3}{c}{\cellcolor{HeaderColor}\textbf{Linearized Josephson eigenmodes}}\tabularnewline[4pt]
\hline~$m,\;M$ &  & The label~$m$ indexes sets associated with the eigenmodes of the
linearized Hamiltonian~$\mathcal{H}_{\mathrm{lin}}$. We consider~$m\in\left\{ 1,\,\ldots,\,M\right\} $,
where~$M$ is the total number of eigenmodes of relevance determined
by the context. That is,~$M$ is either the total number of eigenmodes
of~$\mathcal{H}_{\mathrm{lin}}$ or of the relevant eigenmodes included
in the finite-element simulation.\tabularnewline~$\mathcal{H}_{\mathrm{full}}$ & $\mathcal{H}_{\mathrm{lin}}+\mathcal{H}_{\mathrm{nl}}$ & Hamiltonian function of the Josephson system. It can be partitioned
into~$\mathcal{H}_{\mathrm{lin}}$, a part comprised of purely linear
terms (quadratic in~$\phit$, or equivalently~$\phim$), and~$\mathcal{H}_{\mathrm{nl}}$,
a part generally comprised of non-linear terms (higher-than quadratic
in~$\phit$, or equivalently~$\phim$). \tabularnewline~$\phim,\,\m Q_{m}$ &  & Column vectors (of length~$M$) whose elements are the eigenmode flux~$\Phi_{m}$ (considered as the generalized position) and charge~$Q_{m}$
(considered as the generalized momentum) canonical variables, associated
with the Hamiltonian~$\mathcal{H}_{\mathrm{lin}}$.\tabularnewline~$\omega_{m}$ &  & Eigenfrequency of the~$m$-th mode of~$\mathcal{H}_{\mathrm{lin}}$,
carries a dimension of circular frequency.\tabularnewline~$\m{\Omega}$ & $\Diag\left(\omega_{1},\omega_{2},\ldots\right)^{\intercal}$ & Diagonal~$M\times M$ matrix comprising the eigenfrequencies. \tabularnewline~$\phit$ & $\m E\phim$ & The spanning-tree branch fluxes~$\phit$ are a linear combination
of the eigenmode~$\phim$ fluxes. The two are related by an affine
transformation given by the properly-constructed eigenvector matrix~$\m E$,
see Eqs.~\eqref{app:eigen:E defn} and~\eqref{app:eigen modes:phib to phim}.
\tabularnewline
\noalign{\vskip8pt}
\multicolumn{3}{c}{\textbf{\cellcolor{HeaderColor}Quantum operators and energy-participation
ratios}}\tabularnewline[4pt]
\hline~$\hat{H}_{\mathrm{lin}}$ & $\sum_{m=1}^{M}\hbar\omega_{m}\hat{a}_{m}^{\dagger}\hat{a}_{m}$ & Hamiltonian operator corresponding to the Hamiltonian function~$\mathcal{H}_{\mathrm{lin}}$,
expressed in second quantization with respect to the eigenmodes of~$\mathcal{H}_{\mathrm{lin}}$. The eigenmodes (expressed in terms
of~$\phim$ and~$\m Q_{m}$) define the bosonic amplitude (lowering)
operators~$\hat{a}_{m}$.\tabularnewline~$\hat{\varphi}_{j}$ & $\sum_{m=1}^{J}\varphi_{mj}\hat{a}_{m}+\text{H.c.}$ & Operator corresponding to the reduced flux~$\varphi_{j}$ of Josephson
dipole~$j$; $\hat{\varphi}_{j}$ admits a linear decomposition in
terms of the mode operators~$\hat{a}_{m}$. The coefficients of this
expansion are the quantum zero-point fluctuations~$\varphi_{mj}$.\tabularnewline~$\varphi_{mj}$ & $s_{mj}\sqrt{\frac{\hbar\omega_{m}}{E_{j}}p_{mj}}$ & The quantum zero-point fluctuation (ZPF) of the reduced magnetic flux~$\hat{\phi}_{j}$
of Josephson dipole~$j$ due to mode~$m$; i.e., for a given mode,
the ZPF magnitude gives the non-zero standard deviation of the magnetic
flux in the ground state. The amplitude of the ZPF~$\varphi_{mj}$
is determined by the physical structure of mode~$m$, and can be
understood in terms of the energy-participation ratio~$p_{mj}$ and
its sign~$s_{mj}$.\tabularnewline~$p_{mj},\;\pml$ &  & Energy-participation ratios (EPR) of Josephson dipole~$j$ and lossy
element~$l$ in mode~$m$, comprised between zero and unity, $0\leq p_{mj},\pml\leq1$.\tabularnewline~$s_{mj},\;s_{ml}$ & $\ $ & Sign~$s_{mj}$ (resp., $s_{ml}$) of the energy-participation ratio~$p_{mj}$
(resp., lossy EPR~$p_{ml}$). The value of the sign is either~$+1$
or~$-1$. See Supplementary Figure~\ref{fig:port-signs}.\tabularnewline
\end{longtable}} 
\twocolumngrid 

In this section, we derive the EPR method from first principles. We
first review quantum electromagnetic circuit theory (Sec.~\ref{app:Review-of-electrical-theory}),
then use it to find the quantum eigenmodes of the circuit (Secs.~\ref{app:Josephson-system}\textendash \ref{app:quantize-modes}).
In Sec.~\ref{app:EPR-defn}, we define the EPR of a Josephson dipole
in a quantum mode, and use it to find the quantum zero-point fluctuations
(ZPF). The universal properties and sum rules of the EPRs are detailed
in Sec.~\ref{app:universal-constraints-epr}.

In Section~\ref{app:equilibrium-point-of-circuit}, we detail use
of the EPR method for biased systems\textemdash those that incorporate
active elements (such as current and voltage sources) or external
bias conditions that result in persistent currents (such as a frustration
by an external magnetic flux).

The EPR derivation consists of a series of exact transformation. No
approximation are made in deriving the EPR or in using it to find
the quantum Hamiltonian of a general Josephson system. In this sense,
the results are universal. Approximation are made in practice when
using numerical methods.

\subsection{Review of electrical circuit theory, and the Josephson junction\label{app:Review-of-electrical-theory}}

We briefly recount the basic formulation of classical electrical theory.
This formulation will be used as a stepping stone in the following
derivation. We focus on the lumped-element regime. An electrical element
is said to be in the lumped-element regime when its \textit{\emph{physical
dimensions are negligible with respect to the electromagnetic wavelengths
considered in the analysis. In other words, self-resonances and parasitic
internal dynamical degrees of freedom of the element are neglectable.}}

Electrical theory describes the physical laws governing four basic
manifestations of electricity. For each element, these are captured
in the variables for voltage~$v\left(t\right)$, current~$i\left(t\right)$,
charge~$Q\left(t\right)$, and magnetic flux~$\Phi\left(t\right)$,
as functions of time~$t$. Maxwell's equations and conservation of
charge lead to the six universal relationships presented in Eqs.~\eqref{eq:universal-v-phi}\textendash \eqref{eq:universal-energy-power}
\citep{feynmanLec-Vol-II,Pozar}.

The time-instantaneous voltage~$v\left(t\right)$ and generalized
magnetic flux across an element are related by the lumped-element
version of Faraday's law of induction \citep{Yurke1984,Devoret1995,feynmanLec-Vol-II,Pozar},
\begin{equation}
v\left(t\right)=\ddt\Phi\left(t\right)\quad\text{and}\quad\Phi\left(t\right)=\int_{-\infty}^{t}v\left(\tau\right)\,\mathrm{d}\tau+\Phi\left(-\infty\right)\;.\label{eq:universal-v-phi}
\end{equation}
By convention, the reference orientation of voltage is opposite to
that of current. This is not the convention typically used in Lenz's
law; in electromagnetism, the current-density vector~$\vec{J}$ and
the electric-field vector~$\vec{E}$ are usually projected on the
same orientation.

From charge conservation, it follows that the time-instantaneous current~$i\left(t\right)$
and the charge~$Q\left(t\right)$ having passed through the element,
obey a relation similar to that of the voltage and flux, 
\begin{equation}
i\left(t\right)=\ddt Q\left(t\right)\quad\text{and}\quad Q\left(t\right)=\int_{-\infty}^{t}i\left(\tau\right)\,\mathrm{d}\tau+Q\left(-\infty\right)\;.\label{eq:universal-i-q}
\end{equation}
All four variables have their support on a non-compact set; i.e.,
$i,\,v,\,Q,\,\Phi\in\left[-\infty,\infty\right]$.

\paragraph{Reference state and initial conditions.}

In Eqs.~\eqref{eq:universal-v-phi} and~\eqref{eq:universal-i-q},
we now assume zero-valued initial conditions,~$\Phi\left(-\infty\right)=Q\left(-\infty\right)=0$.
In the case of a circuit frustrated by sources such that the equilibrium
system state is non-zero, we can define the variables~$i,\,v,\,Q$,
and~$\Phi$ to denote deviations away from the global equilibrium
of the circuit, as discussed in more detail in Sec.~\ref{app:equilibrium-point-of-circuit}.
By way of analogy, imagine a mechanical spring. The spring is stretched
from its rest position to a new equilibrium by a second stretched
spring. The deviations of the spring are measured from the equilibrium
position of the combined system (this is what we mean by global equilibrium),
not the spring in isolation.

\paragraph{Power and energy.}

The instantaneous power~$p\left(t\right)$ delivered\emph{ to} an
element and the total energy absorbed by the element~$\mathcal{E}\left(t\right)$
are
\begin{equation}
\ddt\mathcal{E}\left(t\right)=p\left(t\right)=v\left(t\right)i\left(t\right)\quad\text{and}\quad\mathcal{E}\left(t\right)=\int_{-\infty}^{t}p\left(t\right)\,\mathrm{d}\tau\;.\label{eq:universal-energy-power}
\end{equation}
Given our convention for the orientation of~$i$ and~$v$, power
delivered to the element is positive if~$p\left(t\right)$ is positive.
Passive elements (i.e., non-source elements) obey~$\mathcal{E}\left(t\right)-\mathcal{E}\left(-\infty\right)\geq0$
for all~$t$, and lossless elements convert all of their energy into
stored electric or magnetic energy.

\paragraph{Capacitive and inductive elements.}

A capacitive (resp., inductive) element is described by an algebraic
relationship between charge and voltage (resp., flux and current).
Let us introduce the simplest linear, passive, time-invariant elements.
The simplest capacitor is defined by the constitutive relationship~$Q\left(t\right)=Cv\left(t\right)$, where~$C$ is its capacitance\textemdash a
positive, real constant. The dual relationship~$\Phi\left(t\right)=Li\left(t\right)$
defines the simplest inductor, where~$L$ is its inductance, also
a positive, real constant. In terms of the flux across the element,
the energies of the simple capacitor and inductor are
\begin{equation}
\mathcal{E}_{\mathrm{cap}}\left(\dot{\Phi}\right)=\frac{1}{2}C\dot{\Phi}^{2}\quad\text{and}\quad\mathcal{E}_{\mathrm{ind}}\left(\Phi\right)=\frac{1}{2L}\Phi^{2}\;,\label{eq:E-cap-ind-in-phi}
\end{equation}
respectively (and assuming~$\Phi\left(-\infty\right)=Q\left(-\infty\right)=0$).
Since~$C,L\geq0$ and~$\Phi\left(t\right)\in\mathbb{R}$ for all~$t$,
we can verify that the total energy gained by these elements is always
positive, as required for passive elements.

\paragraph{Josephson tunnel junction.}

The chief non-linear element used in circuit quantum electrodynamics
is the Josephson tunnel junction \citep{Josephson1962,Devoret1995,Girvin2014},
characterized by the flux-controlled inductive relationship~$i\left(t\right)=I_{0}\sin\left(\Phi\left(t\right)/\phi_{0}\right)$,
where~$\phi_{0}\isdef\hbar/2e$ is the reduced magnetic flux quantum
and~$I_{0}$ is the junction critical current. The energy function
of the junction in terms of flux is

\begin{equation}
\mathcal{E}_{J}\left(\Phi\right)=E_{J}\left(1-\cos\left(\Phi/\phi_{0}\right)\right)\,,\label{eq:jj-energy-func}
\end{equation}
where~$E_{J}\isdef I_{0}\phi_{0}$ denotes the Josephson energy.
For small deviations about~$\Phi=0$, ignoring the constant energy
offset, 
\begin{equation}
\mathcal{E}_{J}\left(\Phi\right)\approx\frac{1}{2}E_{J}\left(\Phi/\phi_{0}\right)^{2}-\frac{1}{4!}E_{J}\left(\Phi/\phi_{0}\right)^{4}+\mathcal{O}\left(\Phi^{6}\right)\;.\label{eq:Ej-jj-expansion-simple}
\end{equation}
To lowest order, the junction responds as a linear inductor with inductance~$L_{J}\isdef E_{J}/\phi_{0}^{2}$
{[}compare this to Eq.~\eqref{eq:E-cap-ind-in-phi}{]} It is useful
to introduce the reduced magnetic flux~$\varphi\isdef\Phi/\phi_{0}$
of the junction.

\paragraph{Single- vs.~multi-valued energy functions.}

The Josephson junction exhibits a fundamental asymmetry. Inverting
the current-flux relationship leads to a multi-valued function with
an infinite number of branches; i.e., $\varphi=\sin^{-1}\left(i\left(t\right)/I_{0}\right)+2\pi k$
or~$\varphi=\pi-\sin^{-1}\left(i\left(t\right)/I_{0}\right)+2\pi k$,
where~$k$ is some integer. For flux-controlled inductors, the multi-valued
situation can be avoided by favoring a description in terms of flux,
rather than charge.

\paragraph{Junction in a frustrated circuit.}

Embedding the junction in a frustrated circuit can lead to a non-zero
equilibrium value for~$\varphi$. For deviations~$\varphi-\varphi_{\mathrm{eq}}$
away from the equilibrium value~$\varphi_{\mathrm{eq}}$, see also
Eq.~\eqref{eq:frustrated-jj-energy},
\begin{multline}
\mathcal{E}_{J}\left(\varphi\right)\approx E_{J}\bigg[\sin\left(\varphi_{\mathrm{eq}}\right)\left(\varphi-\varphi_{\mathrm{eq}}\right)+\frac{1}{2}\cos\left(\varphi_{\mathrm{eq}}\right)\left(\varphi-\varphi_{\mathrm{eq}}\right)^{2}\\
-\frac{1}{6}\sin\left(\varphi_{\mathrm{eq}}\right)\left(\varphi-\varphi_{\mathrm{eq}}\right)^{3}\bigg]+\mathcal{O}\left(\varphi^{4}\right)\;.\label{eq:Ej-jj-frustrated}
\end{multline}
We can identify the differential inductance of the junction at~$\varphi=\varphi_{\mathrm{eq}}$
to be~$L_{J}\left(\varphi_{\mathrm{eq}}\right)=L_{J}/\cos\left(\varphi_{\mathrm{eq}}\right)$.
In Sec.~\ref{app:equilibrium-point-of-circuit}, we discuss how~$\varphi$
can be taken as a deviation away from the equilibrium value~$\varphi_{\mathrm{eq}}$,
in effect mapping~$\varphi-\varphi_{\mathrm{eq}}\mapsto\varphi$,
see Eq.~\eqref{eq:simple-flux-deviation}. Also in Sec.~\ref{app:equilibrium-point-of-circuit},
we discuss sources terms such as the one presented by~$E_{J}\sin\left(\varphi_{\mathrm{eq}}\right)\left(\varphi-\varphi_{\mathrm{eq}}\right)$.
Such energy terms linear in~$\varphi$ turn the junction into an
active component, capable of supplying current; i.e., we observe that
the current relationship of the junction to lowest order~$I\left(t\right)=\frac{\partial\mathcal{E}_{J}}{\partial\Phi}=I_{0}\sin\left(\varphi_{\mathrm{eq}}\right)+L_{J}^{-1}\left(\varphi_{\mathrm{eq}}\right)\Phi\left(t\right)+\mathcal{O}\left(\Phi^{2}\right)$
contains the constant term~$I_{0}\sin\left(\varphi_{\mathrm{eq}}\right)$,
which acts like a current source.

\paragraph{Relationship between the gauge-invariant superconducting phase difference
of a tunnel junction and the reduced magnetic flux~$\varphi$ (compact
vs.~non-compact variables).}

In superconductivity, the Josephson energy coupling two small superconducting
islands has a cosine dependence on the gauge-invariant phase difference~$\theta_{j}$
of the superconducting phases of the two islands~\citep{tinkham2004-book,Clarke2004}.
This macroscopic variable is a phase angle\textemdash a compact variable
in the half-open interval~$\theta\in[0,2\pi[$; in contrast with
the non-compact variable~$\varphi\in\left[-\infty,+\infty\right]$,
which must be used in circuits where a superconducting wire connects
the two sides of the junction.

In our treatment of the Josephson circuits so far, we have completely
ignored this subtlety. Rather, we have based our discussion on the
non-compact variable~$\varphi$. The relationship between these two
collective, macroscopic variables is~$\theta=\Phi/\phi_{0}\mod2\pi$.
Although the variable~$\varphi$ is non-compact, the associated wavefunction~$\psi\left(\varphi\right)$
is submitted in practice to constraints (like confinement in one or
a few potential wells) such that it is decomposed onto a basis set
of wavefunctions that are indexed by a single discrete (rather than
continuous) index; for example, the Fock basis~$\psi_{n}\left(\varphi\right)$,
with~$n\in\mathbb{N}$. Quantum-mechanically, the representation
of~$\psi\left(\varphi\right)$ and~$\psi\left(\theta\right)$ is
therefore not very different since~$\psi\left(\theta\right)$ is
represented by the discrete rotor basis,~$\psi_{k}\left(\theta\right)$,
with~$k\in\mathbb{Z}$. 

\paragraph{A broken symmetry.}

The compact support of~$\theta$ corresponds to a symmetry that is
usually broken in most circuits\textemdash that of the impossibility
to distinguish between different values of~$\varphi$ differing by~$2\pi$.
Losses associated with the junction, or coupling to other elements,
such as in the RF-SQUID or fluxonium qubit~\citep{Manucharyan2009},
render~$2\pi$ turns of~$\varphi$ macroscopically distinguishable\textemdash hence
demanding a description in terms of non-compact support corresponding
to a point on an open-ended line rather than a circle. For certain
special cases, however, it may be advantageous to retain the compact
support version. But even in such cases, one can start with the non-compact
version of~$\varphi$ and recover the compact version by a limit
procedure~\citep{Koch2007}. Thus, in this article, we use only gauge-invariant
phases with non-compact support; i.e.,~$\varphi\in\left[-\infty,+\infty\right]$.

\paragraph{Flux-controlled inductor.}

To generalize from the Josephson junction and introduce more general
non-linear elements, consider the flux-controlled inductor. It is
defined by an algebraic relationship~$i\left(t\right)=h\left(\Phi\left(t\right),t\right)$,
where~$h$ is a single-valued function. If~$h$ is time-invariant,
such as for the case of the Josephson tunnel junction, the energy
function {[}see Eq.~\eqref{eq:universal-energy-power}{]} is
\begin{equation}
\mathcal{E}_{\mathrm{ind}}\left(\Phi\right)=\int_{\Phi\left(-\infty\right)}^{\Phi\left(t\right)}h\left(\Phi'\right)\,\mathrm{d}\Phi'\;.\label{eq:Eind-flux-control}
\end{equation}
Similar results can be obtained for current controlled inductors and
for generalized capacitors.

\paragraph{A network of circuit elements.}

An electrical circuit is an interconnected collection of circuit elements.
The connectivity of the elements can be described by an oriented graph.
Each branch in the graph corresponds to one element. The~$b$-th element
is associated with the instantaneous voltage~$v_{b}\left(t\right)$,
current~$i_{b}\left(t\right)$, charge~$Q_{b}\left(t\right)$, flux~$\Phi_{b}\left(t\right)$,
and reduced flux~$\varphi_{b}\left(t\right)$. The universal relationships
Eqs.~\eqref{eq:universal-v-phi}\textendash \eqref{eq:universal-energy-power}
link the variables at each branch. Variables across different branches
are linked by Kirchhoff's laws.

\paragraph{Kirchhoff's two universal circuit laws.}

The following two laws are universal and topological in nature. They
describe relationship among the branch variables, independent of the
constitution of the branch elements. In other words, they apply to
nonlinear, time-dependent, and even hysteric elements.

\emph{Kirchhoff's voltage law} (KVL) is the lumped-element manifestation
of the Maxwell-Faraday equation, $\nabla\times\vec{E}=-\frac{\partial\vec{B}}{\partial t}$.
By applying Stokes' theorem to the Maxwell-Faraday equation along
an oriented loop of lumped elements (and a surface associated with
the closed loop), one finds a relationship valid for any closed circuit
loop: the oriented sum of the fluxes along the~$l$-th loop is equal
to the external applied flux~$\Phi_{l}^{\mathrm{ext}}\left(t\right)$
threading the loop,
\begin{equation}
\sum_{b\in\mathrm{loop}_{l}}\pm\Phi_{b}(t)=\Phi_{l}^{\mathrm{ext}}\left(t\right)\;,\label{eq:KVL}
\end{equation}
where the sum runs over all branches~$b$ that form the~$l$-th loop.
For a given branch~$b$, the positive (resp., negative) sign in Eq.~\eqref{eq:KVL}
is selected if its flux reference direction aligns (resp., is opposite
to) the loop orientation. Algebraically, KVL leads to a set of constraints
among the network variables. Thus, in the context of a Lagrangian
description of the circuit in which the generalized position variables
are taken to be fluxes~$\Phi_{b}$, the KVL conditions express a
set of holonomic constraints that need to be eliminated in order to
obtain a Lagrangian of the second kind \citep{Landau1982}.

\emph{Kirchhoff's current law} (KCL) is a statement of the conservation
of charge: at every node in the circuit, the algebraic sum of all
the current leaving or entering the node is equal to zero. Recast
another way, for all branches~$b$ connected to node~$n$, 
\begin{equation}
\sum_{b\in\mathrm{node}_{n}}\pm i_{b}(t)=0\;.\label{eq:KCL}
\end{equation}
The negative sign is chosen for branches whose current reference direction
points toward the~$n$-th node. In the flux-based Lagrangian description
of the circuit, the KCL algebraic conditions become the Lagrangian
equations of motion.

\paragraph{Eliminating the KVL constraints using the method of the minimum spanning-tree.}

The set of KVL algebraic equations defined in Eq.~\eqref{eq:KVL}
reduce the number of independent branch fluxes~$\Phi_{b}$. We can
systematically choose a minimal set of independent branch fluxes using
the minimum spanning-tree graph method \citep{Devoret1995,Burkard2004,Girvin2014,Vool2017}.
For our derivation, it is not necessary to explicitly construct the
tree. A set of branches from the graph can be selected to form a complete
and minimum spanning tree. In general, there are many satisfactory
tree sets of branches. Different trees are related by a simple algebraic
transformation; similar to a basis change. The branches that belong
to the spanning tree can be labeled~$\mathrm{t}_{1},\,\mathrm{t}_{2},\,\ldots$
The flux of the~$k$-th spanning-tree branch is denoted~$\Phi_{\mathrm{t}_{k}}\left(t\right)$.
In subscripts, roman (resp., italic) symbols denote labels (resp.,
variables). The spanning-tree can be organized in a column vector
\begin{equation}
\ensuremath{\phit\left(t\right)\isdef}\begin{pmatrix}\Phi_{t_{1}}\left(t\right)\\
\Phi_{t_{2}}\left(t\right)\\
\vdots
\end{pmatrix}\;,\label{eq:Phi-spanning-tree-vector}
\end{equation}
which serves the purposes of a basis for the description of the circuit.
The branch-fluxes \emph{not }in the spanning tree (\emph{links }or\emph{
chords} of the graph) are obtained by a linear transformation of~$\phit$.
We define the energy functions and Lagrangian of the system in terms
of~$\phit$ in the Sec.~\ref{app:energy-and-lagrangian}.

\subsection{The Josephson system and its non-linear Josephson dipoles\label{app:Josephson-system}}

\textit{\emph{In }}the main text, under \emph{Quantizing the general
Josephson system}, we introduced the notion of a \emph{Josephson system\textemdash }a
general electromagnetic environment that incorporates nonlinear devices,
\textit{\emph{referred to as}}\textit{ Josephson dipoles. }\textit{\emph{The
}}Josephson\textit{ }\textit{\emph{system is }}treated as a distributed
black-box structure.

\paragraph{Discretization.}

We aim to model the Josephson system as realistically as possible.
To account in detail for its physical layout, materials, boundary
conditions, and dipole structures, we aim to leverage conventional
electromagnetic analysis techniques, such as the finite-element (FE)
method. The FE method subdivides the physical layout of the system.
Using a set of basis functions, the electromagnetic circuit is discretized
\citep{Louisell1973,Jin2014}. The discretized circuit can be represented
by a lumped-element model. In principle, we can take the limit of
infinite subdivision. The Josephson dipoles are assumed to be the
only non-linear elements in the circuit. All other elements are linear,
as representative of the linear nature of Maxwell's equations.

\paragraph{Dissipation.}

\textit{\emph{As the object of interest is the control of quantum
information in the system, in this section, we focus on systems with
low dissipation. This condition requires that the quality factor of
all modes of relevance is high, }}$Q_{m}\gg1$, where~$m$ is the
mode index. In Sec.~\ref{app:dissipation}, we treat dissipation
as a perturbation to the lossless solutions. 

\paragraph{Josephson dipole.}

The Josephson dipole was introduced in the main text, under \emph{Quantizing
the general Josephson system}. For simplicity of discussion, here,
we treat the dipole as a lumped, two-terminal, flux-controlled element.
The~$j$-th Josepson dipole in the system is fully specified by its
energy function~$\mathcal{E}_{j}\left(\Phi_{j};\Phi_{j,\mathrm{ext}}\right)$,
see Eq.~\eqref{eq:Eind-flux-control}. The energy depends only on
the magnetic flux~$\Phi_{j}\left(t\right)$ across the terminals
of the dipole and on any external parameters~$\Phi_{j,\mathrm{ext}}$
that control the energy landscape; these can include a voltage bias,
a current bias, and an external magnetic field bias. The index~$j$
runs from unit to the total number~$J$ of Josephson dipoles in the
circuit.

This formulation is rather general. It encapsulate the wide span of
Josephson dipoles discussed in the main text. These include simple
devices, such as the Josephson tunnel junction or a nanowire, and
also composite devices, such as SQUIDs, SNAILs, or more general sub-circuits.
The underlying physical phenomenon giving rise to the low-loss, non-linearity
of the dipole is immaterial.

\paragraph{Partition of the Josephson dipole energy-function.}

It is always possible to partition~$\mathcal{E}_{j}\left(\Phi_{j};\Phi_{j,\mathrm{ext}}\right)$
into a linear and non-linear part,
\begin{equation}
\mathcal{E}_{j}\left(\Phi_{j};\Phi_{j,\mathrm{ext}}\right)=\ejlin\left(\Phi_{j};\Phi_{j,\mathrm{ext}}\right)+\ejnl\left(\Phi_{j};\Phi_{j,\mathrm{ext}}\right)\;.\label{eq:app:circuit: defn of junc energy}
\end{equation}
This division is purely a conceptual one\textemdash the Josephson
dipole cannot be physically divided into a linear and nonlinear part.
By selecting an equilibrium point of the circuit and defining the
branch fluxes~$\Phi_{j}$ as deviations away from the equilibrium
{[}discussed in more detail in Sec.~\eqref{app:equilibrium-point-of-circuit}{]},
the partitions take the concrete form 
\begin{subequations}
\label{eq:Ej-dipole-defined} 
\begin{eqnarray}
\ejlin\left(\Phi_{j}\right) & \isdef & \frac{1}{2}E_{j}\left(\Phi_{j}/\phi_{0}\right)^{2}\;,\label{eq:Ej-lin-defn}\\
\ejnl\left(\Phi_{j}\right) & \isdef & \mathcal{E}_{j}\left(\Phi_{j}\right)-\ejlin\left(\Phi_{j}\right)\label{eq:Ej-nl-defn-series}\\
 & = & E_{j}\sum_{p=3}^{\infty}c_{jp}\left(\Phi_{j}/\phi_{0}\right)^{p}\;,\label{eq:Ej-nl-defn-series-taylor}
\end{eqnarray}
\end{subequations}
where~$E_{j}$ is an overall scaling factor of the energy function,
defined in Eq.~\eqref{eq:Ej-lin-defn}; we refer to it as the Josephson
dipole energy scale. In Eq.~\eqref{eq:Ej-nl-defn-series-taylor},
we have introduced the Taylor series expansion of~$\ejnl$ around
the equilibrium state of the circuit and have introduced the \textit{dimensionless}
expansion coefficients~$c_{jp}$. We stress that the expansion is
not needed for the EPR method. It is merely a convenient tool for
working analytically with weakly non-linear circuits, such as the
transmon qubit. For notation simplicity, in Eq.~\eqref{eq:Ej-dipole-defined}
the dependance of~$\ejlin,\ejnl,E_{j}$, and~$c_{jp}$ on the external
bias parameter~$\Phi_{j,\mathrm{ext}}$ is made implicit, and we
will continue to do so henceforth; in other words, keep in mind that~$E_{j}\isdef E_{j}\left(\Phi_{j,\mathrm{ext}}\right)$
and~$c_{jp}\isdef c_{jp}\left(\Phi_{j,\mathrm{ext}}\right)$.

\paragraph{Example of a Josephson dipole: the Josephson junction.}

We illustrate the partitioning construction defined in Eq.~\eqref{eq:Ej-dipole-defined}
using the example of the Josephson tunnel junction. For an un-frustrated
junction, it follows from Eq.~\eqref{eq:jj-energy-func} that
\begin{subequations}
\label{eq:Ej-jj-dipole} 
\begin{eqnarray}
\ejlin\left(\Phi_{j}\right) & \isdef & \frac{1}{2}E_{j}\left(\Phi_{j}/\phi_{0}\right)^{2}\;,\label{eq:Ej-jj-dipole-linear}\\
\ejnl\left(\Phi_{j}\right) & \isdef & -E_{j}\left[\cos\left(\Phi_{j}/\phi_{0}\right)+\frac{1}{2}\left(\Phi_{j}/\phi_{0}\right)^{2}\right]\;,\label{eq:Ej-jj-dipole-nl}
\end{eqnarray}
\end{subequations}
where~$E_{j}$ is the Josephson energy. The energy function~$\ejlin$
is associated with the linear response of the junction. It presents
the inductance~$L_{j}=\phi_{0}^{2}/E_{j}$. The energy~$\ejnl$
is associated with the response of non-linear inductor. The expansion
coefficient of~$\ejnl$ as defined in Eq.~\eqref{eq:Ej-nl-defn-series-taylor}
are
\begin{equation}
c_{jp}=\begin{cases}
\frac{\left(-1\right)^{p/2+1}}{p!} & \text{for even }p\;,\\
0 & \text{for odd }p\;.
\end{cases}\label{eq:app:Jos potentual Ujn}
\end{equation}
In partitioning~$\mathcal{E}_{j}$, we included all of the linear
response of the junction in~$\ejlin$, leaving none for~$\ejnl$;
i.e., $\ejnl$ lacks quadratic terms in~$\Phi_{j}$. However, this
is not required. There are certain cases for which retaining some
part of the linear response in~$\ejnl$ is advantageous.

For notational ease, we now introduce the reduced flux~$\varphi_{j}\left(t\right)\isdef\Phi\left(t\right)/\phi_{0}$.

\paragraph{Example of a Josephson dipole in a frustrated circuit.}

Imagine a Josephson tunnel junction incorporated in a closed loop
of several circuit elements. Suppose the loop supports the flow of
a direct current. A current source in the path of the loop (or perhaps
an external magnetic flux threading the loop) establishes a persistent
current. The equilibrium flux~$\varphi_{j}$ of the junction shifts
to a non-zero equilibrium value~$\varphi_{\mathrm{eq},j}$, as determined
by the circuit equilibrium considerations (see Sec.~\ref{app:equilibrium-point-of-circuit}).
Applying the partition defined in Eq.~\eqref{eq:Ej-jj-dipole} to
Eq.~\eqref{eq:Ej-jj-frustrated}, we find that in terms of the out-of-equilibrium
flux deviation~$\varphi_{j}$, 
\begin{multline}
\mathcal{E}_{j}\left(\varphi_{j};\varphi_{\mathrm{eq},j}\right)=\frac{1}{2}E_{j}\left(\varphi_{\mathrm{eq},j}\right)\varphi_{j}^{2}\\
+E_{j}\left(\varphi_{\mathrm{eq},j}\right)c_{j3}\left(\varphi_{\mathrm{eq},j}\right)\varphi_{j}^{3}+\cdots\;,\label{eq:Ej-jj-frustrated-partition}
\end{multline}
where~$E_{j}\left(\varphi_{\mathrm{eq},j}\right)=E_{J}/\cos\left(\varphi_{\mathrm{eq},j}\right)$,
$E_{J}$ is the~$j$-th junction Josephson energy, and~$c_{j3}\left(\varphi_{\mathrm{eq},j}\right)=-\frac{1}{6}\sin\left(\varphi_{\mathrm{eq},j}\right)$.
We emphasize that~$\varphi_{j}$ denotes deviations away from the
equilibrium; compare Eq.~\eqref{eq:Ej-jj-frustrated-partition} to
Eq.~\eqref{eq:Ej-jj-frustrated}.

\subsection{Energy of the Josephson circuit and its Lagrangian\label{app:energy-and-lagrangian}}

\paragraph{Capacitive energy.}

The total capacitive energy~$\mathcal{E}_{\mathrm{cap}}$ of the
Josephson system is simply the algebraic sum of the total energy of
all its capacitive elements. Using Eq.~\eqref{eq:E-cap-ind-in-phi},
and summing over all capacitive branches, $\mathcal{E}_{\mathrm{cap}}\isdef\sum_{b\in\mathrm{cap.}}\frac{1}{2}C_{b}\dot{\Phi}_{b}^{2}$,
where~$C_{b}$ is the capacitance of branch~$b$. Each of these
fluxes can be expressed in terms of the linearly-independent spanning-tree
fluxes~$\phit$. The energy function is quadratic in~$\phit$,
\begin{equation}
\mathcal{E}_{\mathrm{cap}}\left(\phitd\right)=\frac{1}{2}\phitd^{\mathrm{T}}\m C\phitd\;,\label{eq:W_C}
\end{equation}
where~$\m C$ is the capacitance matrix of the circuit \citep{Yurke1984,Devoret1995,Girvin2014}.
It follows from KVL and the constitutive relationships of the capacitors
that~$\m C$ is a positive-definite, real, symmetric (PDRS) matrix.
In the continuous limit of space, the total capacitive energy~$\mathcal{E}_{\mathrm{cap}}$
can be found using Eq.~\eqref{eq:E_elec-fields}.

\paragraph{Inductive energy.}

The total inductive energy in the circuit~$\mathcal{E}_{\mathrm{ind}}$
is similarly the algebraic sum of the total energy of all circuit
inductive branches; i.e., $\mathcal{E}_{\mathrm{ind}}\isdef\sum_{b\in\mathrm{ind.}}\mathcal{E}_{b}\left(\Phi_{b}\right)$.
In the Josephson system, inductive branches come in two distinct flavors:
linear and non-linear. Physically, linear inductive branches are associated
with the geometry and magnetic fields. We denote their total energy~$\mathcal{E}_{\mathrm{mag}}$.
The non-linear inductive branches (Josephson dipoles) are generally
associated with the kinetic inductance of electrons. We denote their
total energy~$\mathcal{E}_{\mathrm{kin}}$. Hence, 
\begin{equation}
\mathcal{E}_{\mathrm{ind}}\left(\phit\right)=\mathcal{E}_{\mathrm{mag}}\left(\phit\right)+\mathcal{E}_{\mathrm{kin}}\left(\phit\right)\;.\label{eq:E-tot-ind}
\end{equation}
The energy of the linear branches~$\mathcal{E}_{\mathrm{mag}}$ is
the dual of Eq.~\eqref{eq:W_C}. It is also a quadratic form,
\begin{equation}
\mathcal{E}_{\mathrm{mag}}\left(\phit\right)=\frac{1}{2}\phit^{\mathrm{T}}\m L{}_{\text{mag}}^{-1}\phit\;,\label{eq:E_magnetic}
\end{equation}
where the inductance matrix~$\m L{}_{\text{mag}}^{-1}$ completely
describes all linear, magnetic-in-origin inductances in the circuit
\citep{Yurke1984,Devoret1995,Girvin2014}. Due to its nature,~$\m L{}_{\text{mag}}^{-1}$
is PDRS. In the continuous limit of space, the total magnetic inductive
energy~$\mathcal{E}_{\mathrm{mag}}$ can be found using Eq.~\eqref{eq:E_mag-fields}.

The total inductive kinetic energy of the circuit, associated with
the non-linear dipoles, is
\begin{multline}
\mathcal{E}_{\mathrm{kin}}=\sum_{j=1}^{J}\mathcal{E}_{j}\left(\Phi_{j}\right)=\sum_{j=1}^{J}\ejlin\left(\Phi_{j}\right)+\sum_{j=1}^{J}\ejnl\left(\Phi_{j}\right)\;.\label{eq:E_kinetic}
\end{multline}
This energy is \textit{not} stored in the magnetic fields. However,
we can group the magnetic energy and the linear part of~$\mathcal{E}_{\mathrm{kin}}$
together to express the inductive energy as a partition of linear
and non-linear contributions
\begin{subequations}
\label{eq:E-ind-total}
\begin{align}
\mathcal{E}_{\mathrm{ind}}\left(\phit\right) & =\frac{1}{2}\phit^{\mathrm{T}}\m L{}^{-1}\phit+\mathcal{E}_{\mathrm{kin}}(\phit)\;,\label{eq:e-ind-split}\\
\m L{}^{-1} & \isdef\m L{}_{\text{mag}}^{-1}+\frac{1}{2}\sum_{j=1}^{J}E_{j}\left(\Phi_{j}/\phi_{0}\right)^{2}\;,\label{eq:L-inv-total}\\
\mathcal{E}_{\mathrm{nl}}\left(\phit\right) & \isdef\sum_{j=1}^{J}\ejnl\left(\Phi_{j}\right)\;,\label{eq:E-nl-defn}
\end{align}
\end{subequations}
where we have introduced the total inductance matrix of the circuit~$\m L{}^{-1}$
and the total non-linear energy function of the circuit~$\mathcal{E}_{\mathrm{nl}}$.
For later use, we can obtain the series expansion of~$\mathcal{E}_{\mathrm{nl}}$
in terms of that of~$\ejnl$, defined in Eq.~\eqref{eq:Ej-nl-defn-series},
\begin{equation}
\mathcal{E}_{\mathrm{nl}}\left(\phit\right)=\sum_{j=1}^{J}\sum_{p=3}^{\infty}E_{j}c_{jp}\left(\Phi_{j}/\phi_{0}\right)^{p}\;.\label{eq:E-nl-total-in-Phi}
\end{equation}

\paragraph{Generalized coordinates for the Lagrangian.}

We have expressed~$\mathcal{E}_{\mathrm{cap}}$ and~$\mathcal{E}_{\mathrm{ind}}$
in terms of the independent set of spanning-tree fluxes~$\phit$.
We could equivalently have expressed~$\mathcal{E}_{\mathrm{cap}}$
and~$\mathcal{E}_{\mathrm{ind}}$ in terms of the charge variables~$Q_{b}$.
However, as discussed in Sec.~\ref{app:Review-of-electrical-theory},
the flux-controlled Josephson dipoles present an asymmetry which favors
treatment in the flux basis. We hence employ~$\phit$ as the generalized
position coordinates in the Lagrangian description of the circuit,
and~$\phitd$ as the generalized velocity.

\paragraph{Lagrangian of the Josephson circuit.}

The Lagrangian function of the Josephson circuit follows from KCL.
Energy functions with~$\phit$ as their argument (resp.,~$\phitd$)
play the role of potential (resp., kinetic) energies. The system Lagrangian
is the difference of the total kinetic and potential energy functions,
\begin{subequations}
\label{eq:L-full}
\begin{align}
\mathcal{L}_{\mathrm{full}}\left(\phit,\phitd\right) & \isdef\mathcal{E}_{\mathrm{cap}}\left(\phitd\right)-\mathcal{E}_{\mathrm{ind}}\left(\phit\right)\label{eq:L-full-1}\\
 & =\mathcal{\mathcal{L}_{\mathrm{lin}}}\left(\phit,\phitd\right)+\mathcal{L}_{\mathrm{nl}}\left(\phit\right)\;,
\end{align}
\end{subequations}
where we have partitioned the Lagrangian into a linear~$\mathcal{L}_{\mathrm{lin}}$
and nonlinear~$\mathcal{L}_{\mathrm{nl}}$ part. Substituting in
Eqs.~\eqref{eq:W_C} and~\eqref{eq:E-ind-total}, 
\begin{subequations}
\label{eq:L-partitioned}
\begin{eqnarray}
\mathcal{L}_{\mathrm{lin}}\left(\phit,\phitd\right) & = & \frac{1}{2}\phitd^{\mathrm{T}}\m C\phitd-\frac{1}{2}\phit^{\mathrm{T}}\m L^{-1}\phit\;,\label{app:L-lin}\\
\mathcal{L}_{\mathrm{nl}}\left(\phit\right) & = & -\mathcal{E}_{\mathrm{nl}}\left(\phit\right)=-\sum_{j=1}^{J}\ejnl\left(\Phi_{j}\right)\;.\label{app:L-nl}
\end{eqnarray}
\end{subequations}

Equation~\eqref{eq:L-partitioned} explicitly constructs the Josephson
system Lagrangian. It partitions it into a linear and nonlinear part.
In Sec.~\ref{app:diagonalize-eigenmodes}, we diagonalize~$\mathcal{L}_{\mathrm{lin}}$
to find the eigenmodes of the linearized system. In Sec.~\ref{app:nonlinear-interactions},
we treat the effect of~$\mathcal{L}_{\mathrm{nl}}$.

\subsection{Eigenmodes of the linearized Josephson circuit\label{app:diagonalize-eigenmodes}}

In this sub-sectino, we diagonalize~$\mathcal{L}_{\mathrm{lin}}$
and find its eigenmodes, eigenfrequencies~$\omega_{m}$ and eigenvectors
(i.e., spatial-mode profiles). These intermediate result provide a
key stepping stone on our path to quantizing the Josephson system
and treating~$\mathcal{L}_{\mathrm{nl}}$. The process conceptually
parallels that taken by the finite-element (FE) electromagnetic (EM)
solver in an eigenanalysis of the linearized Josephson system.

The Lagrangian~$\mathcal{L}_{\mathrm{lin}}$ is the sum of two quadratic
forms, see Eq.~\eqref{eq:L-partitioned}. We use the standard method
for their simultaneous diagonalization \citep{Landau1982}, based
on a series of principle-axis transforms. We then transform the Lagrangian
into a diagonalized Hamiltonian.

\paragraph{Diagonalizing the inductance matrix.}

Since the inverse inductance matrix~$\m L^{-1}$ is a PDRS matrix,
we can diagonalize it with a real orthogonal matrix~$\m O_{\m L}$,
obeying~$\m O_{\m L}\m O_{\m L}^{\intercal}=\m O_{\m L}^{\intercal}\m O_{\m L}=\m I$,
where~$\m I$ is the identity matrix,%
\begin{equation}
\m O_{\m L}^{\mathrm{T}}\m L^{-1}\m O_{\m L}=\m{\Lambda}_{\m L}^{-1}\mathbb{\m I}_{\m L}^{-1}\;,\label{eq:diagonalize-L-inv}
\end{equation}
where~$\m{\Lambda}_{\m L}^{-1}$ is a diagonal matrix comprising
the (dimensionless) eigenvalue magnitudes, and~$\mathbb{\m I}_{\mathrm{\m L}}$
is the identity matrix with physical dimensions of inductance. The
eigenvectors of~$\m L^{-1}$ form the columns of~$\m O_{\m L}$.

Employing Eq.~\eqref{eq:diagonalize-L-inv} with~Eq.~\eqref{eq:L-partitioned},
the system Lagrangian takes the suggestive form
\begin{multline}
\mathcal{L}_{\mathrm{full}}\left(\phit,\phitd\right)=\frac{1}{2}\phitd^{\mathrm{T}}\m C\phitd+\mathcal{L}_{\mathrm{nl}}\left(\phit\right)\\
-\frac{1}{2}\left(\phit^{\mathrm{T}}\m O_{\m L}\m{\Lambda}_{\m L}^{-1/2}\right)\mathbb{\m I}_{\mathrm{\m L}}^{-1}\left(\m{\Lambda}_{\m L}^{-1/2}\m O_{\m L}^{\mathrm{T}}\phit\right)\;,
\end{multline}
which motivates the principle-axis transformation of the magnetic
flux defined by 
\begin{equation}
\breve{\m{\Phi}}\isdef\m{\Lambda}_{\m L}^{-1/2}\m O_{\m L}^{\mathrm{T}}\phit\;,\label{eq:L-transform-1}
\end{equation}
where~$\breve{\m{\Phi}}$ is a rotated and then scaled version of~$\phit$.
Under this transformation, the transformed Lagrangian function becomes
\begin{equation}
\begin{split}\breve{\mathcal{L}}_{\mathrm{full}}\left(\breve{\m{\Phi}},\dot{\breve{\m{\Phi}}}\right)\isdef\frac{1}{2}\dot{\breve{\m{\Phi}}}^{\mathrm{T}}\m{\breve{C}}\dot{\breve{\m{\Phi}}}-\frac{1}{2}\breve{\m{\Phi}}^{\mathrm{T}}\mathbb{\m I}_{\mathrm{\m L}}^{-1}\breve{\m{\Phi}}+\breve{\mathcal{L}}_{\mathrm{nl}}\left(\breve{\m{\Phi}}\right)\;,\end{split}
\label{eq:L-full-breve}
\end{equation}
where~$\breve{\mathcal{L}}_{\mathrm{nl}}\left(\breve{\m{\Phi}}\right)\isdef\mathcal{L}_{\mathrm{nl}}\left(\m{\Phi}\left(\breve{\m{\Phi}}\right)\right)$
and 
\begin{equation}
\m{\breve{C}}\isdef\left(\m{\Lambda}_{\m L}^{1/2}\m O_{\m L}^{\mathrm{T}}\right)\m C\left(\m O_{\m L}\m{\Lambda}_{\m L}^{1/2}\right)\;.\label{eq:C-breve}
\end{equation}
Since the capacitance matrix~$\m C$ is PDRS and it is transformed
by a rotation and then a dilation, it follows that~$\m{\breve{C}}$
is also PDRS. More generally, the eigenvalues of a matrix are invariant
under a similarity transform, such as the one employed in Eq.~\eqref{eq:C-breve}.

\paragraph{Diagonalizing the capacitance matrix.}

Since~$\m{\breve{C}}$ is PDRS, we diagonalize it with a real, orthogonal
transformation~$\m O_{\m{\breve{C}}}$, such that 
\begin{equation}
\m O_{\m{\breve{C}}}^{\mathrm{T}}\m{\m{\breve{C}}}\m O_{\breve{\m C}}=\m{\Lambda}_{\m{\breve{C}}}\mathbb{\m I}_{\m C}\;,\label{eq:diagonalize-C-breve}
\end{equation}
where~$\m{\Lambda}_{\m{\breve{C}}}$ is the dimensionless, diagonal
matrix constructed from the eigenvalues of~$\m{\m{\breve{C}}}$ and~$\mathbb{\m I}_{\m C}$
is the identity matrix with physical dimensions of capacitance.

Employing the orthogonality transformation~$\m O_{\m{\breve{C}}}^{\mathrm{T}}$
in a manner similar to the one used with~$\m O_{\m L}$, we rotate
(but do not scale) the coordinates for a second time. Under this second
principle-axis transform, we define the eigenmode magnetic flux variable
\begin{equation}
\phim\isdef\m O_{\breve{\m C}}^{\mathrm{T}}\breve{\m{\Phi}}\;,\label{eq:L-transform-2}
\end{equation}
in terms of which the Lagrangian~$\mathcal{L}_{\mathrm{lin}}$ is
diagonal,
\begin{multline}
\tilde{\mathcal{L}}_{\mathrm{full}}\left(\phim,\phimd\right)\isdef\frac{1}{2}\phimd^{\mathrm{T}}\m{\Lambda}_{\m{\breve{C}}}\mathbb{\m I}_{\m C}\phimd-\frac{1}{2}\phim^{\mathrm{T}}\mathbb{\m I}_{\m L}^{-1}\phim\\
+\tilde{\mathcal{L}}_{\mathrm{nl}}\left(\phim\right)\;,\label{eq:deriv-bare:L-twidle}
\end{multline}
where the nonlinear part under the transformation is 
\begin{equation}
\tilde{\mathcal{L}}_{\mathrm{nl}}\left(\phim\right)\isdef\mathcal{L}_{\mathrm{nl}}\left(\phit\left(\phim\right)\right)\;.
\end{equation}

\paragraph{Equations of motion.}

The Lagrangian equations of motion, $\frac{\partial\tilde{\mathcal{L}}}{\partial\phim}-\frac{\mathrm{d}}{\mathrm{d}t}\frac{\partial\tilde{\mathcal{L}}}{\partial\phimd}=\m 0$,
yield the harmonic eigenvalue equation~$\frac{\mathrm{d}^{2}}{\mathrm{d}t^{2}}\phim+\m{\Omega}^{2}\phim=\m 0$,
where~$\m 0$ is the column vector of all zero elements and~$\m{\Omega}^{2}\isdef\m{\Lambda}_{\m{\breve{C}}}^{-1}\mathbb{\m I}_{\m{\omega}}^{2}$.
The identity matrix~$\mathbb{\m I}_{\m{\omega}}$ has physical dimensions
of circular frequency. The generalized momentum canonical to~$\phim$
is the vector of charge variables~$\m Q_{\mathrm{m}}\isdef\frac{\partial\tilde{\mathcal{L}}}{\partial\phimd}=\m{\Lambda}_{\m{\breve{C}}}\mathbb{\m I}_{\m C}\phimd$.

\paragraph{Diagonalized Hamiltonian of the Josephson system.}

The system Hamiltonian follows from the Legendre transform on~$\tilde{\mathcal{L}}_{\mathrm{full}}$,
$\mathcal{H}_{\mathrm{full}}\left(\phim,\m Q_{\mathrm{m}}\right)=\left(\phimd(\m Q_{\mathrm{m}})\right)^{\mathrm{T}}\m Q_{\mathrm{m}}-\tilde{\mathcal{L}}_{\mathrm{full}}$,
which we can partition into
\begin{equation}
\mathcal{H}_{\mathrm{full}}\left(\phim,\m Q_{\mathrm{m}}\right)=\mathcal{H}_{\mathrm{lin}}\left(\phim,\m Q_{\mathrm{m}}\right)+\mathcal{H}_{\mathrm{nl}}\left(\phim,\m Q_{\mathrm{m}}\right)\;,\label{eq:app:H_full-linearized-modes}
\end{equation}
where the linear and nonlinear parts of the Hamiltonian expressed
in the eigenmode coordinates are
\begin{subequations}
\label{eq:H-classical-edecomp}
\begin{align}
\mathcal{H}_{\mathrm{lin}}\left(\phim,\m Q_{\mathrm{m}}\right) & \isdef\frac{1}{2}\m Q_{\mathrm{m}}^{\mathrm{T}}\m{\Omega}^{2}\mathbb{\m I}_{\m L}\m Q_{\mathrm{m}}+\frac{1}{2}\phim^{\mathrm{T}}\mathbb{\m I}_{\m L}^{-1}\phim\label{eq:Hlin-classical-eigendecomp}\\
 & =\sum_{m=1}^{M}\frac{1_{L}}{2}\omega_{m}^{2}Q_{m}^{2}+\frac{1}{2}1_{L}\Phi_{m}^{2}\;,\\
\mathcal{H}_{\mathrm{nl}}\left(\phim,\m Q_{\mathrm{m}}\right) & \isdef-\tilde{\mathcal{L}}_{\mathrm{nl}}\left(\phim\right)\;,\label{eq:Hnl-classical-eigendecomp}
\end{align}
\end{subequations}
where the diagonal \emph{eigenfrequency matrix~$\m{\Omega}$} of
the Hamiltonian~$\mathcal{H}_{\mathrm{lin}}$ is
\begin{align}
\m{\Omega} & \isdef\m{\Lambda}_{\m{\breve{C}}}^{-1/2}\mathbb{\m I}_{\m{\omega}}=\begin{pmatrix}\omega_{1}\\
 & \ddots\\
 &  & \omega_{M}
\end{pmatrix}\;,\label{app:Omega-eval-matrix-defin}
\end{align}
and~$1_{L}$ is unity carrying physical dimensions of inductance.
The entries of~$\m{\Omega}$ are the eigenmode frequencies~$\omega_{m}$
of the linearized circuit. These correspond to the eigenfrequencies
solved for by the FE eigenanalysis. 

\paragraph{Eigenvectors of the Josephson system.}

The \emph{eigenvector matrix~}$\m E$ relates the spanning-tree fluxes~$\phit$
to the eigenmode ones~$\phim$, 
\begin{equation}
\phit=\m E\phim\;.\label{app:eigen modes:phib to phim}
\end{equation}
It is found by concatenating the principle-axis transformations defined
in Eqs.~\eqref{eq:L-transform-1} and~\eqref{eq:L-transform-2},
\begin{equation}
\m E\isdef\m O_{\m L}\m{\Lambda}_{\m L}^{1/2}\m O_{{\breve{\m C}}}\;.\label{app:eigen:E defn}
\end{equation}
The eigenvector matrix~$\m E$ is real and positive-definite, since
it is the produce real, positive-definite transforms. It is dimensionless
and, in general, non-symmetric. It is related to the square root of
the inductance matrix,~$\m E\m E^{\intercal}=\m L\m I_{\mathrm{H}}^{-1}$~and~$\left(\m E^{-1}\right)^{\intercal}\m E^{-1}=\m L^{-1}\m I_{\mathrm{H}}$.
 The eigenvector matrix~$\m E$ represents the eigenfield solutions
found in the FE~analysis. It is key in determining the quantum zero-point
fluctuations of the mode and dipole fluxes, as shown in the following
section.

\subsection{Quantizing the Josephson circuit\label{app:quantize-modes}}

We quantize~$\mathcal{H}_{\mathrm{full}}$ using Dirac's canonical
approach \citep{Dirac1982-Book}. Before taking passage from classical
to quantum, we introduce the complex mode amplitude operator~$\alpha_{m}$\textemdash the
classical analog of the bosonic amplitude operator~$\hat{a}_{m}$.
This provides a direct path to second quantization in the eigenmode
basis of~$\mathcal{H}_{\mathrm{lin}}$.

\paragraph{Complex action-angle variables.}

We define the vector of action-angle variables~$\bm{\alpha}=\left(\alpha_{1},\ldots,\alpha_{M}\right)^{\mathrm{T}}$
by the non-canonical, complex transformation
\begin{equation}
\bm{\alpha}\left(t\right)\isdef\frac{1}{\sqrt{2\hbar\m{\Omega}}}\left(\phim\left(t\right)1_{\mathrm{H}}^{-1/2}+i\m{\Omega}\mathbf{Q}_{\mathrm{m}}\left(t\right)1_{\mathrm{H}}^{1/2}\right)\;,\label{eq:alpha-action-angle}
\end{equation}
where~$1_{\mathrm{H}}$ is unity with dimensions of inductance. The
normalization~$1/\sqrt{2\hbar\m{\Omega}}$ is chosen so that the
Poisson bracket of the action-angles is~$\left\{ \alpha_{m},\alpha_{m'}^{*}\right\} _{\mathrm{P}}=1/(i\hbar)\delta_{mm'}$.
In terms of the action-angles, the Hamiltonian remain diagonal, 
\begin{equation}
\mathcal{H}_{\mathrm{lin}}=\frac{\hbar}{2}\left(\bm{\alpha}^{T}\m{\Omega}\bm{\alpha}^{*}+\bm{\alpha}^{*T}\m{\Omega}\bm{\alpha}\right)\;.\label{eq:Hlin-action-angle}
\end{equation}
We have symmetrized~$\mathcal{H}_{\mathrm{lin}}$ to avoid operator
order ambiguity. The flux is~$\phim=\sqrt{\frac{\hbar\m{\Omega}1_{\mathrm{H}}}{2}}\left(\bm{\alpha}^{*}+\bm{\alpha}\right)$.

\paragraph{Quantizing the action-angle variables.}

Following Dirac's prescription \citep{Dirac1982-Book}, we supplant
Poisson brackets by commutators and promote observables to operators.
The action-angles~$\alpha_{m}$ and~$\alpha_{m}^{*}$ promote into
the ladder annihilation~$\hat{a}_{m}$ and creation~$\hat{a}_{m}^{\dagger}$
operators. Their commutator follows from the Poisson bracket \citep{Louisell1961,Yurke1984,Devoret1995},
$i\hbar\times\{\alpha_{m},\alpha_{m}^{*}\}\mapsto\left[\hat{a}_{m},\hat{a}_{m}^{\dagger}\right]$,
\begin{equation}
\left[\hat{a}_{m},\hat{a}_{k}^{\dagger}\right]=\delta_{mk}\hat{I}\;,\label{eq:commutations}
\end{equation}
where~$\left[\hat{A},\hat{B}\right]\isdef\hat{A}\hat{B}-\hat{B}\hat{A}$
is the commutator,~$\delta_{mk}$ is the Kronecker delta function,
and~$\hat{I}$ is the identity operator. Particles of the electromagnetic
field (photons) are distinguishable bosons \citep{Devoret1995}; symmetrization
of the many-body wavefunction is not required. Operators are in the
Schr\"odinger picture, unless otherwise indicated by an explicit
time argument.

\paragraph{Hamiltonian.}

The quantized form of~$\mathcal{H}_{\mathrm{lin}}$, see Eq.~\eqref{eq:Hlin-action-angle},
is
\begin{equation}
\boxed{\hat{H}_{\mathrm{lin}}=\sum_{m=1}^{M}\hbar\omega_{m}\hat{a}_{m}^{\dagger}\hat{a}_{m}\;.}\label{eq:app:Hlin-multi}
\end{equation}
This is Eq.~(13) of the main text. The quantized form of~$\mathcal{H}_{\mathrm{nl}}$
follows from combining Eqs.~\eqref{eq:E-ind-total}, \eqref{eq:E-nl-total-in-Phi},
\eqref{app:L-nl} and~\eqref{eq:Hnl-classical-eigendecomp},

\begin{equation}
\boxed{\hat{H}_{\mathrm{nl}}=\sum_{j=1}^{J}\mathcal{E}_{j}^{\mathrm{nl}}\left(\hat{\varphi}_{j}\right)=\sum_{j=1}^{J}\sum_{p=3}^{\infty}E_{j}c_{jp}\hat{\varphi}_{j}^{p}\;.}\label{eq:Hnl-quantum}
\end{equation}
where~$\hat{\varphi}_{j}\isdef\hat{\Phi}_{j}/\phi_{0}$ is the reduced
magnetic-flux operator for Josephson dipole~$j$. This is Eq.~(14)
of the main text. In the following, we decompose~$\hat{\varphi}_{j}$
in terms of~$\hat{a}_{m}$; i.e., in second quantization with respect
to the eigenmodes of~$\mathcal{H}_{\mathrm{lin}}$. In writing Eq.~\eqref{eq:app:Hlin-multi},
we have omitted the~$\frac{1}{2}\hbar\omega_{m}$ ground-state energy
of every mode. In the limit of infinite discretization, the sum of
these ground energies tends to infinity, a standard conceptual difficulty
in quantum field theory. Since physical experiments observe changes
in the energy of the field, the vacuum energy can be neglected (except
for special cases; e.g., Casimir effect). To proceed, we introduce
helpful notation.

\paragraph{Notation for vectors of operators.}

A bold symbol typeset in roman, such as~$\phit$, denotes a vector
or matrix, whose elements are constants or variables. Since variable,
such as~$\Phi_{\mathrm{t}_{1}}$ are promoted to quantum operators,~$\hat{\Phi}_{\mathrm{t}_{1}}$,
we can accommodate vectors of operators in our notation with a hat
symbol; e.g., 
\begin{equation}
\phithat=\begin{pmatrix}\hat{\Phi}_{\mathrm{t}_{1}}\\
\hat{\Phi}_{\mathrm{t}_{2}}\\
\vdots
\end{pmatrix}\quad\text{and}\quad\ensuremath{\phimhat=}\begin{pmatrix}\hat{\Phi}_{\mathrm{m}_{1}}\\
\vdots\\
\hat{\Phi}_{\mathrm{m}_{M}}
\end{pmatrix}\;.
\end{equation}
The spanning-tree flux operator~$\hat{\Phi}_{\mathrm{t}_{k}}$ corresponding
to the~$k$-th spanning-tree-branch flux variable~$\Phi_{\mathrm{t}_{k}}$.
Similarly, the eigenmode flux operator~$\hat{\Phi}_{\mathrm{m}_{k}}$
corresponding to the~$k$-th eigenflux variable~$\Phi_{\mathrm{m}_{k}}$.

\paragraph{Zero-point fluctuations of the eigenoperators.}

Inverting Eq.~\eqref{eq:alpha-action-angle}, one finds the vectors
of eigenflux and eigencharge operators, 
\begin{subequations}
\label{eq:Phi-andQ-operators}
\begin{eqnarray}
\mh{\Phi}_{\mathrm{m}} & \isdef & \mathbf{\Phi_{\mathrm{m}}^{ZPF}}\left(\mh a^{\dagger}+\mh a\right)\;,\label{eq:quantize:phim}\\
\mh Q_{\mathrm{m}} & \isdef & i\mathbf{Q_{\mathrm{m}}^{ZPF}}\left(\mh a^{\dagger}-\mh a\right)\;,\label{eq:quantize:Q}
\end{eqnarray}
\end{subequations}
 respectively, where the diagonal matrices of the quantum ZPF of the
operators are 
\begin{subequations}
\begin{eqnarray}
\mathbf{\Phi_{\mathrm{m}}^{ZPF}} & \isdef & \sqrt{\frac{\hbar}{2}}\m{\Omega}^{1/2}\mathbb{\m I}_{\mathrm{H}^{1/2}}\label{eq:app:quantize:phi_zpf}\\
 & = & \begin{pmatrix}\sqrt{\frac{\hbar\omega_{1}}{2}1_{\mathrm{H}}}\\
 & \ddots\\
 &  & \sqrt{\frac{\hbar\omega_{1}}{2}1_{\mathrm{H}}}
\end{pmatrix}\;,\nonumber \\
\mathbf{Q_{\mathrm{m}}^{ZPF}} & \isdef & \sqrt{\frac{\hbar}{2}}\m{\Omega}^{-1/2}\mathbb{\m I}_{\mathrm{H}^{-1/2}}\;.\label{eq:app:quantize:phi_zpf Q}
\end{eqnarray}
\end{subequations}
We recall that the operator~$\hat{\Phi}_{m}=\Phi_{m}^{\mathrm{ZPF}}\left(\hat{a}_{m}^{\dagger}+\hat{a}_{m}\right)$
has a zero-mean, gaussian-distributed distribution in the ground state;
i.e., its mean is~$\braOket 0{\hat{\Phi}_{m}}0=0$ and its variance
is~$\braOket 0{\left(\hat{\Phi}_{m}\right)^{2}}0=\left(\Phi_{m}^{\mathrm{ZPF}}\right)^{2}$.
The non-zero variance is representative of the ground state energy
and the quantum zero-point fluctuations of the flux. The ZPF saturate
the Heisenberg uncertainty bound, $\mathbf{\Phi_{\mathrm{m}}^{ZPF}}\mathbf{Q_{\mathrm{m}}^{ZPF}}=\frac{\hbar}{2}\m I$.
All mode ZPFs are positive,~$\Phi_{m}^{\mathrm{ZPF}},Q_{m}^{\mathrm{ZPF}}\in\mathbb{R}_{>0}$.

\paragraph{Interpretation of the eigenmode ZPFs and effective mode inductances,
and impedances.}

The mode impedance~$Z_{m}=\Phi_{m}^{\mathrm{ZPF}}/Q_{m}^{\mathrm{ZPF}}=\omega_{m}1_{\mathrm{H}}$
and ZPF~$\Phi_{m}^{\mathrm{ZPF}}=\sqrt{\hbar\omega_{m}/2\times1_{\mathrm{H}}}$
depend only on~$\omega_{m}$. These quantities are in general removed
from direct physical meaning. Physical meaning is extracted by using
them as computational tools to calculate the quantum ZPF of the branch
fluxes and charges. The EPR method negates the need to explicitly
compute them; rather, we only directly work with physically measure
quantities, such as the eigenfields of the modes and the Josephson
dipole EPRs, as discussed in the next sub-section.

\paragraph{From abstract to physical ZPFs.}

The zero-point fluctuations of the spanning-tree fluxes follow from
Eqs.~\eqref{app:eigen:E defn} and~\eqref{eq:app:quantize:phi_zpf},
\begin{equation}
\mathbf{\Phi_{\mathrm{t}}^{ZPF}}=\m E\mathbf{\Phi_{\mathrm{m}}^{ZPF}}\;.\label{eq:appdx:ZPF of bare in ZPF dressed}
\end{equation}
The~$\left(k,m\right)$ element of~$\phit^{\mathrm{ZPF}}$ is the
ZPF of the~$k$-th spanning-tree branch due to mode~$m$ and is either
positive or negative. The overall sign is arbitrary. Hover, the relative
sign between two branches~$k$ and~$k'$ in the same mode determines
if they are excited by the mode in-phase or out-of phase with each
other. The eigenvalue matrix~$\m E$ is constructed canonically so
as to correctly relate the fluctuations. The eigenvector matrix of
the standard product~$\m C^{-1}\m L^{-1}$, obtained in the Lagrangian
equations of motion, is not canonical and cannot be used in Eq.~\eqref{eq:appdx:ZPF of bare in ZPF dressed}.

\paragraph{Second quantization of the dipole flux operator.}

Since~$\m E$ is real, the reduced-magnetic-flux ZPF amplitudes~$\varphi_{mj}$
can be chosen to be real-valued numbers; Eq.~14(c) of the main text
follows, 
\begin{equation}
\boxed{\hat{\varphi}_{j}\isdef\hat{\Phi}_{j}/\phi_{0}=\sum_{m=1}^{M}\varphi_{mj}\left(\hat{a}_{m}+\hat{a}_{m}^{\dagger}\right)\;.}\label{eq:appdx:phi_j defn}
\end{equation}

\subsection{Energy-participation ratio (EPR)\label{app:EPR-defn}}

We define energy-participation ratio in the context of quantum circuits.
We motivate the omission of vacuum energy contributions and use the
EPR to find the quantum zero\textendash point fluctuations~$\varphi_{mj}$.

\paragraph{Definition of EPR in plain english.}

The EPR~$p_{mj}$ of Josephson dipole~$j$ in eigenmode~$m$ is
the fraction of inductive energy allocated to the Josephson dipole
when mode~$m$ is excited.

\paragraph{Interpretation of the energy-participation ratio.}

The EPR quantifies how much of the inductive energy of a mode is allocated
to a Josephson dipole. The lowest possible participation is zero\textemdash the
Josephson dipole inductor does not participate in the mode. When the
mode is excited, none of the excitation energy flows to the dipole.
The largest possible participation is unity. When the mode is excited,
all of its excitation flows to the Josephson dipole; none of it is
distributed to any other inductor. As we show in the following, a
larger participation leads to larger quantum vacuum fluctuations.

\paragraph{Transmon-qubit example.}

Consider the transmon qubit coupled to a readout cavity mode (see
Methods). The Josephson junction has an EPR of near unity in the
qubit eigenmode, since nearly all of the inductance of the transmon
is due to the kinetic inductance of the tunnel junction. On the other
hand, the EPR of the junction in the cavity mode is on the order of~$10^{-2}$.
The junction kinetic inductance contributes only on the order of one
percent to the total inductance of the cavity, associated with the
large cavity box. This leads to the large ZPF of the junction flux
in the transmon mode,~$\varphi_{q}^{\mathrm{ZPF}}\sim1$, and the
therefore much smaller fluctuations of the junction in the cavity
mode~$\varphi_{q}^{\mathrm{ZPF}}\sim10^{-1}$ (see also Sec.~\ref{app:universal-constraints-epr}).

\paragraph{Energy offset.}

The EPR is defined in terms of the energy excitation of a mode, rather
than in terms of its absolute energy. 

\paragraph{Vacuum energy.}

Mathematically, the vacuum energy is a consequence of the non-commutativity
of~$\hat{a}$ and~$\hat{a}^{\dagger}$, which in effect adds a constant
offset to the energy of each mode. In calculating the EPR, we take
the reference energy of an element relative to its vacuum energy.

\paragraph{Equipartition theorem.}

Classically, for a linear circuit, the energy of an eigenmode oscillates
in time between inductive and capacitive energy. At periodic intervals
given by~$\pi/\omega_{m}$, the inductive energy of the mode is exactly
zero. For this reason, we use the time-average energy of the Josephson
dipole and the eigenmode. We recall that the time-averaged energy
is half of the peak energy. This can be exploited in expressing the
total inductive energy as half the total mode energy,
\begin{equation}
\overline{\left\langle \hat{\mathcal{E}}_{\mathrm{ind}}\right\rangle }=\frac{1}{2}\overline{\left\langle \hat{H}_{\mathrm{lin}}\right\rangle }=\frac{1}{2}\sum_{m}\hbar\omega_{m}\overline{\left\langle \hat{a}_{m}^{\dagger}\hat{a}_{m}\right\rangle }\;.\label{eq:inductive-is-half-total}
\end{equation}
We will discuss what states are used in calculating the expectation
values below. The ground state energy is taken to be zero.

\paragraph{Calculating the EPR.}

These ideas lead us to the following definition of the EPR in the
quantum setting,
\begin{equation}
p_{mj}\isdef\overline{\left<\hat{\mathcal{E}}_{j\mathrm{,lin}}\right>}/\overline{\left<\hat{\mathcal{E}}_{\mathrm{ind}}\right>}\;,\label{eq:pmj-defn-app}
\end{equation}
where the over-line denotes a time average and the expectation value
is taken over a state with an excitation only in mode~$m$. As discussed
above, all energies are referenced to their ground-state expectation
values, to disregard vacuum-energy contributions (for special notation
to accommodate this, see remark on Wick ordering at the end of this
section). Using Eqs.~\eqref{eq:inductive-is-half-total} and~\eqref{eq:app:Hlin-multi},
we simplify the denominator,~$\overline{\left<\hat{\mathcal{E}}_{\mathrm{ind}}\right>}=\sum_{m}\frac{1}{2}\hat{a}_{m}^{\dagger}\hat{a}_{m}$.
The operator for the junction energy follows from Eq.~\eqref{eq:Ej-lin-defn},~$\hat{\mathcal{E}}_{j\mathrm{,lin}}=\frac{1}{2}E_{j}\left(\hat{\varphi}_{j}^{2}-\left\langle \hat{\varphi}_{j}^{2}\right\rangle _{0}\right)$,
where~$\left\langle \,\right\rangle _{0}$ indicates an expectation
value over the ground state. Thus,
\begin{equation}
p_{mj}=\frac{\overline{\left\langle \frac{1}{2}E_{j}\hat{\varphi}_{j}^{2}\right\rangle }-\overline{\left\langle \frac{1}{2}E_{j}\hat{\varphi}_{j}^{2}\right\rangle _{0}}}{\frac{1}{2}\sum_{m}\hbar\omega_{m}\left\langle \hat{a}_{m}^{\dagger}\hat{a}_{m}\right\rangle }\;.\label{eq:pmj-from-Ej-Hlin}
\end{equation}
Using Eq.~\eqref{eq:appdx:phi_j defn},
\begin{multline}
\hat{\varphi}_{j}^{2}-\left\langle \hat{\varphi}_{j}^{2}\right\rangle _{0}=\sum_{m=1}^{M}\varphi_{mj}^{2}\left(2\hat{a}_{m}^{\dagger}\hat{a}_{m}+\hat{a}_{m}^{2}+\hat{a}_{m}^{\dagger2}\right)+\\
\sum_{m'\neq m}^{M}\varphi_{mj}\varphi_{m'j}\left(\hat{a}_{m}\hat{a}_{m'}+\hat{a}_{m}\hat{a}_{m'}^{\dagger}+\hat{a}_{m}^{\dagger}\hat{a}_{m'}+\hat{a}_{m}^{\dagger}\hat{a}_{m'}^{\dagger}\right)\;.\label{eq:phi_mj-squared-wick}
\end{multline}

\paragraph{EPR for a Fock state excitation.}

We can take expectation value in Eq.~\eqref{eq:pmj-from-Ej-Hlin}
for a Fock excitation of~$n$ photons in mode~$m$, denoted~$\ket{n_{m}}$.
We recall that the many-body vacuum state is~$\ket 0\isdef\ket{\mathrm{vac}}\isdef\prod_{k=1}^{M}\otimes\ket 0_{k}$,
where~$\ket 0_{k}$ denotes the single-particle, zero Fock state of
mode~$k$; thus,~$\ket{n_{m}}=\frac{1}{\sqrt{n_{m}!}}\left(\hat{a}_{m}^{\dagger}\right)^{n_{m}}\ket{\mathrm{vac}}$.
To take the expectation value of the Josephson dipole flux~$\left\langle \hat{\varphi}_{j}^{2}\right\rangle~$,
we recall that Fock states have trivial time dynamic, and use Eq.~\eqref{eq:phi_mj-squared-wick}.
We observe that the expectation value of any product of two amplitude
operators from different modes is zero. The only non-zero term in
Eq.~\eqref{eq:phi_mj-squared-wick} is the~$\hat{a}_{m}^{\dagger}\hat{a}_{m}$
term. Thus, the EPR of the Josephson dipole for the Fock state~$\ket{n_{m}}$
is
\begin{equation}
p_{mj}=\frac{E_{j}\varphi_{mj}^{2}}{\frac{1}{2}\hbar\omega_{m}}\;.\label{eq:epr-Fock-state}
\end{equation}
The EPR is independent of the excitation amplitude~$n_{m}$.

\paragraph{EPR for a coherent state excitation.}

A coherent state excitation of mode~$m$ is denoted~$\ket{\beta_{m}}\isdef\hat{D}\left(\beta_{m},\hat{a}_{m}\right)\ket{\mathrm{vac}}$,
where~$\hat{D}$ is the displacement operator of the~$m$th mode,
$\hat{D}\left(\beta_{m},\hat{a}_{m}\right)\isdef\exp\left(\beta_{m}\hat{a}_{m}^{\dagger}-\beta_{m}^{\ast}\hat{a}_{m}\right)$,
and~$\beta$ is a non-zero complex number. The coherent state time-evolution
is a rotation,~$\ket{\beta_{m}\left(t\right)}=\ket{\beta_{m}\left(0\right)e^{-i\omega_{m}t}}$.
Using Eq.~\eqref{eq:pmj-from-Ej-Hlin}, we find the EPR to again
be excitation independent and exactly equal to that given in Eq.~\eqref{eq:epr-Fock-state}.

\paragraph{Quantum fluctuations in terms of EPR.}

The EPR for a Fock or coherent state excitation is given by Eq.~\eqref{eq:epr-Fock-state}.
Inverting this expression, we find the ZPF in terms of the EPR,
\begin{equation}
\boxed{\varphi_{mj}^{2}=p_{mj}\frac{\hbar\omega_{m}}{2E_{j}}\;.}\label{eq:phi-ZPF-from-EPR}
\end{equation}
In the case of a single Josephson dipole in the circuit, Eq.~\eqref{eq:phi-ZPF-from-EPR}
reduces to Eq.~(6) of the main text. The left-hand side of Eq.~\eqref{eq:phi-ZPF-from-EPR}
gives the root-mean-square deviation of the quantum fluctuations of
the reduced magnetic flux of Josephson dipole~$j$, referenced to
its equilibrium value. The right-hand side of Eq.~\eqref{eq:phi-ZPF-from-EPR}
comprises three classically-known parameters: the eigenmode frequency~$\omega_{m}$,
the Josephson dipole energy scale~$E_{j}$, and the EPR~$p_{mj}$.

\paragraph{Sign of the EPR.}

Solving Eq.~\eqref{eq:phi-ZPF-from-EPR} explicitly, 
\begin{equation}
\varphi_{mj}=s_{mj}\sqrt{p_{mj}\frac{\hbar\omega_{m}}{2E_{j}}}\;,\label{eq:phimj-general}
\end{equation}
where~$s_{mj}$ is the EPR sign,~$s_{mj}\in\left\{ -1,+1\right\} $.
In practice, the sign is calculated using Eq.~\eqref{eq:Smj-calc}.
Algebraically, the sign can be found from the eigenvector matrix~$\m E$.
If the~$j$th row of the eigenvector matrix corresponds to the~$j$th
spanning-tree branch flux and the~$m$th column corresponds to the~$m$th eigenmode, then~~$s_{mj}=\operatorname{sign}\left(\left[\m E\right]_{mj}\right)$,
where~$\left[\m E\right]_{mj}$ is the~$\left(m,j\right)$th entry
of the matrix.

\paragraph{EPR sign: sign freedom.}

The value of an individual sign~$s_{mj}$ is completely arbitrary.
It does not have measurable consequences in the same way that the
amplitude of a standing mode can be taken to be positive or negative,
indicating a~$\pi$-phase shift. The sign~$s_{mj}$ derives its
physical meaning in relationship to other signs~$s_{mj'}$ of the
same mode but different Josephson dipoles (see Supplementary Figure~\ref{fig:port-signs}).
If the signs of two dipoles are the same (resp., different), then
the two dipoles oscillate in-phase (resp., out-of-phase). The sign
of the EPR can flipped by either flipping the reference direction
of the junction or the overall phase of the mode.

\paragraph{EPR for circuit design.}

The EPR~$p_{mj}$ in Eq.~\eqref{eq:phi-ZPF-from-EPR} is essentially
the only free parameter in engineering the quantum ZPF and the amplitudes
of the non-linear couplings. It is readily calculated from the classical
eigenmode FE simulation of the distributed physical layout of the
circuit, as detailed in Sec.~\ref{app:FE-sims}. In this way, the
EPR serves as a bridge between the linearized classical circuits treated
with FE solvers and the Josephson circuit in the quantum domain.

\paragraph{Remark: Equivalent formulation using Wick normal-ordered expectation
value for energies.}

Mathematically enforcing the energy to be referenced to that of the
vacuum state can be equivalently accomplished by disregarding the
commutation relationships in expectation values of the energy. The
compact notation that accomplishes this was introduced in quantum
optics and is also used in quantum field theory: for an operator~$\hat{O}$,
one takes the (Wick) normal-ordered form of the operator~\citep{Gerry2005},
denoted by a colon on either side of the operator, $\normord O$.
Thus, $\normord{\hat{a}\hat{a}^{\dagger}}=\hat{a}^{\dagger}\hat{a}$
and~$\normord{\left(\hat{a}^{\dagger}\hat{a}\right)^{2}}=\hat{a}^{\dagger2}\hat{a}^{2}$.
The normal-ordered form of the operators should not be confused with
the normal-ordering transformation~$\text{\ensuremath{\mathscr{N}}}$
applied to the operator, which takes the commutation relations into
account; e.g.,~$\text{\ensuremath{\text{\ensuremath{\mathscr{N}}}}}\left[\hat{a}\hat{a}^{\dagger}\right]=\hat{a}^{\dagger}\hat{a}+\hat{I}$
and~$\text{\ensuremath{\mathscr{N}}}\left[\left(\hat{a}^{\dagger}\hat{a}\right)^{2}\right]=\hat{a}^{\dagger2}a^{2}+\hat{a}^{\dagger}\hat{a}$.
In the presence of non-linear interactions, the vacuum energy leads
to observable effects, such as the Lamb shift, introduced in Eq.~(7)
of the main text.  The definition given in Eq.~\eqref{eq:pmj-defn-app}
can be equivalently stated using Wick notation:
\begin{equation}
p_{mj}\isdef\overline{\left<\normord{\hat{\mathcal{E}}_{j\mathrm{,lin}}}\right>}/\overline{\left<\normord{\hat{\mathcal{E}}_{\mathrm{ind}}}\right>}\;.\label{eq:pmj-Wick-Notation}
\end{equation}

\subsection{Universal EPR properties\label{app:universal-constraints-epr}}

The energy-participation ratios obey four universal properties. These
properties are valid regardless of the circuit topology and nature
of the Josephson dipoles. They follow directly from the normal-mode
structure of the eigenmodes of~$H_{\mathrm{lin}}$, obtained in Sec.~\ref{app:diagonalize-eigenmodes}
and are linked to the ZPF, as discussed in Sec.~\ref{app:EPR-defn}.
In the following,~$M$ denotes the total number of modes. 

\paragraph{The EPR is bounded.}

The EPR is an energy fraction. It follows from its definition in Eq.~\eqref{eq:pmj-defn-app}
that it is a real number comprised between zero and one\textemdash since
the Josephson dipole energy is always positive and equal to or smaller
than the total inductive energy of the mode, 

\begin{equation}
0\leq p_{mj}\leq1\;.\label{eq:pmj-bounded}
\end{equation}

\paragraph{The total EPR of a dipole follows a sum rule\textemdash it is conserved
while being diluted among the modes.}

The total EPR of a dipole is conserved\textemdash it is exactly unity
across all modes, 
\begin{equation}
\sum_{m=1}^{M}p_{mj}=1\quad\text{for}\ j\in\left\{ 1,\ldots,J\right\} \;,\label{eq:sum_pj-is-unity}
\end{equation}
which follows from Eqs.~\eqref{app:eigen:E defn} and~\eqref{eq:pmj-from-Ej-Hlin}.
Increasing the EPR of a Josephson dipole in one mode will proportionally
reduce its participation across other modes. The EPR is neither created,
nor destroyed. It is distributed among the modes. Adding additional
modes to the circuit or removing modes from the circuit does not increase
or decrease the total EPR of a dipole. 

\paragraph{The total EPR of a mode is at most unity.}

A single mode~$m$ can at most have a total EPR of unity, and no
less than zero,
\begin{equation}
0\leq\sum_{j=1}^{J}p_{mj}\leq1\quad\text{for}\ m\in\left\{ 1,\ldots,M\right\} \;.
\end{equation}
The upper bound is saturated when there are no linear inductors excited
in the mode.

\paragraph{The vector EPRs of two dipoles are orthogonal.}

For each dipole, we can define a vector EPR by the components~$\left\{ s_{mj}\sqrt{p_{mj}}\,\vert\,m=1,\ldots,M\right\} $.
The vector EPRs of two dipoles are orthogonal in the following sense
\begin{align}
\sum_{m=1}^{M}s_{mj}s_{mj'}\sqrt{p_{mj}p_{mj'}}\, & =0\quad\text{for}\ j\neq j'\;,\label{eq:epr-orthogonality}
\end{align}
where~$s_{mj}$ is the EPR sign.

\paragraph{Remark: Use of these four properties in quantum circuit design.}

These universal EPR constraints are useful in the design of quantum
circuits, especially for designs that require weak and strong nonlinear
interactions simultaneously. For example, see Methods, where we discuss
the impossibility of a design we targeted due to the EPR constraints.
We subsequently use the constraints to obtain a best approximation
of the design, and to gain insight into the range of possible non-linear
couplings. In our experience, the EPR has allowed us to thus circumvent
the need to run a prohibitively expensive finite-element sweep to
explore the many possible design-parameter regimes.

\subsection{The biased Josephson system: equilibrium state in the presence of
persistent currents\label{app:equilibrium-point-of-circuit}}

\begin{figure}
\begin{centering}
\includegraphics{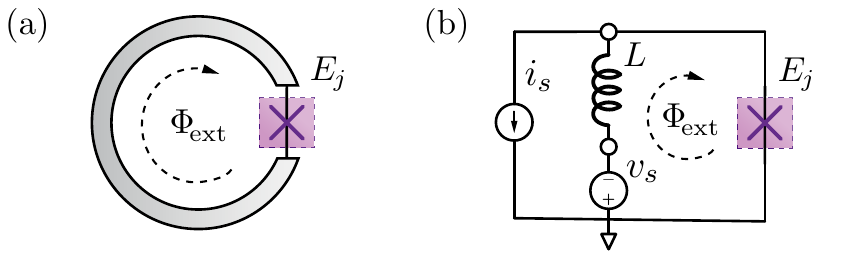}
\par\end{centering}
\caption{\label{fig:Example-junction-in}\textbf{Example of a Josephson dipole
embedded in frustrated loop.} (a) Illustration of a Josephson tunnel
junction (with Josephson energy~$E_{j}$) embedded in a distributed,
superconducting ring subjected to an external magnetic-flux bias~$\Phi_{\mathrm{ext}}$.
(b) Lumped-element model a junction in an inductive ring (with inductance~$L$)
frustrated by an external flux bias~$\Phi_{\mathrm{ext}}$, a voltage
source~$v_{s}$, and a current source~$i_{s}$.}
\end{figure}

The magnetic flux across a Josephson dipole~$\Phi_{j}$, introduced
in Sec.~\ref{app:Review-of-electrical-theory}, is defined as a \emph{deviation}
away from the equilibrium configuration of the Josephson system. A
system that incorporates an active element\textemdash one that can
act as a voltage or current source\textemdash or a system subjected
to a frustrated constraint\textemdash such as found in a conducting
loop frustrated by an external magnetic flux\textemdash can have a
non-zero equilibrium state. Referencing the Josephson dipole and spanning-tree
fluxes~$\phit$ from equilibrium guarantees that the total inductive
energy~$\mathcal{E}_{\mathrm{ind}}\left(\phit\right)$ is at a minimum
for~$\phit=\m 0$.

\paragraph{Static-equilibrium conditions.}

In static equilibrium, the net generalized forces (currents) at each
node in the system vanish and, consequently, the generalized acceleration
vanishes, $\frac{\mathrm{d}^{2}\phit}{\mathrm{d}t^{2}}=\m 0$. The
corresponding magnetic flux of the spanning-tree branches in static
equilibrium~$\m{\Phi}_{\mathrm{t,eq}}$ can be found by extremizing
the Lagrangian with respect to the generalized position variables
\citep{Landau1982}, 
\begin{equation}
-\frac{\partial\mathcal{L}}{\partial\phit}\left(\m{\Phi}_{\mathrm{t,eq.}}\right)=\frac{\partial\mathcal{E}_{\mathrm{ind}}}{\partial\phit}\left(\m{\Phi}_{\mathrm{t,eq}}\right)=\m 0\;.\label{eq:Lagrnagian equilibrium}
\end{equation}
Solving these conditions for the equilibrium flux~$\m{\Phi}_{\mathrm{t,eq}}$
amounts to solving for the direct-current (dc) operating point of
the circuit\textemdash a standard, classical problem; although one
that is non-local in nature and, in the presence of Josephson dipoles,
one that involves solutions to non-linear equations. In general, classical
numerical methods can be used to find the equilibrium of the circuit,
especially when the equilibrium equations are transcendental. However,
for many practical situations, Eq.~\eqref{eq:Lagrnagian equilibrium}
simplifies or need not be evaluated at all, as discussed in Sec.~\ref{subsec:The-biased-Josephson-simplications}.

\paragraph{Equilibrium flux of a Josephson Dipole in isolation vs.~in a circuit.}

In general, a Josephson dipole has a different equilibrium flux across
its terminals when in isolation vs.~when embedded in a system. When
in isolation, the native equilibrium flux of the Josephson dipole
is found from the simple local condition~$\frac{\partial\mathcal{E}_{j}}{\partial\Phi_{j}}=0$,
where~$\mathcal{E}_{j}$ is the energy function of the Josephson
dipole and~$\Phi_{j}$ is the magnetic flux across the dipole. When
embedded in a system, the equilibrium flux of the Josephson dipole,
found from Eq.~\eqref{eq:Lagrnagian equilibrium}, is in general
not a local property of the dipole anymore\textemdash but one of the
system, as illustrated by the following example.

\begin{figure}
\begin{centering}
\includegraphics{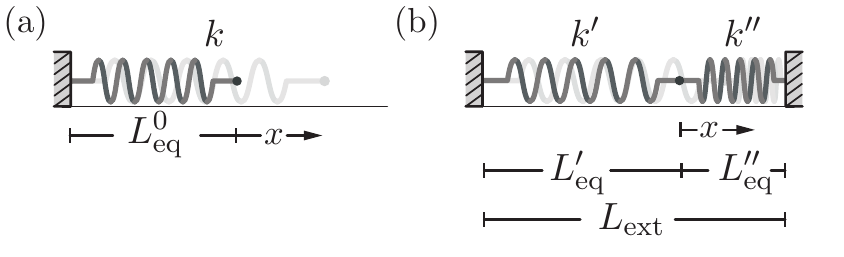}
\par\end{centering}
\caption{\label{fig:Mechanical-analogy}\textbf{Mechanical analogy of a Josephson
dipole in isolation vs.~in a frustrated system.} (a) Depiction of
a mechanical spring in isolation. Without applied forces on the spring,
the native equilibrium length of the spring is~$L_{\mathrm{eq}}^{0}$.
Stretching or compressing the spring from equilibrium is measured
by the deviation~$x$ away from the native equilibrium. The spring
constant is~$k$. (b) Depiction of two connected springs, with spring
constant~$k'$ and~$k''$, constrained between two walls separated
by a fixed distance~$L_{\mathrm{ext}}$. The equilibrium lengths
of the two springs in the system are~$L_{\mathrm{eq}}'$ and~$L_{\mathrm{eq}}''$.
These differ from their equilibrium lengths in isolation; i.e.,~$L_{\mathrm{eq}}\protect\neq L_{\mathrm{eq}}'$.
Deviations from the system equilibrium are denoted by~$x$.}
\end{figure}

\paragraph{Example: Josephson junction in a frustrated ring.}

Consider a Josephson tunnel junction in isolation\textemdash not connected
to any other elements. It's energy function, see Eq.~\eqref{eq:jj-energy-func},
is~$\mathcal{E}_{i}\left(\Phi_{i}\right)=-E_{J}\cos\left(\Phi_{i}/\phi_{0}\right)$,
where~$\Phi_{i}$ is the total magnetic flux across the junction
in isolation. (Here, the subscript~$i$ denotes the junction in isolation.)
The energy is minimized at the native equilibrium-flux value~$\Phi_{\mathrm{eq}}^{\mathrm{nat}}=0$.
Now, consider embedding this junction in a superconducting ring frustrated
by an external magnetic flux~$\Phi_{\mathrm{ext}}$, as depicted
in Supplementary Figure~\ref{fig:Example-junction-in}(a). Suppose
the geometric ring inductance is~$L$, see Supplementary Figure~\ref{fig:Example-junction-in}(b).
The equilibrium condition of the circuit, given by Eq.~\eqref{eq:Lagrnagian equilibrium},
reduces to Kepler's transcendental equation
\begin{equation}
L^{-1}\Phi_{i}+\phi_{0}^{-1}E_{J}\sin\left(\Phi_{i}/\phi_{0}\right)=i_{s}^{\mathrm{eff}}\;,\label{eq:Keplers-equation}
\end{equation}
where the effective dc current sourced to the loop is~$i_{s}^{\mathrm{eff}}=L^{-1}\Phi_{\mathrm{ext}}$.
In the additional presence of a voltage source~$v_{s}$ and current
source~$i_{s}$ as depicted in see Supplementary Figure~\ref{fig:Example-junction-in}(b),
the net effective current is~$i_{s}^{\mathrm{eff}}=L^{-1}\Phi_{\mathrm{ext}}+i_{s}$.

A solution~$\Phi_{\mathrm{eq}}\left(\Phi_{\mathrm{ext}},\Phi_{s},i_{s}\right)$
of Eq.~\eqref{eq:Keplers-equation} can be obtained numerically or
using Lagrange inversion. In the EPR method, the canonical flux deviation
away from the selected equilibrium~$\Phi_{\mathrm{eq}}$ is then
\begin{equation}
\Phi_{j}\isdef\Phi_{i}-\Phi_{\mathrm{eq}}\;.\label{eq:simple-flux-deviation}
\end{equation}
The energy function of the junction now in terms of the deviation~$\Phi_{j}$
is
\begin{equation}
\mathcal{E}_{j}\left(\Phi_{j}\right)\isdef-E_{J}\cos\left(\left(\Phi_{\mathrm{eq}}+\Phi_{j}\right)/\phi_{0}\right)\;.\label{eq:frustrated-jj-energy}
\end{equation}
Using the series expansion of Eq.~\eqref{eq:frustrated-jj-energy},
given in Eq.~\eqref{eq:Ej-jj-frustrated}, the Josephson dipole energy
of the junction, as defined in Eq.~\eqref{eq:Ej-dipole-defined},
is~$E_{J}\left(\Phi_{\mathrm{ext}},\Phi_{s},i_{s}\right)\isdef E_{J}\cos\left(\Phi_{\mathrm{eq}}/\phi_{0}\right)$
and the leading-order non-linear coefficient is~$c_{j3}\left(\Phi_{\mathrm{ext}},\Phi_{s},i_{s}\right)=-\frac{1}{6}\tan\left(\Phi_{\mathrm{eq}}/\phi_{0}\right)$.

\paragraph{Mechanical analogy: spring in isolation.}

By way of analogy, a Josephson dipole is like a mechanical spring.
Imagine a spring in isolation, see Supplementary Figure~\ref{fig:Mechanical-analogy}(a).
No external forces act on it. Its native equilibrium length (when
the spring is in isolation) is~$L_{\mathrm{eq}}^{0}$. Stretching
the spring to length~$L$ results in the force~$F=-k\left(x_{i}\right)$
on its ends, where~~$x_{i}\isdef L-L_{\mathrm{eq}}^{0}$ denotes
a deviation away from equilibrium. In general,~$k\left(x_{i}\right)$
is a non-linear function; however, in equilibrium,~$k\left(0\right)=0$.
For a linear spring, the energy,~$\mathcal{E}_{i}\left(x_{i}\right)=\frac{1}{2}kx_{i}^{2}$,
is minimized for~$x=0$.

\paragraph{Mechanical analogy: spring in a frustrated system.}

Imagine embedding the spring in a simple system as depicted in Supplementary
Figure~\ref{fig:Mechanical-analogy}(b). The system imposes the KVL-like
constraint condition~$L'+L''=L_{\mathrm{ext}}$, where~$L'$ is
the length of the second spring, and~$L_{\mathrm{ext}}$ is the total
distance between the two walls. For linear springs, the length of
the spring in the equilibrium \emph{of the system} is~$L_{\mathrm{eq}}=\left[kL_{\mathrm{eq}}'+k'\left(L_{\mathrm{ext}}-L_{\mathrm{eq}}''\right)\right]/\left(k+k'\right)$,
where~$k'$ and~$L_{\mathrm{eq}}''$ are the spring constant and
native equilibrium length of the second spring, respectively. The
energy of the spring is~$\mathcal{E}\left(x\right)=\frac{1}{2}kx^{2}$,
where we omit constant energy terms and use~$x\isdef L-L_{\mathrm{eq}}$
to denote deviations away from the equilibrium length of the spring
in the system (not the spring in isolation).

\paragraph{Remark: voltage source and equilibrium.}

The voltage bias~$v_{s}$ does not appear in the effective current
bias~$i_{s}$ of the junction. The capacitor isolates it in dc from
the rest of the circuit and prevents it from establishing a dc current
in any loop. In Sec.~\ref{subsec:The-biased-Josephson-simplications},
we discuss further situation in which a bias has no effect on the
equilibrium.

\paragraph{Remark: voltage source and dc loops.}

A dc voltage source is absent from the dc-conducting superconducting
loop formed by the junction and inductor, since its magnetic flux
grows linearly in time,~$\Phi_{s}=v_{s}t$, and it would hence supply
a current to the loop that grows infinitely large. At some time, this
would break the abstraction of the ideal voltage source or quench
the superconductivity.

\subsection{The biased Josephson system: simplifications\label{subsec:The-biased-Josephson-simplications}}

\begin{figure}
\begin{centering}
\includegraphics{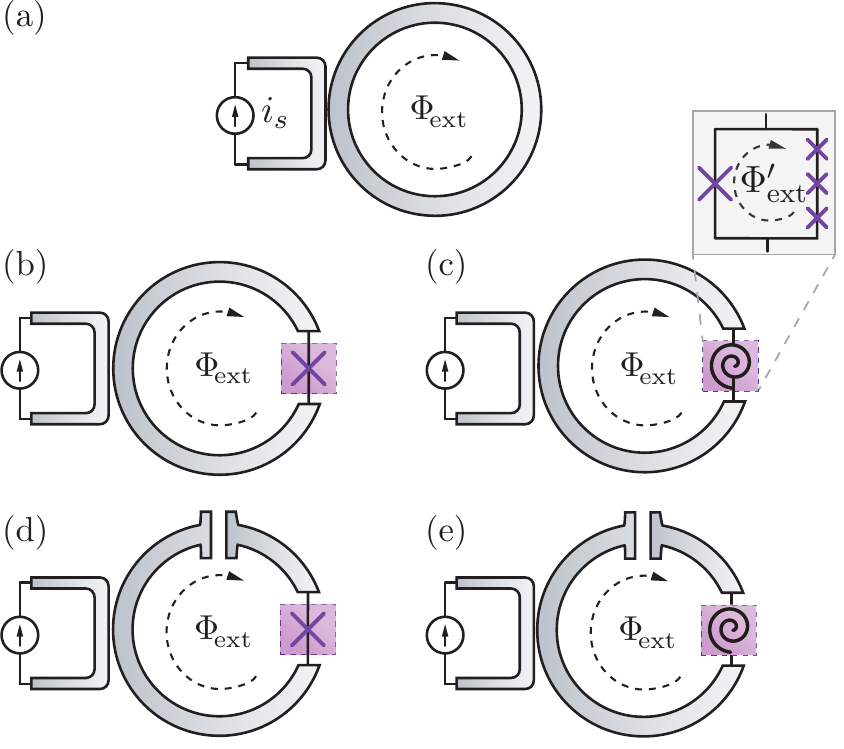}
\par\end{centering}
\caption{\label{fig:Example-frustrated-sys-simplify}\textbf{Schematic of five
types of elementary flux-biased Josephson circuits. }(a) Example of
a fully linear circuit (no Josephson dipoles).\textbf{ }A superconducting
ring is frustrated by an external magnetic flux~$\Phi_{\mathrm{ext}}$
and, through a mutual inductive coupling, a current source~$i_{s}$.
(b) Example of system incorporating a Josephson dipole (here, a Josephson
tunnel junction) in a dc-conducting distributed loop frustrated by~$\Phi_{\mathrm{ext}}$
and~$i_{s}$. (c) Same as panel~(b), but with a Superconducting
Nonlinear Asymmetric Inductive eLement (SNAIL) instead of a junction
in the loop. The SNAIL, a composite Josephson dipole, contains internal
nodes and loops. Its internal loop is frustrated by the external magnetic
flux~$\Phi_{\mathrm{ext}}'$. (d) Example of Josephson tunnel junction
embedded in a distributed non-dc-conducting loop. In the region of
the gap, we define the segment of the line-contour for the magnetic
flux~$\Phi_{\mathrm{ext}}$ to be the a minimal-length one across
the gap\textemdash i.e., in terms of electrical schematics, we take
the flux across the effective capacitance associated with the gap
in the loop. (e) Example similar to (d) but with a SNAIL element instead
of tunnel junction.}
\end{figure}

In Sec.~\ref{app:equilibrium-point-of-circuit}, we reviewed the
equilibrium conditions of the Josephson system. Here, we develop several
common situations in which the analysis of these conditions simplifies
from a global to a local one or one that need not be done at all.

\paragraph{A frustrated loop not incorporating Josephson dipoles.}

If an external bias sets up a persistent current involving only linear
parts of the circuit {[}no Josephson dipoles; see Supplementary Figure~\ref{fig:Example-frustrated-sys-simplify}(a){]},
the equilibrium of the circuit does not need to be calculated, since
the dc current will not affect the eigenmodes of the circuit. Put
another way, the alternating-current (ac) response of a \emph{linear}
sub-circuit is independent of its dc configuration.

\paragraph{An open loop incorporating a Josephson dipole without internal loops.}

If a Josephson dipole is only a member of loops in the circuit that
cannot \emph{not }support a dc~current {[}see Supplementary Figure~\ref{fig:Example-frustrated-sys-simplify}(d){]},
any external frustration or bias of the loop is impotent. Thus, the
equilibrium flux of the Josephson dipole in isolation~$\Phi_{\mathrm{eq}}^{\mathrm{nat}}$
can be taken as the equilibrium flux of the Josephson dipole in the
circuit. In other words, we can treat the dipole locally in the steady-state
analysis\textemdash i.e., we can forget about the rest of the distributed
circuit in which it is embedded. This is one of the most common cases
of practical interest, and is the one of the transmon and SQUID-based
qubit.

\paragraph{An open loop incorporating a Josephson dipole with internal loops.}

A Josephson dipole may have frustrated loops internal to its structure
{[}see Supplementary Figure~\ref{fig:Example-frustrated-sys-simplify}(e){]}.
For example, a lumped-element SQUID, RF-SQUID, or SNAIL is subjected
to an external magnetic flux~$\Phi_{\mathrm{ext}}'$. However, since
a Josephson dipole is purely inductive and if it is not part of a
global conducting loop, the equilibrium conditions of its internal
nodes are solved locally.

\paragraph{A dc-conducting loop incorporating a Josephson dipole with no internal
loops.}

If a simple Josephson dipole with no loops internal to it (such as
the Josephson tunnel junction; not a SQUID or SNAIL), is embedded
in a dc-conducting distributed and frustrated loop, see Supplementary
Figure~\ref{fig:Example-frustrated-sys-simplify}(b), then the flux
of the dipole in the equilibrium of the total circuit cannot be determined
locally. Rather, calculating the equilibrium requires incorporating
information about the geometric-inductance response of the distributed
loop at dc. Classical simulation methods can be used to obtain an
equilibrium point.

\paragraph{A dc-conducting loop incorporating a Josephson dipole with internal
loops.}

In the most general case, a Josephson dipole is part of a dc-conducting
loop; see Supplementary Figure~\ref{fig:Example-frustrated-sys-simplify}(c).
If the loop is not frustrated, then the equilibrium of flux of the
dipole can be taken as the native equilibrium flux of the dipole in
isolation. However, in practice, even if no frustration is intended,
there may be spurious magnetic flux that will frustrate the loop.
If the loop is frustrated by sources (which can include other Josephson
dipoles able to source current, such as a biased SNAIL), the equilibrium
condition of the global loop should be calculated accounting for the
geometric inductances.

\paragraph{Voltage offset.}

A voltage offset will have also have a frustrating effect on the eigenspectrum
of the Hamiltonian. However, for the purposes of calculating the potential
energy minima for use in the EPR linearization, it is sufficient to
consider only the frustration due to currents.

\paragraph{Multiple equilibrium states.}

The equilibrium conditions can yield multiple solutions that locally
minimize the energy function~$\mathcal{E}_{\mathrm{ind}}$. It is
preferable to choose a lowest-energy, stable equilibrium state for
the operating point of the circuit. In principle, any of the minima
can be used in the EPR method as the operational point; the physics
of all other minima can be reconstructed from the operating point.
However, we note that the presence of multiple minima can lead to
phase slips \citep{Matveev2002,Manucharyan2012,Pop2012,Masluk2012,Astafiev2012,Rastelli2013}.

\section{Nonlinear interactions, effective Hamiltonians, and the EPR\label{app:nonlinear-interactions}}

In this section, we find the amplitude of a general nonlinear interaction
due to~$\hat{H}_{\mathrm{nl}}$ explicitly. In Sec.~\ref{app:excitation-conserving-mixing},
we find the effective interaction Hamiltonian of weakly non-linear
systems in the dispersive regime to leading-order. In Sec.~\ref{app:EPR-matrix},
to more systematically handle large systems, we introduce the EPR
matrix~$\m P$ and use it find the Kerr matrix and the vectors of
anharmonicities and Lamb shifts. In Sec.~\ref{app:general-interactions},
we find the general, normal-ordered form of~$\hat{H}_{\mathrm{nl}}$
and an expression to calculate the amplitude of any many-body interaction
contained within. In Sec.~\ref{app:pumping-tones}, we use these
results to describe the parametric activation of nonlinearities in
a pumped Josephson circuit.

\subsection{Excitation-number-conserving interactions of weakly-nonlinear systems\label{app:excitation-conserving-mixing}}

Josephson circuits that are weakly non-linear and in the dispersive
regime, such as the transmon-cavity and transmon-transmon circuits,
have, in the absence of drives, dominant non-linear interactions that
conserve excitation numbers.

\paragraph{Perturbative and dispersive.}

For such weakly nonlinear systems with small zero-point fluctuations~$\varphi_{mj}\ll1$,
for all modes~$m$ and junctions~$\text{j}$, the Hamiltonian~$\hat{H}_{\mathrm{nl}}$
exerts only a perturbative effect on the eigenspectrum of~$\hat{H}_{\mathrm{lin}}$.
Hence, we treat its effect order-by-order using the expansion~$\hat{H}_{\mathrm{nl}}=\sum_{j=1}^{J}\sum_{p=3}^{\infty}E_{j}c_{jp}\hat{\varphi}_{j}^{p}$,
obtained in Eq.~\eqref{eq:Hnl-quantum}. For this approach, we focus
on the regime in which the energy difference between two modes is
much larger than the phase excitation of the Josephson dipoles,~$\hbar\left(\omega_{k}-\omega_{m}\right)\gg E_{j}c_{jp}\left|\hat{\varphi}_{j}^{p}\right|$
for all~$k,m,j$, and~$p$.

\paragraph{Example: transmon coupled to a readout cavity mode.}

Recall the simple example circuit quantized at the beginning of the
main text. A qubit ($q$) is coupled to a cavity ($c$). To leading
order in~$p$, the circuit Hamiltonian~$\hat{H}_{\mathrm{nl}}$
is, using Eq.~\eqref{eq:Hnl-quantum},

\begin{eqnarray}
\hat{H}_{\mathrm{nl}} & = & -\frac{E_{J}}{24}\left(\varphi_{c}\hat{a}_{c}+\varphi_{q}\hat{a}_{q}+\mathrm{H.c.}\right)^{4}+\mathcal{O}\left(\hat{\varphi}_{J}^{6}\right)\;,\label{eq:Hnl-cav-qubit}
\end{eqnarray}
where~$E_{J}$ and~$\hat{\varphi}_{J}$ are the Josephson energy
and flux operator of the tunnel junction, and~$\varphi_{q}$ and~$\varphi_{c}$
are the qubit and cavity mode ZPF, respectively.

\paragraph{Expanding the multinomial.}

Expanding the~$p=4$ multinomial term in Eq.~\eqref{eq:Hnl-cav-qubit},
we find a weighted sum of all possible four-body interactions between
the qubit and cavity; example terms include~$\hat{a}_{q}\left(\hat{a}_{q}^{\dagger}\right)^{3}$
and~$\hat{a}_{c}\hat{a}_{c}^{\dagger}\hat{a}_{q}\hat{a}_{q}^{\dagger}$.
Using the commutation relations {[}Eq.~\eqref{eq:commutations}{]},
we normal order the polynomial. For example, the qubit-only operators,
this yields
\begin{multline*}
\hat{a}_{q}^{4}+4\hat{a}_{q}^{\dagger}\hat{a}_{q}^{3}+6\hat{a}_{q}^{\dagger2}\hat{a}_{q}^{2}+4\hat{a}_{q}^{\dagger3}\hat{a}_{q}+\hat{a}_{q}^{\dagger4}\\
+\text{\ensuremath{6\hat{a}_{q}^{2}+12\hat{a}_{q}^{\dagger}\hat{a}_{q}+6\hat{a}_{q}^{\dagger2}}}+3\hat{I}\;.
\end{multline*}

\paragraph{Higher-order nonlinearity yields lower-order coupling.}

Due to the non-commutativity of the operators, the normal-ordering
procedure results in terms that have lower polynomial order than those
in original unordered expression; these new terms include~$\hat{a}_{q}^{\dagger}\hat{a}_{q}$
and~$3\hat{I}$. Such quadratic terms (e.g.,~$\hat{a}_{q}^{\dagger}\hat{a}_{q}$,
$\hat{a}_{c}^{\dagger}\hat{a}_{c}$, and~$\hat{a}_{c}^{\dagger}\hat{a}_{q}$)
dress the modes of~$\hat{H}_{\mathrm{lin}}$ in a linear manner\textemdash both
renormalizing their frequencies and hybridizing them. These new linear
couplings cannot be a-priori straightforwardly included in~$\hat{H}_{\mathrm{lin}}$,
since they occur only as a non-classical consequence of the ZPF in
the non-linearity; they do not appear in a classical treatment of
the Josephson system, which lacks the non-commuting operator aspect.
Hence, their amplitudes are determined by the quantum ZPF of the eigenmodes.
Remark: A self-consistent approach to include these linear couplings
in~$H_{\mathrm{lin}}$ is possible under certain assumptions \citep{Solgun2017}.

\paragraph{Rotating-wave approximation in the dispersive regime.}

To leading-order, under the rotating-wave approximation (RWA), only
excitation-number conserving interactions in~$\hat{H}_{\mathrm{nl}}$
(those with an equal number of raising and lowering operators in each
mode) contribute. Thus, Eq.~\eqref{eq:Hnl-cav-qubit} reduces to
the effective, dispersive qubit-cavity Hamiltonian 
\begin{equation}
\begin{split}\hat{\overline{H}}_{p=\mathrm{4}}=\sum_{m}-\hbar\Delta_{m}\hat{a}_{m}^{\dagger}\hat{a}_{m}-\frac{1}{2}\hbar\alpha_{m}\hat{a}_{m}^{\dagger2}\hat{a}_{m}^{2}\\
-\sum_{n\neq m}\frac{1}{2}\hbar\chi_{mn}\hat{a}_{m}^{\dagger}\hat{a}_{m}\hat{a}_{n}^{\dagger}\hat{a}_{n}\;,
\end{split}
\label{eq:app:mixing:energy-conserving:H4-bar}
\end{equation}
where~$m\in\left\{ q,c\right\} $ is the mode label,~$\Delta_{m}$
is the effective Lamb shift,~$\alpha_{m}$ is the anharmonicity,
and~$\chi_{mn}$ is the total cross-Kerr frequency shift induced
between modes~$m$ and~$n$. The bar over~$H$ denotes the RWA;
the subscript~$p=4$ denotes the highest the power of the Taylor expansion
included in the effective Hamiltonian. The excitation spectrum of
this Hamiltonian is illustrated in Supplementary Figure~\ref{fig:transmon-cavity-spectrum}.
Remark: Using first-order time-independent perturbation theory in
place of the RWA yields the same result as the RWA.

\paragraph{Hamiltonian parameters.}

The Hamiltonian parameters are obtained from the normal ordering and
using Eq.~\eqref{eq:phi-ZPF-from-EPR}, 
\begin{subequations}
\label{eq:Hconsr-params}
\begin{align}
\chi_{mn} & =\sum_{j=1}^{J}\hbar^{-1}E_{j}\varphi_{mj}^{2}\varphi_{nj}^{2}=\sum_{j=1}^{J}\frac{\hbar\omega_{m}\omega_{n}}{4E_{j}}p_{mj}p_{nj}\;,\label{eq:app:conserve-E:chimn_zpf}\\
\Delta_{m} & =\frac{1}{2}\sum_{n=1}^{M}\chi_{mn}\;,\\
\alpha_{m} & =\frac{1}{2}\chi_{mm}\;.
\end{align}
\end{subequations}

\paragraph{Constraints in the Hamiltonian.}

First, the structure of the non-linearity imposes certain constraints.
For example,~$\chi_{mn}\leq2\sqrt{\alpha_{m}\alpha_{n}}$, where
the equality holds only for the one-junction case, $J=1$. Second,
the universal EPR properties impose a strict limit on physically realizable
designs, subject to Eqs.~\eqref{eq:pmj-bounded}\textendash \eqref{eq:epr-orthogonality}.

\paragraph{Degeneracies.}

So far, we implicitly assumed the system does not have accidental
degeneracies, such as those that occur when the frequency of a mode
is an integer multiple of that of another,~$\omega_{k}\approx z\times\omega_{m}$,
where~$z\in\mathbb{Z}$. In the case of such a degeneracy, additional
terms survive the RWA; for example, if~$\omega_{c}=3\omega_{q}$,
then the non-excitation conserving term~$\hat{a}_{q}^{3}\hat{a}_{c}^{\dagger}$
survives the RWA.

\paragraph{EPR sign \& example.}

Due to the even-power nature of the nonlinear interactions, the EPR
sign drops out altogether from Eq.~\eqref{eq:Hconsr-params}. The
interaction strength between the qubit and cavity modes only depends
on the overlap int the EPRs of the two modes alone. The significance
of this is curiously showcased by Device~DT3, presented in Methods.
The qubit eigenmodes of DT3 are the equally-hybridized symmetric (bright,
B) and antisymmetric (dark, D) combinations of the two bare transmon
qubit modes. Naively, using the analogy of atoms in a cavity, one
reasons that the symmetric mode couples to the cavity~(C),~$\left|\chi_{\mathrm{BC}}\right|\gg0$,
and the antisymmetric modes does not,~$\left|\chi_{\mathrm{DC}}\right|=0$,
due to the out-of-phase oscillation of the junction current dipole\textemdash a
cancelation effect. However, from the EPR expression, it is seen that,
to the contrary, the signs are irrelevant\textemdash no such cancelation
is possible. Rather, according to the EPR method, since both modes
have equal participation in the two junctions ($p_{D1}=p_{D2}=p_{B1}=p_{B2}$),
the couplings to the cavity are essentially equal,~$\chi_{\mathrm{DC}}\approx\chi_{\mathrm{BC}}$;
indeed, this is observed in the experiment (see the corresponding
data table in the Methods section).

\paragraph{Generalizing to many modes.}

Equations~\eqref{eq:app:mixing:energy-conserving:H4-bar} and~\eqref{eq:Hconsr-params}
were derived for the example of a qubit-cavity circuit; however, they
generalize straightforwardly to~$M$ modes under the simple extension~$m\in\left\{ 1,\ldots,M\right\} $.

\begin{figure}
\begin{centering}
\includegraphics{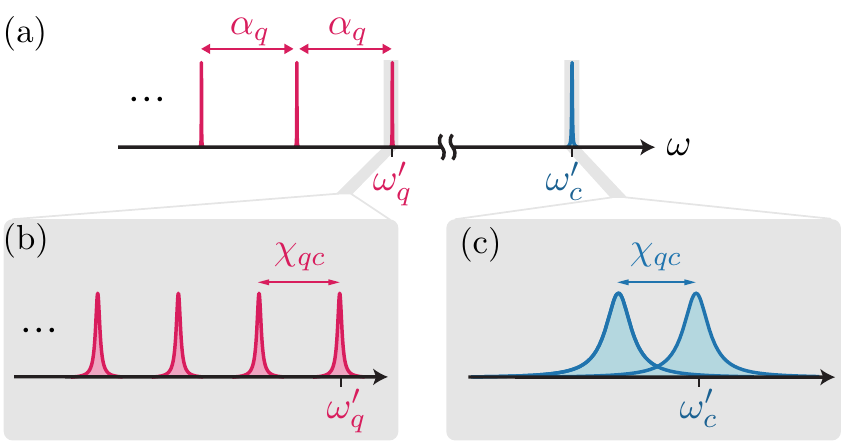}
\par\end{centering}
\caption{\label{fig:transmon-cavity-spectrum}\textbf{Excitation spectrum }(not-to-scale)
of a transmon-cavity circuit, corresponding to the effective dispersive
Hamiltonian~$\hat{H}_{\mathrm{lin}}+\hat{\overline{H}}_{p=4}$, see
Eq.~\eqref{eq:app:mixing:energy-conserving:H4-bar}. (a) The cavity
and qubit frequencies in the ground state are~$\omega_{c}'$ and~$\omega_{q}'$,
respectively. The qubit anharmonicity is~$\alpha_{q}$, and the qubit-cavity
dispersive cross-Kerr shift is~$\chi_{qc}$. (b) Zoom-in on the qubit
spectrum in the vicinity of~$\omega_{q}'$. Each photon in the resonator
loads the qubit frequency down by~$\chi_{qc}$. (c) Zoom-in on the
cavity spectrum in the vicinity of~$\omega_{c}'$.}
\end{figure}

\subsection{EPR matrices and many-body interactions\label{app:EPR-matrix}}

To more easily and systematically handle large-scale circuits and
higher-order non-linearities, we introduce the EPR matrix~$\m P$,
comprising the~$M\times J$ EPRs~$p_{mj}$ as its entries, 
\begin{equation}
\m P\isdef\left(\begin{array}{ccc}
p_{11} & \cdots & p_{1J}\\
\vdots & \ddots & \vdots\\
p_{M1} & \cdots & p_{MJ}
\end{array}\right)\;.\label{ex:P-mat-defn}
\end{equation}
The cross-Kerr interaction amplitudes~$\chi_{mn}^{(p)}$ due to the~$p$-th order terms of~$\hat{H}_{\mathrm{nl}}$ {[}for~$p=4$, see
Eq.~\eqref{eq:app:conserve-E:chimn_zpf}{]} can similarly be organized
in an~$M\times M$ matrix,
\begin{equation}
\boldsymbol{\chi}_{p}\isdef\left(\begin{array}{ccc}
\chi_{11}^{(p)} & \cdots & \chi_{1M}^{(p)}\\
\vdots & \ddots & \vdots\\
\chi_{M1}^{(p)} & \cdots & \chi_{MM}^{(p)}
\end{array}\right)\;.\label{eq:Kerr-matrix}
\end{equation}
It follows from Eq.~\eqref{eq:app:conserve-E:chimn_zpf} that the
leading-order Kerr matrix is 
\begin{equation}
\boxed{\boldsymbol{\chi}_{4}=\frac{\hbar}{4}\left(\m{\Omega}\m P\right)\m E_{j}^{-1}\left(\m{\Omega}\m P\right)^{\mathrm{T}}\;,}\label{eq:Kerr-matrix-p4}
\end{equation}
where~$\m{\Omega}$ is the diagonal eigenfrequency matrix and 
\begin{equation}
\m E_{j}^{-1}\isdef\begin{pmatrix}E_{1}^{-1}\\
 & \ddots\\
 &  & E_{J}^{-1}
\end{pmatrix}\label{app:Eji-matrix}
\end{equation}
is the diagonal matrix of inverse Josephson energies.

To leading order, $p=4$, the vector of anharmonicities is the diagonal
of~$\boldsymbol{\chi}_{4}$, 
\begin{equation}
\bm{\alpha}_{4}\isdef\left(\alpha_{1}^{(4)},\ldots,\alpha_{M}^{(4)}\right)^{\mathrm{T}}=\frac{1}{2}\mathrm{diag}\,\boldsymbol{\chi}_{4}\;,\label{eq:anharmonicity-vector-p4}
\end{equation}
and the Lamb shift of mode~$m$ is the~$m$-th row sum of the Kerr
matrix. 
\begin{equation}
\m{\Delta}_{4}\isdef\left(\Delta_{1}^{(4)},\ldots,\Delta_{M}^{(4)}\right)^{\mathrm{T}}=\frac{1}{2}\boldsymbol{\chi}_{4}\m 1_{M}\;,\label{eq:Lamb-shift-vec-p4}
\end{equation}
where~$\m 1_{M}$ denotes a column vector of length~$M$ with all
entries equal to unity.

\paragraph{Dilution of the nonlinearity.}

The dilution of the nonlinearity of the Josephson dipole elements
among the eigenmodes is neatly expressed in Eq.~\eqref{eq:Kerr-matrix-p4}.
The Josephson dipole energies~$\m{E_{J}}^{-1}$ are diluted by~$\m{\Omega}\m P$
through a congruence transform to the Kerr coefficients. While the
eigenfrequencies~$\m{\Omega}$ weighs the contribution to each mode,
the participation matrix~$\m P$ dictates the dilution of the junction
energies and their nonlinearity among the modes.

\paragraph{Higher-order nonlinear corrections and dilution.}

Using the results of Sec.~\ref{app:general-interactions}, the~$p$-th
other correction to the Kerr matrix is 
\begin{equation}
\boldsymbol{\chi}_{p}=\hbar c_{p}\left(\m{\Omega}\m P\right)\m E_{j}^{-1}\bm{\varphi}_{\mathrm{tot}}^{p-4}\left(\m{\Omega}\m P\right)^{\mathrm{T}}\;,\label{eq:chi-mat-pth-order}
\end{equation}
where~$\bm{\varphi}_{\mathrm{tot}}=\operatorname{Diag}\left(\varphi_{1,\mathrm{tot}},\ldots,\varphi_{J,\mathrm{tot}}\right)$
is the diagonal matrix of the total ZPF fluctuation of the Josephson
dipole reduced fluxes, defined in Eq.~\eqref{eq:app:arb:phi_tot}.
The Kerr matrix incorporating corrections due to all orders of the
nonlinearity, see Eq.~\eqref{eq:Hnl-general-coefficient}, is~$\boldsymbol{\chi}\isdef\sum_{p=3}^{\infty}\boldsymbol{\chi}_{p}$.
The congruence-transformation form of Eq.~\eqref{eq:chi-mat-pth-order}
is identical to that of Eq.~\eqref{eq:Kerr-matrix-p4}; it governs
the dilution of the nonlinearity in the same manner, subject to~$\m{\Omega}\m P$.

Similarly to the results of Eqs.~\eqref{eq:anharmonicity-vector-p4}
and~\eqref{eq:Lamb-shift-vec-p4}, the~$p$-th order corrections
to the anharmonicity and Lamb-shift vectors are 
\begin{equation}
\boldsymbol{\alpha}_{p}=\frac{1}{2}\mathrm{diag}\,\boldsymbol{\chi}_{p}\quad\text{and}\quad\m{\Delta}_{p}=c_{p}\left(\bm{\Omega}\m P\right)\bm{\varphi}_{\mathrm{tot}}^{p-2}\bm{1}_{M}\;.\label{eq:alpha-delta-vec}
\end{equation}
In general, the Lamb shift correction depends on the total ZPF frustration
of the junctions.

\subsection{General many-body interactions\label{app:general-interactions}}

So far, we explicitly calculated the leading-order correction on the
spectrum of~$\hat{H}_{\mathrm{lin}}$ due to~$\hat{H}_{\mathrm{nl}}$.
We expressed the eigenmode interactions using normal-ordered many-body
terms, such as~$\chi\hat{a}_{q}^{\dagger}\hat{a}_{c}^{\dagger}\hat{a}_{q}\hat{a}_{c}$.
Here, we extend the analysis and compute the amplitude of any general
normal-ordered term in~$\hat{H}_{\mathrm{nl}}$.

\paragraph{The form a general many-body interaction.}

A general normal-ordered many-body interaction has the form 
\begin{equation}
C_{\bm{\alpha},\bm{\beta}}^{p}\hat{a}_{1}^{\dagger\beta_{1}}\cdots\hat{a}_{M}^{\dagger\beta_{M}}\hat{a}_{1}^{\alpha_{1}}\cdots\hat{a}_{M}^{\alpha_{M}}\eqqcolon C_{\bm{\alpha},\bm{\beta}}^{p}\bm{\hat{a}^{\dagger\beta}}\bm{\hat{a}^{\alpha}}\;,\label{eq:many-body-mixing-form}
\end{equation}
where~$C_{\bm{\alpha},\bm{\beta}}^{p}$ is an energy-dimensioned
amplitude, determined by the~$p$-th-order of~$\hat{H}_{\mathrm{nl}}$
{[}see Eqs.~\eqref{eq:Hnl-quantum} and~\eqref{eq:appdx:phi_j defn}{]},
and the multi-index~$M$-tuples
\begin{equation}
\bm{\alpha}\isdef\left(\alpha_{1},\ldots,\alpha_{M}\right)\quad\text{and}\quad\bm{\beta}\isdef\left(\beta_{1},\ldots,\beta_{M}\right)
\end{equation}
describe the distribution of annihilation and creation operators involved
in the interaction among the~$M$ modes, respectively. The entires
of the multi-index tuples~$\bm{\alpha}$ and~$\bm{\beta}$ are non-negative
integers,~$\alpha_{m},\beta_{m}\in\mathbb{Z}_{\geq0}$. The right-hand
side of Eq.~\eqref{eq:many-body-mixing-form} introduces the multi-index
shorthand notation for the interaction terms.

\paragraph{Multi-index shorthand and operator powers.}

The total number of lowering and raising operators in the expression
of Eq.~\eqref{eq:many-body-mixing-form} is equal to the 1-norm of~$\bm{\alpha}$
and~$\bm{\beta}$, 
\begin{equation}
\left|\bm{\alpha}\right|\isdef\sum_{m=1}^{M}\alpha_{m}\quad\text{and}\quad\left|\bm{\beta}\right|\isdef\sum_{m=1}^{M}\beta_{m}\;,\label{eq:beta-norm}
\end{equation}
respectively. For a given power~$p$ of~$\hat{H}_{\mathrm{nl}}$,
the total number of resulting operators is bounded, $\left|\bm{\alpha}\right|+\left|\bm{\beta}\right|\leq p$.

\paragraph{Expanding the~$\hat{H}_{\mathrm{nl}}$ multinomials.}

To arrive at Eq.~\eqref{eq:many-body-mixing-form} from Eq.~\eqref{eq:Hnl-quantum},
we first group the sum of the annihilation operators for a Josephson
dipole, see Eq.~\eqref{eq:appdx:phi_j defn}, and define 
\begin{align}
\hat{A}_{j} & \isdef\sum_{m=1}^{M}\varphi_{mj}\hat{a}_{m}\;;\label{eq:Aj-sum-operators}\\
\hat{H}_{\mathrm{nl}} & =\sum_{p=3}^{\infty}\sum_{j=1}^{J}E_{j}c_{jp}\left(\hat{A}_{j}+\hat{A}_{j}^{\dagger}\right)^{p}\;.\label{eq:Hnl-using-A}
\end{align}
Importantly, the commutator~$\left[\hat{A}_{j},\hat{A}_{j}^{\dagger}\right]=\sum_{m=1}^{M}\left|\varphi_{mj}\right|^{2}$
is scalar-valued, which allows us to use the normal-ordering non-commutative
binomial theorem to expand the~$p$-th power term of~$\hat{H}_{\mathrm{nl}}$
\citep{Cohen1966,Blasiak2005,Reagor2015}. Using the non-commuting
expansion, 
\begin{equation}
\begin{split}\left(\hat{A}_{j}+\hat{A}_{j}^{\dagger}\right)^{p}=\sum_{k=0}^{\mathrm{floor}\frac{p}{2}}\sum_{i=0}^{p-2k}\frac{p!}{2^{k}k!i!(p-2k-i)!}\times\\
\left(\varphi_{j,\mathrm{tot}}^{2}\right)^{k}\left(\hat{A}_{j}^{\dagger}\right)^{i}\left(\hat{A}_{j}\right)^{p-2k-i}\;,
\end{split}
\label{eq:binomial-expansion-nc}
\end{equation}
where~$\mathrm{floor}\frac{p}{2}$ gives the greatest integer that
is less than or equal to~$\frac{p}{2}$ and the total variance of
the ZPF of the~$j$-th dipole is 
\begin{align}
\varphi_{j,\mathrm{tot}}^{2}\isdef & \left[\hat{A},\hat{A}^{\dagger}\right]=\frac{\hbar}{2}E_{j}^{-1}\sum_{m=1}^{M}p_{mj}\omega_{m}\;.\label{eq:app:arb:phi_tot}
\end{align}
Since~$\hat{A}_{j}$ is the sum of operators that commute, see Eq.~\eqref{eq:Aj-sum-operators},
we can now expand the powers of~$\hat{A}_{j}$ using the classical
multinomial theorem,
\begin{eqnarray}
\left(\hat{A}_{j}\right)^{i} & = & \sum_{\left|\bm{\alpha}\right|=i}\left(\begin{array}{c}
\left|\bm{\alpha}\right|\\
\bm{\alpha}
\end{array}\right)\varphi_{mj}^{\bm{\alpha}}\bm{\hat{a}}^{\bm{\alpha}}\;,\label{eq:Aj-power-expasion}
\end{eqnarray}
where the multi-index shorthand~$\left(\varphi_{mj}\right)^{\bm{\alpha}}\isdef\prod_{m=1}^{M}\varphi_{mj}^{\alpha_{m}}$,
the multinomial coefficient is 
\begin{equation}
\left(\begin{array}{c}
\left|\bm{\alpha}\right|\\
\bm{\alpha}
\end{array}\right)\isdef\frac{\left|\bm{\alpha}\right|!}{\alpha_{1}!\cdots\alpha_{M}!}\;,
\end{equation}
and the sum condition~$\left|\bm{\alpha}\right|=i$ means that the
sum includes terms with all possible tuples~$\bm{\alpha}$ such that
their 1-norm has the value~$i$.

\paragraph{The general normal-ordered form.}

Combining Eqs.~\eqref{eq:Hnl-using-A}\textendash \eqref{eq:Aj-power-expasion},
we find the general normal-ordered many-body form of~$\hat{H}_{\mathrm{nl}}$
to all orders in~$p$ and without approximations,
\begin{equation}
\boxed{\begin{split}\hat{H}_{\mathrm{nl}}=\sum_{p=3}^{\infty}\sum_{k=0}^{\mathrm{floor}\frac{p}{2}}\sum_{i=0}^{p-2k}\sum_{\substack{\abs{\bm{\beta}}=i,\\
\abs{\bm{\alpha}}=p-2k-i
}
}C_{\bm{\alpha},\bm{\beta}}^{p}\bm{\hat{a}^{\dagger\bm{\beta}}}\bm{\hat{a}}^{\bm{\alpha}}\;.\end{split}
}\label{eq:Hnl-general-many-body-form}
\end{equation}
The energy-dimensioned amplitude of an interaction due to the~$p$-th
power of~$\hat{H}_{\mathrm{nl}}$ for a general many body mode-interaction
term is the sum of the individual junction contributions, 
\begin{equation}
\boxed{C_{\bm{\alpha},\bm{\beta}}^{p}\isdef\frac{p!}{\bm{\alpha}!\bm{\beta}!k!2^{k}}\,\sum_{j=1}^{J}E_{j}c_{jp}\varphi_{mj}^{\bm{\alpha}}\varphi_{mj}^{\bm{\beta}}\varphi_{j,\mathrm{tot}}^{2k}\;,}\label{eq:Hnl-general-coefficient}
\end{equation}
where~$k\isdef\frac{1}{2}\left(p-\left|\bm{\alpha}\right|-\left|\bm{\beta}\right|\right)$.

Equation~\eqref{eq:Hnl-general-coefficient} provides the amplitude
for any mode interaction. Its exact value is calculated using the
EPR to obtain the ZPF~$\varphi_{mj}$, using Eq.~\eqref{eq:phimj-general}.
Thus, we have analytically fully constructed~$\hat{H}_{\mathrm{nl}}$,
and, individually, every term contained within, from the FE simulations
through the EPR.

\paragraph{Use cases.}

Equation~\eqref{eq:Hnl-general-coefficient} can be used to explicitly
calculate higher-order corrections to an effective Hamiltonian; for
example, see Eqs.~\eqref{eq:chi-mat-pth-order} and~\eqref{eq:alpha-delta-vec}.
Using Eq.~\eqref{eq:Hnl-general-coefficient}, we can calculate mode
parameters even when the numerical diagonalization of~$\hat{H}_{\mathrm{nl}}$
becomes intractable\textemdash an issue that occurs at even moderate
number of modes. Moreover, Eq.~\eqref{eq:Hnl-general-coefficient}
can be used to engineer pumped multi-photon drive processes and to
activate non-RWA interactions \citep{Mundhada2018}, as discussed
in Sec.~\ref{app:pumping-tones}.

\subsection{The driven Josephson system: parametrically-activated interactions\label{app:pumping-tones}}

\begin{figure}
\begin{centering}
\includegraphics{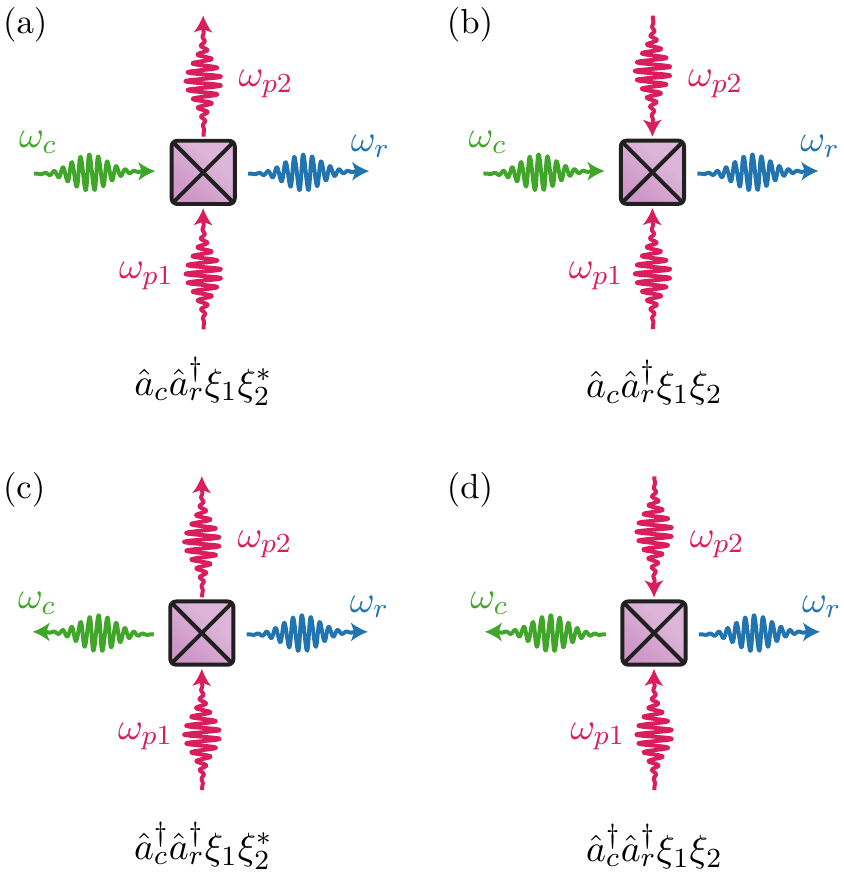}
\par\end{centering}
\caption{\label{fig:pumped-circuit}\textbf{Depiction of four-wave mixing in
a Josephson junction used to parametrically activate a bilinear interaction
between two modes.} Off-resonant pumping of the Josephson circuit
at~$\omega_{p1}$ and~$\omega_{p2}$ with effective straights~$\xi_{1}$and~$\xi_{2}$,
respectively, can activate a specific four-wave-mixing process present
in~$\hat{H}_{\mathrm{nl}}$. The pump frequencies~$\omega_{p1}$
and~$\omega_{p2}$ could be degenerate. The activation condition
is determined by the storage- and readout-mode frequencies~$\omega_{c}$
and~$\omega_{r}$, respectively; their corresponding mode operators
are~$\hat{a}_{c}$ and~$\hat{a}_{q}$. (a), (b) The activated interaction
is a beam-splitter-like conversion process, containing exactly one
mode annihilation and one creation operator. The resonance condition
is determined by~$\omega_{c}-\omega_{r}$. (c), (d) Two-mode squeezing
interaction, contains exactly two creation operators. The resonance
condition is set by~$\omega_{c}+\omega_{r}$. In the first column,
panels (a) and (c), the resonance condition is set by~$\omega_{p1}-\omega_{p2}$,
whereas in the second column, panels (b) and (d), it is set by~$\omega_{p1}+\omega_{p2}$.
Note, for each of the four diagram, the conjugate process (all arrow
directions flipped) is also activated by the pumps. }
\end{figure}

The Josephson system can be subjected to strong external drives used
to parametrically activate or enhance nonlinear couplings. We illustrate
the use of Equation~\eqref{eq:Hnl-general-coefficient} in this context
by calculating the amplitude of a excitation-swapping beam-splitter
interaction.

\paragraph{Example: parametrically-activated beam-splitter interaction.}

Motivated by the setup of device~WG1, we aim to parametrically activate
a beam-splitter (BS) interaction between two non-resonant modes of
a Josephson system; for example, such an interaction can be used as
a Q-switch \citep{Leghtas2015,Pfaff2017}. The system has a high-quality,
storage-cavity mode~($c$) and another low-quality, readout-cavity
mode~($r$). The two modes are far detuned and obey the conditions
outlined at the start of Sec.~\ref{app:excitation-conserving-mixing}.
From the system~$\hat{H}_{\mathrm{nl}}$, we aim to obtain the effective
BS Hamiltonian

\begin{equation}
\hat{H}_{\mathrm{eff}}=\hbar g\hat{a}_{c}^{\dagger}\hat{a}_{r}+\hbar g^{*}\hat{a}_{r}^{\dagger}\hat{a}_{c}\;,\label{eq:app:hb beam split}
\end{equation}
where~$g$ is the rate excitation exchange.

\paragraph{Interaction in the rotating frame.}

In the rotating frame with respect to~$\hat{H}_{\mathrm{lin}}$,
defined by the transform~$\hat{U}\left(t\right)\isdef\exp\left[it\left(\omega_{c}\hat{a}_{c}^{\dagger}\hat{a}_{c}+\omega_{r}\hat{a}_{r}^{\dagger}\hat{a}_{r}\right)\right]$,
the mode operators~$\hat{a}_{m}$ acquire a harmonic time dependence,~$\hat{a}_{m}\mapsto\hat{a}_{m}\left(t\right)\isdef\hat{U}^{\dagger}\left(t\right)\hat{a}_{m}\hat{U}\left(t\right)=\hat{a}_{m}e^{-i\omega_{m}t}$,
where~$m\in\{c,r\}$. While a term in~$\hat{H}_{\mathrm{nl}}$ of
the form~$\hat{a}_{c}^{\dagger}\hat{a}_{r}$ exists, this terms is
non-stationary, $C_{\left(1,0\right),\left(0,1\right)}^{p=4}e^{-it\left(\omega_{r}-\omega_{c}\right)}\hat{a}_{c}^{\dagger}\hat{a}_{r}+\text{H.c.}\;$and
is eliminated in the RWA in deriving the effective Hamiltonian.

\paragraph{Parametric activation and interaction rate.}

A third mode of the system, indexed by~$p$, is off-resonantly driven
at frequencies~$\omega_{P}$ and~$\omega_{P}'$ with amplitudes~$\epsilon_{1}$
and~$\epsilon_{2}$, respectively; within the RWA, the drive Hamiltonian
is

\begin{equation}
\hat{H}_{p}\isdef\epsilon_{p}e^{-i\omega_{P}t}\hat{a}_{p}+\epsilon_{p}^{*}e^{+i\omega_{P}t}\hat{a}_{p}^{\dagger}\;.
\end{equation}
Consider the following term contained in~$\hat{H}_{\mathrm{nl}}$,
see Eq.~\eqref{eq:Hnl-general-many-body-form}, 
\begin{equation}
C_{\left(0,1,2\right),\left(1,0,0\right)}^{p=4}\hat{a}_{r}^{\dagger}\left(t\right)\hat{a}_{c}\left(t\right)\hat{a}_{p}^{2}\;,\label{eq:pumped-hnl-term}
\end{equation}
where the mode label order is~$\left(r,c,p\right)$. Moving into
a displaced and rotating frame defined by~$\hat{D}\left(\xi_{p}\right)\isdef\exp\left(\xi_{p}\hat{a}_{p}^{\dagger}-\xi_{p}^{*}\hat{a}_{p}\right)\exp\left(-i\hbar\omega_{P}t\hat{a}_{p}^{\dagger}\hat{a}_{p}\right)$,
where~$\xi_{P}\isdef\epsilon/\left(\omega_{P}-\omega_{p}\right)$,
the~$p$~mode operator is shifted and rotated,~$\hat{a}_{p}\mapsto\hat{a}_{p}\left(t\right)\isdef\hat{D}\left(\xi_{p}\right)^{-1}\hat{a}_{p}\hat{D}\left(\xi_{p}\right)=\left(\hat{a}_{p}+\xi_{p}\right)e^{-i\omega_{P}t}$
\citep{Leghtas2015,Touzard2019}. In this frame, expanding Eq.~\eqref{eq:pumped-hnl-term}
yields the suggestive interaction term
\begin{equation}
C_{\left(0,1,2\right),\left(1,0,0\right)}^{p=4}\hat{a}_{r}^{\dagger}\hat{a}_{c}\xi_{p}^{2}e^{-it\left(\omega_{r}-\omega_{c}-2\omega_{P}\right)}\;,\label{eq:BS-H-raw}
\end{equation}
which is resonantly activated when~$\omega_{r}-\omega_{c}-2\omega_{P}=0$,
see Supplementary Figure~\ref{fig:pumped-circuit}. Under this condition,
this term survives the RWA when deriving the effective Hamiltonian.
We find the effective BS rate by casting Eq.~\eqref{eq:BS-H-raw}
in the canonical BS form, Eq.~\eqref{eq:app:hb beam split}, 
\begin{equation}
\hbar g=C_{\left(0,1,2\right),\left(1,0,0\right)}^{p=4}\xi_{p}^{2}\;.
\end{equation}

\paragraph{General parametrically-activated interaction.}

The procedure illustrated with the BS example generalizes straightforwardly
to the parametric activation of most other interactions and to finding
their rates using Eq.~\eqref{eq:Hnl-general-coefficient}. The procedure
is useful for more complex and even cascaded processes \citep{Mundhada2017,Mundhada2018}
used in dissipation engineering \citep{Kapit2017}. It is reported
that the limit of procedure is typically reached when~$\xi_{P}$
approaches unity, and other activated nonlinear processes become non-negligible
\citep{Verney2019,Lescanne2019,Tripathi2019,Malekakhlagh2020}.

\section{Finite-element electromagnetic-analysis methodology\label{app:FE-sims}}

We detail the EPR methodology for the finite-element (FE) analysis
of the Josephson circuit. In Sec.~\ref{subsec:Modeling-the-Josephson},
we model a Josephson dipole in the FE simulation as a rectangular
sheet with an inductive lumped-element boundary condition with inductance~$L_{j}$.
In Secs.~\ref{app:FE-epr-single-junc} and~\ref{app:FE-multiple-JJ},
we extract the EPR~$p_{mj}$ and~EPR signs~$s_{mj}$ from the result
of an eigenanalysis simulation of the linearized Josephson circuit,
corresponding to~$\hat{H}_{\mathrm{lin}}$. This step completes the
classical analysis part of the EPR method; from here, the Hamiltonian
is fully specified, as described in Secs.~\ref{app:main-derivation}
and~\ref{app:nonlinear-interactions}. The eigenresult also provides
complete information on the dissipation and input-output coupling
of the circuit, as described in Sec.~\ref{app:dissipation}.

These steps, and the calculations detailed in this text, are automated
by the open-source software package \textsc{pyEPR} \footnote{See the \textsc{pyEPR }\citep{pyEPR} code repository at \href{http://github.com/zlatko-minev/pyEPR}{http://github.com/zlatko-minev/pyEPR}},
which we offer to the community.

\subsection{Modeling the Josephson dipole\label{subsec:Modeling-the-Josephson}}

In the FE model of the Josephson circuit, we model a Josephson dipole
as a simple rectangular sheet with a lumped-element boundary condition
\citep{Nigg2012}, see Supplementary Figure~\ref{fig:FE-model-Josephson-dipole}.
The sheet abstracts away the physical layout of the Josephson dipole
and its wiring leads.

\paragraph{Idealization of the deep-sub-wavelength features.}

This idealization of the Josephson dipole is justified in that its
physical size is in the deep-sub-wavelength regime with respect to
the eigenmodes of interest. For example, for the devices described
in the Methods section, the separation between the Josephson dipole
size and the mode wavelengths of interest is approximately five orders
of magnitude. We hence treat a Josephson dipole in the FE model as
lumped-element inductor with inductance~$L_{j}$, given by Eq.~\eqref{eq:Ej-lin-defn};
the linearization is taken with respect to the circuit equilibrium,
see Sec.~\ref{app:equilibrium-point-of-circuit}.

\paragraph{Electromagnetic model of the lumped inductance.}

The~$j$-th Josephson dipole is modeled as a two-dimensional sheet~$S_{j}$,
see Supplementary Figure~\ref{fig:FE-model-Josephson-dipole}. The
sheet is assigned a surface-impedance boundary condition, which imposes~$\vec{E}_{\parallel}=Z_{s}(\hat{n}\times\vec{H}_{\parallel})$
across the sheet, where~$\vec{E}_{\parallel}$ and~$\vec{H}_{\parallel}$
are the tangential electric and magnetic fields of the sheet, respectively,~$\hat{n}$
is the sheet normal vector, and~$Z_{s}$ is the complex-valued surface
impedance corresponding to a total sheet inductance of~$L_{j}$.
The hat symbol over~$n$ denotes a unit vector in the context of
electromagnetic fields; it is not to be confused with the hat notation
used for quantum operators.

\paragraph{Reducing the model complexity: ignoring leads.}

If the geometric inductance of the wire leads connecting a Josephson
dipole to larger distributed surfaces (such as the pads of a transmon
qubit) is negligible in comparison to~$L_{j}$ and the wire leads
are deeply sub-wavelength in size, then the wire leads can be ignored,
as depicted in Supplementary Figure~\ref{fig:FE-model-Josephson-dipole}.
If the wire leads have non-negligible inductance, then they can either
be included in the simulation or their linear inductance can be included
in the definition of the Josephson dipole. In practice, it is preferable
to be able to abstract away the wire leads as much as possible, since
their feature sizes are typically orders of magnitude smaller than
other all other design features. The inclusion of such fine detail
in the model can result in very large geometric aspect ratios, which
in turn result in increased computational costs. However, while this
finer detail is more representative of the physical design, we have
found that it does not lead to noticeable corrections for typical
cQED devices.

\begin{figure}[th]
\centering{}\includegraphics[width=3.375in]{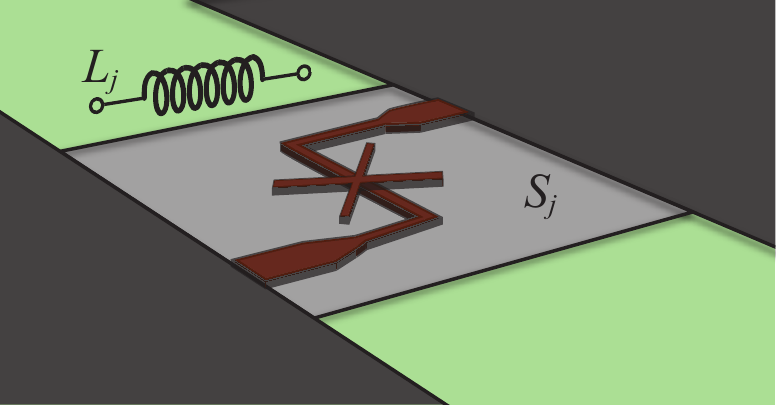} \caption{\label{fig:FE-model-Josephson-dipole}\textbf{Illustration (not-to-scale)
of the representation of a Josephson dipole in a finite-element (FE)
simulation model. }The two large grey rectangles on the edges of the
illustration depict metal pads of a transmon qubit. The light-grey
sheet~$S_{j}$ in the center depicts the sheet used to model the
Josephson dipole (and potentially its leads) in the FE model. The
sheet is assigned a lumped-element inductive boundary condition, with
inductance~$L_{j}$ (sheet inductance symbolized by floating inductor-element
symbol), corresponding to the Josephson dipole inductance with respect
to the operating equilibrium point, see Sec.~\ref{subsec:The-biased-Josephson-simplications}.
The structural details of the Josephson dipole (location marked by
brown cross) and its lead wires (brown wires connected to cross) can
be abstracted away. }
\end{figure}

\paragraph{Performance tip: mesh operations.}

The sheet~$S_{j}$ is generally one of the smallest features of the
FE model. Due to this small size but critical role, one can gently
seed a higher level of mesh on~$S_{j}$ to speed up the eigenanalysis.
However, caution should be used to avoid seeding too heavy of an initial
mesh, which can instead lead to poor convergence of the simulation.
Convergence can be verified by plotting~$p_{mj}$ as a function of
the simulation adaptive pass number and by verifying the conditions
detailed in Sec.~\ref{app:universal-constraints-epr}.

\subsection{Calculating the EPR~$p_{m}$ in the case of a single Josephson dipole\label{app:FE-epr-single-junc}}

If a Josephson circuit incorporates exactly one Josephson dipole,
then we use the global electric and magnetic eigenmode energies to
directly calculate the EPR~$p_{m}$ of the dipole in the mode.

\paragraph{Energy balance.}

The time-averaged electromagnetic energy in a resonantly excited mode
is equally split into an inductive~$\mathcal{E}_{\mathrm{ind}}$
and capacitive~$\mathcal{E}_{\mathrm{cap}}$ contribution \citep{Pozar}.
This detailed balance,~$\mathcal{E}_{\mathrm{ind}}=\mathcal{E}_{\mathrm{cap}}$,
holds even in the presence of dissipation and defines the eigenmode
condition. In the presence of a Josephson dipole, the inductive energy
is split into a magnetic~$\mathcal{E}_{\mathrm{mag}}$ and a kinetic~$\mathcal{E}_{\mathrm{kin}}$
contribution; $\mathcal{E}_{\mathrm{ind}}=\mathcal{E}_{\mathrm{mag}}+\mathcal{E}_{\mathrm{kin}}$.
The magnetic contribution is associated with magnetic fields and geometric
inductance. The kinetic contribution is associated with the Josephson
dipole kinetic inductance and the flow of electrons, and their inertia.
From the point of view of the FE analysis, the magnetic energy is
stored in the magnetic eigenfields~$\vec{H}_{m}$ and the kinetic
energy is stored in the lumped-element boundary condition on~$S_{j}$.
If lumped-element capacitive boundary conditions are absent from the
model, then the capacitive eigenmode energy is stored entirely in
electric eigenfields~$\vec{E}_{m}$, $\mathcal{E}_{\mathrm{cap}}=\mathcal{E}_{\mathrm{elec}}$;
hence, 
\begin{equation}
\mathcal{E}_{\mathrm{elec}}=\mathcal{E}_{\mathrm{cap}}=\mathcal{E}_{\mathrm{ind}}=\mathcal{E}_{\mathrm{mag}}+\mathcal{E}_{\mathrm{kin}}\;.\label{eq:energy-balance}
\end{equation}

\paragraph{Calculating EPR from the energy balance.}

Using Eq.~\eqref{eq:energy-balance} and Eq.~(5) of the main text,~$p_{m}=\mathcal{E}_{\mathrm{kin}}/\mathcal{E}_{\mathrm{ind}}$,
the EPR is calculated from the ratio of global energy quantities,
\begin{equation}
\boxed{p_{m}=\frac{\mathcal{E}_{\mathrm{elec}}-\mathcal{E}_{\mathrm{mag}}}{\mathcal{E}_{\mathrm{elec}}}\;,}
\end{equation}
where the total magnetic- and electric-field energies are computed
from the eigenfields phasors, 
\begin{eqnarray}
\mathcal{E}_{\mathrm{elec}} & = & \frac{1}{4}\mathrm{Re}\int_{V}\vec{E}_{\text{max}}^{*}\overleftrightarrow{\epsilon}\vec{E}_{\text{max}}\,\mathrm{d}v\;,\label{eq:E_elec-fields}\\
\mathcal{E}_{\mathrm{mag}} & = & \frac{1}{4}\mathrm{Re}\int_{V}\vec{H}_{\text{max}}^{*}\overleftrightarrow{\mu}\vec{H}_{\text{max}}\,\mathrm{d}v\;,\label{eq:E_mag-fields}
\end{eqnarray}
where~$\vec{E}_{\mathrm{max}}(x,y,z)$ {[}resp.,~$\vec{H}_{\mathrm{max}}(x,y,z)${]}
is the eigenmode electric (resp., magnetic) phasor, and~$\overleftrightarrow{\epsilon}$
(resp.,~$\overleftrightarrow{\mu}$) denotes the electric-permittivity
(resp., magnetic-permeability) tensor. The spatial integrals are performed
over total volume~$V$ of the device. The electric eigenfield is
related to the phasor by 
\begin{equation}
\vec{E}\left(x,y,z,t\right)=\mathrm{Re}\left[\vec{E}_{\mathrm{max}}\left(x,y,z\right)e^{i\omega_{m}t}\right]\;.
\end{equation}

\begin{figure}[ht]
\centering{}\includegraphics[width=3.375in]{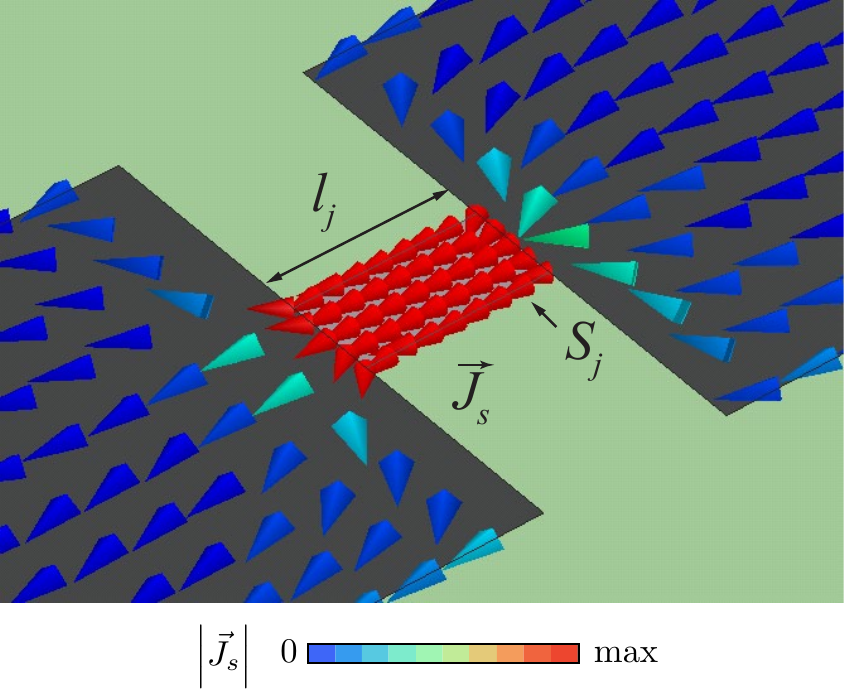} \caption{\label{fig:FE-surf-current-transmon}\textbf{Model of a transmon qubit
}overlaid with the surface-current eigendensity~$\vec{J}_{s}$$\left(\vec{r}\right)$
of the qubit eigenmode, obtained using a FE simulation. Transmon pads
depicted by the two large gray rectangles, separated by distance~$l_{j}$.
Small center rectangle represents the sheet model~$S_{j}$ of the
Josephson junction.}
\end{figure}

\subsection{Calculating the EPR~$p_{mj}$ in the case of multiple Josephson
dipoles\label{app:FE-multiple-JJ}}

\paragraph{EPR~$p_{mj}$ of the~$j$-th Josephson dipole.}

In the case of multiple Josephson dipoles, the total kinetic energy~$\mathcal{E}_{\mathrm{kin}}$
is itself split among the~$J$ dipoles, see Eq.~\eqref{eq:E_kinetic}.
We calculate the EPR~$p_{mj}$ of junction~$j$ in mode~$m$ using
the EPR definition, Eq.~\eqref{eq:pmj-defn-app}, 
\begin{equation}
p_{mj}=\frac{1}{2}L_{j}I_{mj}{}^{2}/\mathcal{E}_{\mathrm{ind}}\;,\label{eq:FE-pmj}
\end{equation}
where~$I_{mj}$ is the peak value of the Josephson dipole current
in mode~$m$. The current~$I_{mj}$ is calculated from the integral
of the mode surface-current density~$\vec{J}_{s,m}\left(x,y,z\right)$
over the dipole sheet~$S_{j}$, 
\begin{equation}
\left|I_{mj}\right|=l_{j}^{-1}\int_{S_{j}}\left|\vec{J}_{s,m}\right|\,\mathrm{d}s\;,\label{eq:FE-ImjAbs}
\end{equation}
where~$l_{j}$ is the length of the sheet, see Supplementary Figure~\ref{fig:FE-surf-current-transmon}.

\paragraph{EPR sign.}

The EPR sign is calculated from the orientation of the current~$I_{mj}$.
The absolute orientation is relative, as described in the main text,
and is determined by defining a directed line~$\mathrm{DL}_{j}$
across~$S_{j}$. If the phasor~$\vec{J}_{s,m}$ is aligned with~$\mathrm{DL}_{j}$
then~$s_{mj}=+1$; otherwise,~$s_{mj}=-1$. Hence, we extract the
sign using
\begin{equation}
s_{mj}=\sign\int_{\mathrm{DL}_{j}}\vec{J}_{s,m}\cdot\mathrm{d}\vec{l}\;.\label{eq:Smj-calc}
\end{equation}
 The direction of the line merely establishes a convention for a positive
EPR sign.

\paragraph{Enforcing energy balance.}

The convergence of~$p_{mj}$, a quantity extracted from the local
eigenfield solutions, can be enhanced by re-normalizing the set of
mode EPR to ensure energy balance, see Eq.~\eqref{eq:energy-balance}.
If lumped-element capacitive boundary conditions are absent, then
the EPR can be renormalized to ensure that their total sum is equal
to the ratio~$\sum_{j=1}^{J}p_{mj}=\mathcal{E}_{\mathrm{kin}}/\mathcal{E}_{\mathrm{ind}}$,
which is a globally calculated quantity, and is therefore expected
to converge quicker.

The above calculations are automated by \textsc{pyEPR} \citep{Note1}.

\subsection{Remarks on the finite-element eigenmode approach}

\paragraph{Finding the eigenfrequencies.}

Rather than searching for the location of an unknown pole in an impedance
response of the circuit and then honing in on it to perform finer
sweeps, the eigenmode analysis returns the lowest~$M$ modes above
a minimum frequency of interest. This typically lifts the requirement
for a prior knowledge of the mode frequencies.

\paragraph{Single simulation.}

The FE eigenmode performs a single simulation from which complete
information of the Josephson circuit is extracted. This is in contrast
to the typical process flow used in an impedance analysis, which requires
that mode frequencies be first identified so that a series of individual
narrow-frequency-range impedance-response sweeps (one for each mode
and junction) can be performed.

\paragraph{Closed-circuit optimization.}

We have found that the above two feature (see remarks) can speed up
the iterative refinement of a quantum design and can circumvents difficulties
associated with finding and fitting unknown, narrow-line poles in
an impedance analysis.

\section{Dissipation budget and input-output coupling\label{app:dissipation}}

In this section, we summarize the methodology used to fully characterize
dissipation and input-output coupling in the Josephson system. Loss
of energy results from material losses and radiative boundaries which
guide energy away from the system. Additionally, control of the system
is achieved by means of the radiative boundaries, such as input-output
(\emph{I-O}) coupling. The dissipation budget, comprising the individual
loss-contribution bound of each lossy and radiative element, is extracted
from the same eigensolution used to calculate the EPR~$p_{mj}$,
in essentially the same way\textemdash by calculating the fraction
of the~$m$-th mode energy stored in the~$l$-th lossy element\textemdash the
lossy EPR~$p_{ml}$. While lossy EPR signs~$s_{ml}$ can also be
calculated for each element, these are not needed for linear dissipation;
however, their role in \emph{I-O} coupling is detailed in Sec.~\ref{app:dissipation:radiative}.

\subsection{Dissipation budget\label{subsec:Breakdown-of-dissipation}}

The dissipation budget comprises the estimated energy-loss-rate contributions
due to each loss mechanism and each lossy object in the Josephson
circuit \citep{Pozar,Gao2008,Wang2009,Wenner2011,Zmuidzinas2012,Geerlings2013,Wang2015,Reagor2016,Gambetta2017a,McRae2020}.
We classily losses as capacitive, inductive, or radiative, and summarize
their calculation here. Each loss mechanism is described by a corresponding
lossy EPR~$p_{ml}$ and an intrinsic quality factor~$Q$. Assuming
the dissipation is linear and the mode of interest is underdamped,
the loss rates of each mechanism simply add; so, the total quality
factor of the~$m$-th mode is \citep{Zmuidzinas2012,Geerlings2013,Reagor2016}
\begin{equation}
\frac{1}{Q_{\text{total}}}=\frac{1}{Q_{\text{cap}}}+\frac{1}{Q_{\text{ind}}}+\frac{1}{Q_{\text{rad}}}\;,
\end{equation}
where~$Q_{\mathrm{cap}}$ and~$Q_{\mathrm{ind}}$ are the total
mode quality factors due to capacitive and inductive losses (i.e.,
those proportional to the intensity of the electric~$\left|\vec{E}\right|^{2}$
and magnetic field~$\left|\vec{H}\right|^{2}$, respectively, see
Secs.~\ref{app:capacitive-losses} and~\ref{subsec:Inductive-losses}),
and~$Q_{\mathrm{rad}}$ is the total radiative mode quality factor,
see Sec.~\ref{app:dissipation:radiative}. In this section, the mode
index~$m$ is implicit.

\paragraph{Remark.}

Beyond these intrinsic mechanisms, extrinsic factors, such as ionizing
radiation, can play a significant role in understanding loss mechanisms,
such as those due to quasiparticle in superconducting quantum circuits
\citep{Vepsalainen2020,Cardani2020}.

\subsection{Capacitive loss\label{app:capacitive-losses}}

The total capacitive mode quality~$Q_{\mathrm{\ensuremath{cap}}}$
is the weighted sum of intrinsic quality factors~$Q_{l}^{\text{cap}}$
(i.e., the inverse of the dielectric loss tangent) of all lossy dielectrics~$l$
in the Josephson circuit \citep{Zmuidzinas2012,Geerlings2013}, 
\begin{equation}
\frac{1}{Q_{\text{cap}}}=\sum_{l}\frac{p_{ml}^{\text{cap}}}{Q_{l}^{\text{cap}}}\;,
\end{equation}
where~$p_{ml}^{\text{cap}}$ is the lossy energy-participation ratio
of the~$l$-th dielectric in the mode\textemdash i.e.,~$p_{ml}^{\text{cap}}$
is the fraction of capacitive energy stored in the dielectric element~$l$.
We classify lossy capacitive elements as either bulk or surface. For
example, the volume of three-dimension dielectric object, such as
a chip substrate, is associated with bulk capacitive loss \citep{Martinis2014,Dial2016,Kamal2016-anneal}.
On the other hand, the surface of the substrate, which may be a surface-dielectric
layer is classified as a surface-loss element \citep{Martinis2014,Wang2015}.
The lossy EPR for bulk capacitive loss is calculated from the eigenfield
solutions, 
\begin{equation}
p_{ml}^{\text{cap}}=\frac{1}{\mathcal{E}_{\mathrm{elec}}}\frac{1}{4}\Re\int_{V_{l}}\vec{E}_{\text{max}}^{*}\overleftrightarrow{\epsilon}\vec{E}_{\text{max}}\,\mathrm{d}v\;,
\end{equation}
where the integral is carried over the volume~$V_{l}$ of the~$l$-th
bulk dielectric object; the total electric energy~$\mathcal{E}_{\mathrm{elec}}$
is defined in Eq.~\eqref{eq:E_elec-fields}. The lossy EPR for a
surface dielectric is approximated by 
\begin{equation}
p_{ml}^{\text{cap,surf}}=\frac{1}{\mathcal{E}_{\mathrm{elec}}}\frac{t_{l}\epsilon_{l}}{4}\Re\int_{\text{surf}_{l}}\abs{\vec{E}_{\text{max}}}^{2}\,\mathrm{d}s\;,
\end{equation}
where the surface-dielectric layer thickness is~$t_{l}$ and its
permittivity is~$\epsilon_{l}$.

\subsection{Inductive loss\label{subsec:Inductive-losses}}

Physically, inductive losses originate from the dissipative flow of
electrical current in metals or through metal-metal seams. Supercurrent
loss due to quasiparticles and vortices can be accounted for in an
effective quality factor of the conducting surface. We denote the
intrinsic inductive quality factor of a lossy object~$Q_{l}^{\mathrm{ind}}$.
The bound on the total inductive-loss quality factor~$Q_{\text{ind}}$
of mode~$m$ is a weighted sum of~$Q_{l}^{\mathrm{ind}}$, 
\begin{equation}
\frac{1}{Q_{\text{ind}}}=\sum_{l}{\frac{p_{ml}^{\mathrm{ind}}}{Q_{l}^{\mathrm{ind}}}}\;,
\end{equation}
where~$p_{ml}^{\mathrm{ind}}$ is the lossy EPR for inductive element~$l$.
We classify inductive losses as either surface, bulk, or seam.

\paragraph{Surface conductive loss (in the skin-depth).}

The fraction of eigenmode energy stored in the skin depth~$\lambda_{0}$
of metal surface~$l$, denoted~$\mathrm{surf}_{l}$, if the lossy
EPR of the surface, and is obtained from the eigenfield solutions,
\begin{equation}
p_{ml}^{\text{ind,surf}}=\frac{1}{\mathcal{E}_{\mathrm{mag}}}\frac{\lambda_{0}\mu_{l}}{4}\Re\int_{\text{surf}_{l}}\abs{\vec{H}_{\text{max},\parallel}}^{2}\,\mathrm{d}s\;,\label{eq:appx:dissip:p ind surf}
\end{equation}
where~$\mu_{l}$ is the magnetic permeability of the surface; typically,~$\mu_{l}=\mu_{0}$,
and~$\vec{H}_{\text{max},\parallel}$ is the magnetic field phasor
parallel to the surface. For superconductors,~$p_{ml}^{\text{ind,surf}}$
is the \emph{kinetic inductance fraction} \citep{Gao2008,Zmuidzinas2012},
commonly denoted~$\alpha$. The total magnetic energy~$\mathcal{E}_{\mathrm{mag}}$
is defined in Eq.~\eqref{eq:E_mag-fields}.

\paragraph{Surface conductive loss: intrinsic quality.}

In the normal state, a metal, such as copper or aluminum, has an intrinsic
inductive quality factor of order unity \citep{Pozar},~$Q_{l}^{\text{ind,surf}}\approx1$.
However, in the superconducting state, the lower bound on the quality
factor is typically found to be in the range of several thousand \citep{Reagor2013},~$Q_{l}^{\text{ind,surf}}\gtrsim10^{3}$.
The lower bound for thin-film superconducting aluminum has been measured
to exceed~$Q_{l}^{\text{ind,surf}}>10^{5}$ \citep{Minev2013}.

\paragraph{Bulk magnetic loss.}

The mode magnetic field can couple to bulk magnetic impurities present
in the volume~$V_{l}$ of lossy object~$l$. The lossy EPR for the
bulk-inductive-loss mechanism of volume~$V_{l}$ is 
\begin{equation}
p_{ml}^{\mathrm{ind,bulk}}=\frac{1}{\mathcal{E}_{\mathrm{mag}}}\frac{1}{4}\Re\int_{V_{l}}\vec{H}_{\text{max}}^{*}\overleftrightarrow{\mu}\vec{H}_{\text{max}}\,\mathrm{d}v\;.
\end{equation}
This coupling is typically negligible in current superconducting quantum
circuits.

\paragraph{Seam loss.}

The seam formed by pressing two metals together provides an electrical
bridge for the current to flow across but also introduces a key dissipation
mechanism \citep{Brecht2015}. A common example of a seam is the one
formed by the two halves of the metal sample holders used to house
a cQED chip. In the FE analysis, we model the seam by a line path~$\text{seam}_{l}$
tracing out the location of the seam at the surface of the two mating
walls. The inductive lossy EPR participation for the seam is 
\begin{equation}
p_{ml}^{\text{ind,seam}}=\frac{1}{\mathcal{E}_{\mathrm{mag}}}\frac{\lambda_{0}t_{l}\mu_{l}}{4}\Re\int_{\text{seam}_{l}}\abs{\vec{H}_{\text{max},\perp}}^{2}\,\mathrm{d}l\;,
\end{equation}
where the seam thickness is denoted~$t_{l}$, its magnetic permeability~$\mu_{l}$,
and its the penetration depth~$\lambda_{0}$. It is convenient to
rewrite the total seam loss due to seam~$l$ in terms of an effective
seam admittance~$g_{\text{seam}}$, defined in Ref.~\onlinecite{Brecht2015},
\begin{equation}
\frac{p_{ml}^{\mathrm{ind,seam}}}{Q_{\text{seam}}}=\frac{1}{g_{\text{seam}}}\frac{\int_{\text{seam}_{l}}\abs{\vec{J}_{s}\times\vec{l}}^{2}\,\mathrm{d}l}{\omega\mu_{0}\int_{\text{all}}\abs{H_{\text{max}}}^{2}\,\mathrm{d}V}\;.
\end{equation}

\subsection{Radiative loss and input-output coupling\label{app:dissipation:radiative}}

\begin{figure}
\centering{}\includegraphics[width=3.375in]{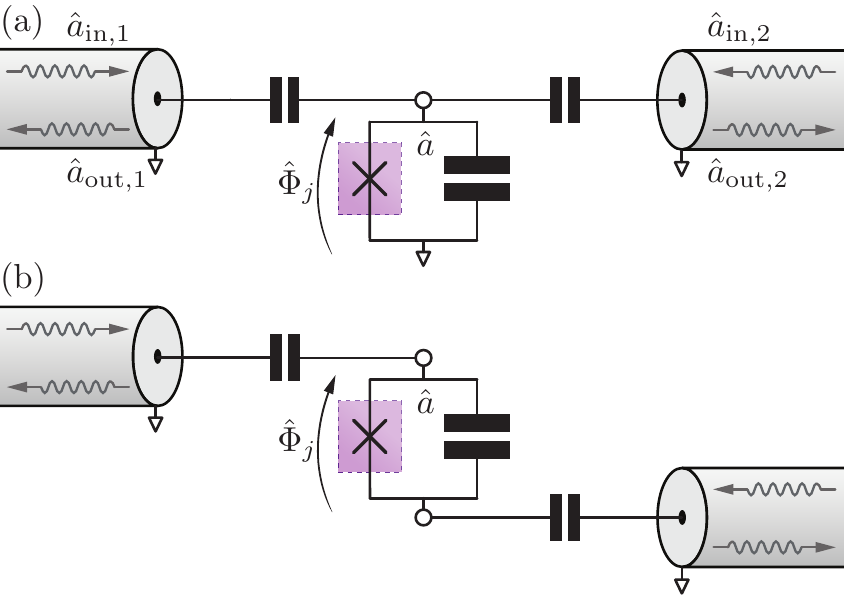}
\caption{\label{fig:port-signs}\textbf{Schematic of a transmon-qubit circuit}
(mode operator~$\hat{a}$), comprising a Josephson tunnel junction
(flux operator~$\hat{\Phi}_{j}$) and a capacitor, coupled to two
input-output ports. The input and output fields of the left (resp.,
right) transmission line are~$\hat{a}_{\mathrm{in},1}$ and~$\hat{a}_{\mathrm{out},1}$
(resp., $\hat{a}_{\mathrm{in},2}$ and~$\hat{a}_{\mathrm{out},2}$).
(a) Ports~1 and~2 both coupled to the top qubit node; hence, both
port EPR signs~$s_{mp}$ are equal,~$s_{q1}=s_{q2}$. (b) Port~1
couples to the top node, but port~2 couples to the bottom qubit node.
The port EPR signs~$s_{mp}$ have opposite signs, $s_{q1}=-s_{q2}$.}
\end{figure}

The Josephson circuit incorporates radiative boundaries, typically
purposefully introduced to serve as input-output ports of the circuit
\citep{Pozar}, thus providing a means to perform system measurement
and control. For cQED devices, the port exposes the circuit to an
external transmission line conduit, such as a coaxial cable or a co-planar
waveguide. The total radiative quality factor~$Q_{\text{rad}}$ of
mode~$m$ is the sum of the individual port contributions~$Q_{\text{rad}}^{-1}=\sum_{p=1}^{P}Q_{mp}^{-1}$,
where~$P$ is the total number of ports and~$Q_{mp}$ is the quality
factor due to port~$p$. Below, we describe how port~$p$ is modeled
in the FE simulation to extract~$Q_{mp}$ from the eigensolutions.

\paragraph{Radiative energy loss.}

Energy stored in mode~$m$ can leak at a rate~$\kappa_{mp}$ through
port~$p$ and be guided away by the transmission line. While for
certain modes, such as readout ones, this coupling is desired, for
other modes, such as a qubit one, this coupling is often considered
spurious. In the case of a qubit mode, the energy loss to a readout
port is seen as a manifestation of the Purcell effect \citep{Houck2008-Purcell};
while, for the readout mode of the same structure, the energy loss
to the readout port sets the rate of information gain \citep{Clerk2010}.
The rate~$\kappa_{mp}$ is calculated from port EPR~$p_{mp}$, as
detailed in the following for both wanted and spurious terms. The
port EPR sign~$s_{mp}$ is calculated concurrently and is important
for the system drive configuration.

\paragraph{FE model of the port.}

In the presence of a port, the boundary of the Josephson circuit is
somewhat ambiguous\textemdash we can include more or less of the port
and conduit structure in the model. We choose to include a minimal
but sufficiently large portion of these to faithfully model the disturbing
effect of the boundary condition on the eigenmodes. For example, in
the case of the qubit-cavity structure of Figure~\ref{fig:ConceptFig}(a)
of the main text, we include a short stub of the \emph{I-O} coaxial
cable in the FE model. The length of the stub can be determined from
the effect of a sweep of its length on the target parameters. We have
found that an alternative heuristic measure is to use the decay of
the eigenfields inside the port structure and to make sure that the
end of the port structure is at least several exponential decay lengths
long (the field in the port structure decays exponentially since the
eigenmodes are generally below its cutoff). The port structure termination
surface~$S_{p}$ is treated as a resistive sheet with to model the
effect of the transmission line guiding waves away from the Josephson
circuit. The sheet effective resistance is~$R_{p}$. In the case
of a 50~$\Omega$ port line,~$R_{p}=50\,\Omega$. While a resistive
boundary condition can be assigned to the sheet, in the case of high-quality
modes, it is possible to use a perturbative approach and to perform
a lossless simulation of the FE model, from which the loss can be
extracted (see also remark at end of this section).

\paragraph{Calculating the input-output (I-O) coupling rate~$\kappa_{mp}$
from the port EPR~$p_{mp}$.}

From the lossless eigensolutions of mode~$m$, the energy lost to
the effective resistor~$R_{p}$ of port~$p$ during one mode oscillation
period~$T_{m}=2\pi/\omega_{m}$ is 
\begin{equation}
P_{mp}=\frac{1}{2}R_{p}I_{mp}^{2}T_{m}\;,\label{eq:app:p_mp}
\end{equation}
where~$I_{mp}$ is the peak current across the port due to the excitation
of mode~$m$; see Eq.~\eqref{eq:FE-ImjAbs}. The total mode energy~$\mathcal{E}_{m}\left(t\right)$
at time~$t$ decays at rate 
\begin{equation}
\frac{d}{dt}\mathcal{E}_{m}=-\sum_{p}{\kappa_{mp}}\mathcal{E}_{m}\;.
\end{equation}
Hence, assuming a high quality mode, the energy loss during one oscillation
period, between times~$t=0$ and~$T_{m}$, is 
\begin{equation}
\mathcal{E}_{m}(T_{m})=\mathcal{E}_{m}(0)-\sum_{p}{T_{m}\kappa_{mp}\mathcal{E}_{m}(0)}\;.\label{eq:Em Tm}
\end{equation}
This shows that the loss to port~$p$ during one period is~$P_{mp}=T_{m}\kappa_{mp}\mathcal{E}_{m}(0)$,
which we equate to the expression of Eq.~\eqref{eq:app:p_mp} to
find the \emph{I-O} coupling rate in terms of quantities calculated
from the eigenfields, 
\begin{equation}
\kappa_{mp}=\frac{\frac{1}{2}RI_{mp}^{2}}{\mathcal{E}_{m}(0)}\;.
\end{equation}
Thus, the mode coupling quality factor is 
\begin{equation}
Q_{mp}\isdef\omega_{m}/\kappa_{mp}=\frac{\omega_{m}\mathcal{E}_{m}(0)}{\frac{1}{2}RI_{mp}^{2}}\;.
\end{equation}

\paragraph{\emph{Sign of the I-O participation.}}

If there are multiple ports, the port EPR sign~$s_{mp}$, calculated
using Eq.~\eqref{eq:Smj-calc}, is important in a manner similar
to that explained for the case of the EPR sign~$s_{mj}$ in the case
of multiple Josephson dipoles; see the main text. To illustrate,
consider a simple transmon-qubit circuit coupled to two transmission
lines in two different ways as depicted in the two panels of Supplementary
Figure~\ref{fig:port-signs}. While in both configuration the eigenmode
frequency and quality factor is identical, the configurations are
inequivalent. Consider driving both transmission lines with the same
amplitude and in-phase, then the circuit of Supplementary Figure~\ref{fig:port-signs}(a)
is excited but that of Supplementary Figure~\ref{fig:port-signs}(b)
is not.

\paragraph{Remark on modeling the port termination as resistive vs.~lossless.}

The port sheet~$S_{p}$ can be treated as either resistive or lossless.
In the former, the sheet is assigned a lumped-element boundary condition
with impedance matching the port input impedance as seen from the
system; typically designed to be~$R_{p}=50\,\Omega$ \citep{Malekakhlagh2016-A2,Malekakhlagh2017-Cutoff-Free}.
The eigenresults fully account for the effect of the dissipation on
the mode profile. These effects are negligible in the case of high-quality
modes, but become significant as~$Q_{m}$ approaches unity. For lossless
treatment of~$S_{p}$, suitable when~$Q_{m}\gg1$, the termination
simply assigned a perfectly conducting boundary condition. In both
treatment, the eigenmode fields can be used to calculate the loss
due to~$R_{p}$. 

\paragraph{Reactive ports.}

In addition to being resistive, ports can have a reactive component,
typically occurring in the presence of non-idealities. For example,
a port coupled to transmission line suffering from a down-line reflection
will have some of the energy it leaks out to the line come back to
it \citep{Burkhart2020}. The reactive part of the port structure
thus houses energy in its internal modes. There can be treated by
modifying the boundary condition on~$S_{p}$ or for a more general
treatment can be accounted for by including the port, line, and scatterer
structure in the FE model. This larger model will account for hybridization
between the line modes and those of the system \citep{Wang2019-cav-atten}.

\bibliographystyle{naturemag}

\end{document}